\documentclass[9pt,tutorial,pubversion]{livecoms}

\usepackage[T1]{fontenc}
\usepackage[utf8]{inputenc}
\usepackage{lmodern}
\usepackage{verbatim}
\usepackage{graphicx}
\usepackage{amsmath}
\usepackage{amssymb}
\usepackage{amsthm}
\usepackage{tabularx}
\usepackage{multirow}
\usepackage{multicol}
\usepackage{fancyvrb}
\usepackage{float}
\usepackage{listings}
\usepackage{xcolor}
\usepackage{array}
\usepackage{booktabs}
\usepackage{times}
\usepackage{subcaption}
\usepackage{mathtools}
\usepackage{menukeys}
\usepackage{microtype}
\usepackage{tcolorbox}
\usepackage{newverbs}
\usepackage[title]{appendix}
\usepackage{csquotes}
\usepackage{hyperref}

\usepackage[
    type={CC},
    modifier={by-nc-sa},
    version={4.0},
]{doclicense}

\usepackage{wrapfig}
\usepackage{geometry}

\usepackage[version=4]{mhchem}
\usepackage{siunitx}
\DeclareSIUnit\Molar{M}
\usepackage[italic]{mathastext}
\graphicspath{{figures/}}

\definecolor{listing}{rgb}{0.95,0.95,0.95}
\definecolor{command}{rgb}{0.15,0.15,0.15}
\definecolor{download}{rgb}{0.039,0.616,0.851}
\definecolor{note}{rgb}{1,0.988,0.93}
\definecolor{note_listing}{rgb}{0.94,0.925,0.86}

\newtcolorbox{note}{
  colback=note, 
  colframe=note, 
  boxrule=0pt, 
  sharp corners, 
  left=5pt, 
  right=5pt, 
  top=5pt, 
  bottom=5pt, 
}

\DisableLigatures{encoding = *, family = *} 

\newcommand{\lmpcmd}[1]{\colorbox{listing}{\textcolor{command}{\small{#1}}}} 
\newcommand{\lmpcmdnote}[1]{\colorbox{note_listing}{\textcolor{command}{\small{#1}}}} 
\newcommand{\flecmd}[1]{\textcolor{command}{\texttt{#1}}} 
\newcommand{\guicmd}[1]{\textcolor{command}{\texttt{«#1»}}} 
\newcommand{\dwlcmd}[1]{\textcolor{download}{\texttt{#1}}} 

\newcommand{\lammpsgui}{\textsf{LAMMPS\textendash GUI}}

\newcommand{\versionnumber}{1.0}  
\newcommand{\githubrepository}{\url{https://github.com/lammpstutorials/lammpstutorials-article}}  

\newcommand{\filepath}{https://raw.githubusercontent.com/lammpstutorials/lammpstutorials-article/main/files/}

\title{A Set of Tutorials for the LAMMPS Simulation Package [Article v\versionnumber]}

\author[1*]{Simon Gravelle}
\affil[1]{University Grenoble Alpes, CNRS, LIPhy, Grenoble, 38000, France}
\corr{simon.gravelle@cnrs.fr}{S.G.}
\orcid{Simon Gravelle}{0000-0003-2149-6706}

\author[2]{Cecilia M.~S.~Alvares}
\affil[2]{Department of Chemistry, University of Warwick, Coventry CV4 7AL, United Kingdom}
\orcid{Cecilia M.~S.~Alvares}{0009-0007-6237-6512}

\author[3]{Jacob R.~Gissinger}
\affil[3]{Stevens Institute of Technology, Hoboken, NJ 07030, USA}
\orcid{Jacob R.~Gissinger}{0000-0003-0031-044X}

\author[4]{Axel Kohlmeyer}
\affil[4]{Institute for Computational Molecular Science, Temple University, Philadelphia, PA 19122, USA}
\orcid{Axel Kohlmeyer}{0000-0001-6204-6475}

\blurb{This LiveCoMS document is maintained online on GitHub at
  \githubrepository; to provide feedback, suggestions, or help improve
  it, please visit the GitHub repository and participate via the issue
  tracker.}

\pubDOI{10.33011/livecoms.6.1.3037}
\pubvolume{6}
\pubissue{1}
\pubyear{2025}
\articlenum{3037}
\datereceived{18 March 2025}
\dateaccepted{29 September 2025}

\begin{document}

\begin{frontmatter}
\maketitle

\begin{abstract}
  The availability of open-source molecular simulation software packages
  allows scientists and engineers to focus on running and analyzing
  simulations without having to write, parallelize, and validate their
  own simulation software.  While molecular simulations thus become
  accessible to a larger audience, the ``black box'' nature of such
  software packages and wide array of options and features can make it
  challenging to use them correctly, particularly for beginners in the
  topic of simulations.  LAMMPS is one such versatile molecular
  simulation code, designed for modeling particle-based systems across a
  broad range of materials science and computational chemistry
  applications, including atomistic, coarse-grained, mesoscale,
  grid-free continuum, and discrete element models.  LAMMPS is capable
  of efficiently running simulations of varying sizes from small desktop
  computers to large-scale supercomputing environments.  Its flexibility
  and extensibility make it ideal for complex and extensive simulations
  of atomic and molecular systems, and beyond.  This article introduces
  a suite of tutorials designed to make learning LAMMPS more accessible
  to new users.  The first four tutorials cover the basics of running
  molecular simulations in LAMMPS with systems of varying complexities.
  The second four tutorials address more advanced molecular simulation
  techniques, specifically the use of a reactive force field, grand
  canonical Monte Carlo, enhanced sampling, and the REACTER protocol.
  In addition, we introduce \lammpsgui{}, an enhanced cross-platform
  graphical text editor specifically designed for use with LAMMPS and
  able to run LAMMPS directly on the edited input.  \lammpsgui{} is used
  as the primary tool in the tutorials to edit inputs, run LAMMPS,
  extract data, and visualize the simulated systems.
\end{abstract}

\end{frontmatter}

\section{Introduction}

Molecular simulations can be used to model a large variety of
atomic and coarse-grained systems, including solids, fluids, polymers,
and biomolecules, as well as complex interfaces and multi-component
systems.  While various molecular modeling methods exist, molecular dynamics (MD) and
Monte Carlo (MC) are most commonly used.  MD is the preferred method for
obtaining the accurate dynamics of a system, as it relies on solving
Newton's equations of motion.  For systems with many degrees of freedom
or complex energy landscapes, the MC method can be a better choice than
MD because it allows for efficiently exploring a configuration space
without being confined by the accessible time scale.  MC involves
performing random changes to the system configuration that are either
accepted or rejected based on energy criteria
\cite{frenkel2023understanding, allen2017computer}.  Molecular simulations allow for
measuring a broad variety of properties, including structural properties
(e.g.,~bond length distribution, coordination numbers, radial
distribution functions), thermodynamic properties (e.g.,~temperature,
pressure, volume), dynamic behaviors (e.g.,~diffusion coefficient,
viscosity), and mechanical responses (e.g.,~elastic constant, Poisson
ratio).  Some of these quantities can be directly compared with
experimental data, enabling the validation of the simulated system,
while others, available only through simulations, are often useful for
interpreting experimental data~\cite{van2008molecular}.

LAMMPS (Large-scale Atomic/Molecular Massively Parallel Simulator)
\cite{lammps_home} is a highly flexible and parallel open-source molecular simulation
tool.  Over the years, a broad variety of particle interaction models
have been implemented in LAMMPS, enabling it to model a wide range of
systems, including atomic, polymeric, biological, metallic, reactive, granular,
mesoscale, grid-free continuum, and coarse-grained systems
\cite{thompson2022lammps}.  LAMMPS can be used on a single CPU core, a
multi-socket and multi-core server, an HPC cluster, or a high-end
supercomputing system.  It can efficiently handle complex and large-scale
simulations, including hybrid MPI-OpenMP parallelization
and MPI + GPU acceleration (for a subset of its functionality).

LAMMPS requires users to write detailed input files, a task that can be
particularly challenging for new users.  Although its documentation
extensively describes all available features~\cite{lammps_docs},
navigating it can be challenging.  Much of the information may be
unnecessary for common use cases, and the detailed manual can often feel
overwhelming.  Beyond the intrinsic complexity of LAMMPS, performing
accurate simulations requires several complex, system-specific decisions
regarding the physics to be modeled, such as selecting the thermodynamic
ensemble (e.g.,~microcanonical, grand canonical), determining the
simulation duration, and choosing the sets of parameters describing the
interactions between atoms (the so-called force field)
\cite{wong2016good, van2018validation, prasad2018best}.  While these
choices are independent of the simulation software, they may
occasionally be constrained by the features available in a given
package.  The tutorials in this article aim to flatten the learning
curve and guide users in performing accurate and reliable molecular
simulations with LAMMPS.

\subsection{Scope}

This set of tutorials consists of eight tutorials arranged in order of
increasing difficulty.  Although each tutorial can be
read independently, information introduced in earlier tutorials is
generally not repeated in detail in later ones. For this reason, we
recommend that beginners follow the tutorials in order.  The novelties
associated with each tutorial are briefly described below.

In \hyperref[lennard-jones-label]{Tutorial 1}, the structure of LAMMPS
input files is illustrated through the creation of a simple atomic
Lennard-Jones fluid system.  Basic LAMMPS commands are used to set up
interactions between atoms, perform an energy minimization, and finally
run a simple MD simulation in the microcanonical (NVE) and canonical (NVT)
ensembles.

In \hyperref[carbon-nanotube-label]{Tutorial 2}, a more complex system
is introduced in which atoms are connected by bonds: a small carbon
nanotube.  The use of both classical and reactive force fields (here,
OPLS-AA~\cite{jorgensenDevelopmentTestingOPLS1996} and
AIREBO~\cite{stuart2000reactive}, respectively) is illustrated.  An
external deformation is applied to the CNT, and its deformation is
measured.  This tutorial also demonstrates the use of an external tool
to visualize breaking bonds, and show the possibility to import
LAMMPS-generated YAML log files into Python.

In \hyperref[all-atom-label]{Tutorial 3}, two components\textemdash liquid water
(flexible three-point model) and a polymer molecule\textemdash are merged and
equilibrated.  A long-range solver is used to handle the electrostatic
interactions accurately, and the system is equilibrated in the
isothermal-isobaric (NPT) ensemble; then, a stretching force is applied
to the polymer.  Through this relatively complex solvated polymer
system, the tutorial demonstrates how to use type labels to make
molecule files more generic and easier to manage~\cite{gissinger2024type}.

In \hyperref[sheared-confined-label]{Tutorial 4}, an electrolyte is
confined between two walls, illustrating the specifics of simulating
systems with fluid-solid interfaces.  With the rigid four-point
TIP4P/2005~\cite{abascal2005general} water model, this tutorial uses a
more complex water model than \hyperref[all-atom-label]{Tutorial 3}.  A
non-equilibrium MD is performed by imposing shear on the fluid through
moving the walls, and the fluid velocity profile is extracted.

In \hyperref[reactive-silicon-dioxide-label]{Tutorial 5}, the ReaxFF
reactive force field, which is specifically designed to simulate chemical
reactions by dynamically adjusting atomic interactions
\cite{van2001reaxff}, is used.  ReaxFF includes charge equilibration (QEq), a
method that allows the partial charges of atoms to adjust according to
their local environment.

In \hyperref[gcmc-silica-label]{Tutorial 6}, the adsorption of a fluid in
silica pores is modeled.  To do so, a Monte Carlo simulation in
the grand canonical ensemble is implemented to demonstrate how LAMMPS
can be used to simulate an open system that exchanges particles with a
reservoir.

In \hyperref[umbrella-sampling-label]{Tutorial 7}, an advanced free
energy method called umbrella sampling is implemented.  By calculating
an energy barrier, this tutorial describes a protocol
for addressing energy landscapes that are difficult to sample using
classical MD or MC methods.

In \hyperref[bond-react-label]{Tutorial 8}, a CNT embedded in
nylon-6,6 polymer melt is simulated.  The
REACTER protocol is used to model the polymerization of nylon, and the formation
of water molecules is tracked over time~\cite{gissinger2020reacter}.

\section{Prerequisites}

\subsection{Background knowledge}

This set of tutorials assumes no prior knowledge of the LAMMPS software
itself.  To complete the tutorials, a text editor and a suitable LAMMPS
executable are required.  We use \lammpsgui{}~\cite{lammps_gui_docs}
here, as it offers features that make it particularly convenient for
tutorials, but other console or graphical text editors, such as GNU
nano, vi/vim, Emacs, Notepad, Gedit, and Visual Studio Code can also be
used.  LAMMPS can be executed either directly from \lammpsgui{}
(\hyperref[using-lammps-gui-label]{Appendix~\ref{using-lammps-gui-label}})
or from a command prompt
(\hyperref[command-line-label]{Appendix~\ref{command-line-label}}), the
latter of which requires some familiarity with executing commands from a
terminal or command-line prompt.

In addition, prior knowledge of the theoretical basics of molecular
simulations and statistical physics is highly beneficial.  Users may
refer to textbooks such as \textit{Understanding Molecular Simulation} by
Daan Frenkel and Berend Smit~\cite{frenkel2023understanding}, as well as
\textit{Computer Simulation of Liquids} by Michael Allen and Dominic
Tildesley~\cite{allen2017computer}.  To better understand
the fundamental concepts behind the soft matter systems simulated in these
tutorials, users can also refer to \textit{Basic Concepts for Simple and
  Complex Liquids} by Jean-Louis Barrat and Jean-Pierre Hansen
\cite{barrat2003basic}, as well as \textit{Theory of Simple Liquids:
  with Applications to Soft Matter} by Jean-Pierre Hansen and Ian Ranald
McDonald~\cite{hansen2013theory}.  For more resources, the SklogWiki
platform provides a wide range of information on statistical mechanics
and molecular simulations~\cite{sklogwiki_main_page}.

\subsection{Software/system requirements}

The LAMMPS stable release version 22Jul2025~\cite{lammps_code}
and the matching \lammpsgui{} software version 1.7.0 are required to
follow the tutorials, as they include features that were first
introduced in these versions.  For Linux (x86\_64 CPU), macOS (BigSur or
later), and Windows (10 and 11) you can download a pre-compiled LAMMPS
package from the LAMMPS release page on
GitHub~\cite{lammps_github_release}.  Select a package with `GUI' in the
file name, which includes both, \lammpsgui{} and the LAMMPS command-line
executable.  These pre-compiled packages are designed to be portable, and
therefore omit support for parallel execution with MPI.  Instructions
for installing \lammpsgui{} and using its most relevant features for the
tutorials are provided in
\hyperref[using-lammps-gui-label]{Appendix~\ref{using-lammps-gui-label}}.

LAMMPS versions are generally backward compatible, meaning that older
input files typically work the same with newer versions of LAMMPS.
However, forward compatibility is not as strong, so input files written
for a newer version may not always work with older versions.  As a
result, it is usually possible to follow this tutorial with more recent
releases of \lammpsgui{} and LAMMPS; older versions may require some
(minor) adjustments.  These tutorials will be periodically updated to
ensure compatibility and benefit from new features in the latest stable
version of LAMMPS.

For some tutorials, external tools are required for plotting and
visualization, as the corresponding functionality in \lammpsgui{} is
limited.  Suitable tools for plotting include Python with
Pandas/Matplotlib~\cite{van1995python,hunter2007Matplotlib}, XmGrace,
Gnuplot, Microsoft Excel, or LibreOffice Calc.  For visualization,
suitable tools include VMD~\cite{vmd_home,humphrey1996vmd} and
OVITO~\cite{ovito_home,stukowski2009visualization}.

\subsection{About \lammpsgui{}}

\lammpsgui{} is a graphical text editor, enhanced for editing LAMMPS
input files and linked to the LAMMPS library, allowing it to run LAMMPS
directly.  The text editor is similar to other graphical
editors, such as Notepad or Gedit, but offers the following enhancements
for running LAMMPS:
\begin{itemize}
  \item Wizard dialogs to set up these tutorials
  \item Auto-completion of LAMMPS commands and options
  \item Context-sensitive online help
  \item Syntax highlighting for LAMMPS input files
  \item Syntax-aware line indentation
  \item Editor switches working directory to that of input file
  \item Visualization using LAMMPS' built-in renderer
  \item Start and stop simulations via mouse or keyboard
  \item Monitoring of simulation progress and parallelization
  \item Dynamic capture of LAMMPS output in a text window
  \item Automatic plotting of thermodynamic data during runs
  \item Capture of ``dump image'' outputs for animations
  \item Export of thermodynamic data for external plotting
  \item Inspection of binary restart files
\end{itemize}
\hyperref[using-lammps-gui-label]{Appendix~\ref{using-lammps-gui-label}}
contains basic instructions for installation and using \lammpsgui{} with
the tutorials presented here.  A complete description of all \lammpsgui{}
features can be found in the LAMMPS manual~\cite{lammps_gui_docs}.

\section{Content and links}

The tutorials described in this article can be accessed at
\href{https://lammpstutorials.github.io}{lammpstutorials.github.io},
where additional exercises with solutions are also provided.  All files
and inputs required to follow the tutorials are available from a
dedicated GitHub organization account,
\href{https://github.com/lammpstutorials}{github.com/lammpstutorials}.
These files can also be downloaded by clicking \guicmd{Start LAMMPS Tutorial X}
(where \texttt{X} = 1...8) from the \guicmd{Tutorials} menu of \lammpsgui{}.

In the following, all LAMMPS input or console commands are formatted
with a \lmpcmd{colored background}.  Keyboard shortcuts and
file names are formatted in \flecmd{monospace}, and \lammpsgui{} options and menus
are displayed in \guicmd{quoted monospace}.

\subsection{Tutorial 1: Lennard-Jones fluid}
\label{lennard-jones-label}

\begin{figure}
\centering
\includegraphics[width=0.65\linewidth]{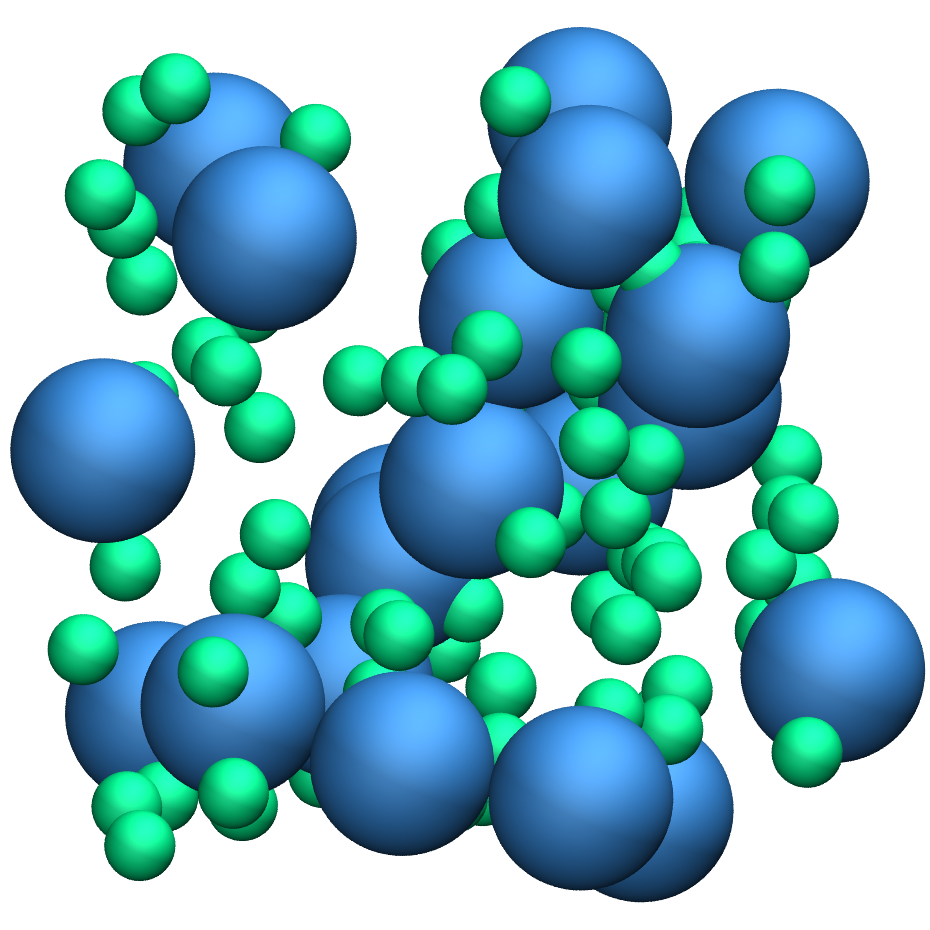}
\caption{The binary mixture simulated in
  \hyperref[lennard-jones-label]{Tutorial 1}, with the atoms of type 1
  represented as small green spheres and the atoms of type 2 as large
  blue spheres.}
\label{fig:LJ-avatar}
\end{figure}

The objective of this tutorial is to perform simple MD simulations
using LAMMPS.  The system consists of a Lennard-Jones fluid composed of
neutral particles with two different effective diameters, contained
within a cubic box with periodic boundary conditions
(Fig.~\ref{fig:LJ-avatar}).  In this tutorial, basic MD simulations in
the microcanonical (NVE) and canonical (NVT) ensembles are performed,
and basic quantities are calculated, including the potential and kinetic
energies.

\subsubsection{My first input}

To run a simulation using LAMMPS, you need to write an input script
containing a series of commands for LAMMPS to execute, similar to Python
or Bash scripts.  For clarity, the input scripts for this tutorial will
be divided into five categories, which will be filled out step by step.
To set up this tutorial, select \guicmd{Start LAMMPS Tutorial 1} from
the \guicmd{Tutorials} menu of \lammpsgui{}, and follow the
instructions.  This will select (or create, if needed) a folder, place
the initial input file \flecmd{initial.lmp} in it, and open the file in
the \lammpsgui{} \guicmd{Editor} window.  It should display the following
content:
\begin{lstlisting}
# PART A - ENERGY MINIMIZATION
# 1) Initialization
# 2) System definition
# 3) Settings
# 4) Monitoring
# 5) Run
\end{lstlisting}
Everything that appears after a hash symbol ($\#$) is a comment
and ignored by LAMMPS.  \lammpsgui{} will color such comments in red.
These five categories are not required in every input script and do not
necessarily need to be in that exact order.  For instance, the \lmpcmd{Settings}
and the \lmpcmd{Monitoring} categories could be inverted, or
the \lmpcmd{Monitoring} category could be omitted.  However, note that
LAMMPS reads input files from top to bottom and processes each command
\emph{immediately}.  Therefore, the \lmpcmd{Initialization} and
\lmpcmd{System definition} categories must appear at the top of the
input, and the \lmpcmd{Run} category must appear at the bottom.  Also, the
specifics of some commands can change after global settings are modified, so the
order of commands in the input script is important.

\paragraph{Initialization}

In the first section of the script, called \lmpcmd{Initialization},
global parameters for the simulation are defined, such as units, boundary conditions
(e.g.,~periodic or non-periodic), and atom types (e.g.,~uncharged point particles
or extended spheres with a radius and angular velocities).  These commands must be
executed \emph{before} creating the simulation box or they will cause
an error.  Similarly, many LAMMPS commands may only be
entered \emph{after} the simulation box is defined.  Only a limited
number of commands may be used in both cases.  Update the \flecmd{initial.lmp} file
so that the \lmpcmd{Initialization} section appears as follows:
\begin{lstlisting}
# 1) Initialization
units lj
dimension 3
atom_style atomic
boundary p p p
\end{lstlisting}

\begin{note}
  Strictly speaking, none of the four commands specified in the
  Initialization section are mandatory, as they correspond to the
  default settings for their respective global properties.  However,
  explicitly specifying these defaults is considered good practice to
  avoid confusion when sharing input files with other LAMMPS users.
\end{note}

The first line, \lmpcmd{units lj}, specifies the use of \emph{reduced}
units, where all quantities are dimensionless.  This unit system is a
popular choice for simulations that explore general statistical
mechanical principles, as it emphasizes relative differences between
parameters rather than representing any specific material.  The second
line, \lmpcmd{dimension 3}, specifies that the simulation is conducted
in 3D space, as opposed to 2D, where atoms are confined to move only in
the xy-plane.  The third line, \lmpcmd{atom\_style atomic}, designates
the atomic style for representing simple, individual point particles.
In this style, each particle is treated as a point with a mass, making
it the most basic atom style.  Other atom styles can incorporate
additional attributes for atoms, such as charges, bonds, or molecule
IDs, depending on the requirements of the simulated model.  The last
line, \lmpcmd{boundary p p p}, indicates that periodic boundary
conditions are applied along all three directions of space, where the
three p stand for $x$, $y$, and $z$, respectively.  Alternatives are
fixed non-periodic (f), shrink-wrapped non-periodic (s), and
shrink-wrapped non-periodic with minimum (m).  For non-periodic
boundaries, different options can be assigned to each dimension, making
configurations like \lmpcmd{boundary p p fm} valid for systems such as
slab geometries.

\begin{figure}
\centering
\includegraphics[width=\linewidth]{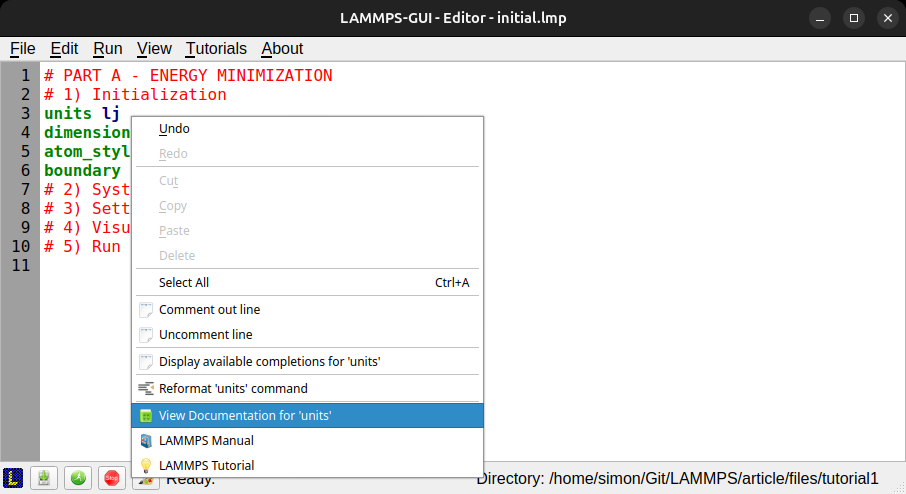}
\caption{Screenshot of the \lammpsgui{} \guicmd{Editor} window during
  \hyperref[lennard-jones-label]{Tutorial 1}.  The pop-up menu is the
  context menu for right-clicking on the \lmpcmd{units lj} command.}
\label{fig:GUI-1}
\end{figure}

\begin{note}
  Each LAMMPS command is accompanied by extensive online documentation
  that lists and discusses the different options for that
  command.  Most LAMMPS commands also have default settings
  that are applied if no value is explicitly specified.  The defaults
  for each command are listed at the bottom of its documentation
  page.  From the \lammpsgui{} editor buffer, you can access the
  documentation by right-clicking on a line containing a command
  (e.g.,~\lmpcmd{units lj}) and selecting \guicmd{View Documentation for
  `units'}.  This action should prompt your web browser to open the
  corresponding URL for the online manual.  A screenshot of this context
  menu is shown in Fig.~\ref{fig:GUI-1}.
\end{note}

\paragraph{System definition}

The next step is to create the simulation box and populate it with
atoms.  Modify the \lmpcmd{System definition} category of
\flecmd{initial.lmp} as shown below:
\begin{lstlisting}
# 2) System definition
region simbox block -20 20 -20 20 -20 20
create_box 2 simbox
create_atoms 1 random 1500 34134 simbox overlap 0.3
create_atoms 2 random 100 12756 simbox overlap 0.3
\end{lstlisting}
The first line, \lmpcmd{region simbox (...)}, defines a region named
\lmpcmd{simbox} that is a block (i.e.,~a rectangular cuboid) extending
from -20 to 20 units along all three spatial dimensions.  The second
line, \lmpcmd{create\_box 2 simbox}, initializes a simulation box based
on the region \lmpcmd{simbox} and reserves space for two types of atoms.
In LAMMPS, every atom is assigned an \emph{atom type}
property.  This property selects which force field parameters (here,
the Lennard-Jones parameters, $\epsilon_{ij}$ and $\sigma_{ij}$) are
applied to each pair of atoms via the \lmpcmd{pair\_coeff} command (see below).
We discuss in \hyperref[carbon-nanotube-label]{Tutorial 2} how this
applies to many-body pair styles, and in
\hyperref[all-atom-label]{Tutorial 3} how this applies to Coulomb
interactions.

\begin{note}
  From this point on, the number of atom types is ``locked
  in'', and any command referencing an atom type larger than 2 will
  trigger an error.  While it is possible to allocate more atom types
  than needed, you must assign a mass and provide force field parameters
  for each atom type.  Failing to do so will cause LAMMPS to terminate
  with an error.
\end{note}

The third line, \lmpcmd{create\_atoms (\dots)}, generates 1500 atoms of
type 1 at random positions within the \lmpcmd{simbox} region.
The integer 34134 is a seed for the internal random
number generator, which can be changed to produce different sequences of
random numbers and, consequently, different initial atom positions.  The
fourth line adds 100 atoms of type 2.  Both \lmpcmd{create\_atoms}
commands use the optional argument \lmpcmd{overlap 0.3}, which enforces
a minimum distance of 0.3 length units between the
randomly placed atoms.  This constraint helps avoid ``close contacts''
between atoms, which can lead to excessively large forces and simulation
instability. Each created atom in LAMMPS is automatically assigned a
unique atom ID, which serves as a numerical identifier to distinguish
individual atoms throughout the simulation.  Atom IDs by default have
the range from 1 to the total number of atoms, but this is not
enforced.  Deleting atoms, for example, causes ``holes'' in the list
of atom IDs.

\begin{note}
  Another way to define a system in LAMMPS, besides the
  \lmpcmd{create\_atoms} commands, is to import an existing topology
  file containing atomic coordinates as well as, optionally, other
  attributes such as atomic velocities and the force field parameters
  using the \lmpcmd{read\_data} command, as shown in
  \hyperref[carbon-nanotube-label]{Tutorial 2}.
\end{note}

\paragraph{Settings}

Next, we specify the settings for the two atom types.  Modify the
\lmpcmd{Settings} category of \flecmd{initial.lmp} as follows:
\begin{lstlisting}
# 3) Settings
mass 1 1.0
mass 2 5.0
pair_style lj/cut 4.0
pair_coeff 1 1 1.0 1.0
pair_coeff 2 2 0.5 3.0
\end{lstlisting}
The two \lmpcmd{mass} commands assign a mass of 1.0 and 5.0 units to the
atoms of type 1 and 2, respectively.  The third line,
\lmpcmd{pair\_style lj/cut 4.0}, specifies that the atoms will be
interacting through a Lennard-Jones (LJ) potential with a cut-off equal
to $r_c = 4.0$ length units~\cite{wang2020lennard,fischer2023history}:
\begin{equation}
E_{ij} (r) = 4 \epsilon_{ij} \left[ \left( \dfrac{\sigma_{ij}}{r} \right)^{12}
  - \left( \dfrac{\sigma_{ij}}{r} \right)^{6} \right], ~ \text{for} ~ r < r_c,
\label{eq:LJ}
\end{equation}
where $r$ is the inter-particle distance, $\epsilon_{ij}$ is the depth
of the potential well that determines the interaction strength, and
$\sigma_{ij}$ is the distance at which the potential energy equals zero.
The indices $i$ and $j$ refer to pairs of atoms with the
corresponding atom types.  The fourth line, \lmpcmd{pair\_coeff 1 1
  1.0 1.0}, specifies the Lennard-Jones coefficients for interactions
between pairs of atoms that both have atom type 1: the
energy parameter $\epsilon_{11} = 1.0$ and the distance parameter
$\sigma_{11} = 1.0$.  Similarly, the last line sets the Lennard-Jones
coefficients for interactions between atoms of type 2,
$\epsilon_{22} = 0.5$, and $\sigma_{22} = 3.0$.

\begin{note}
  By default, LAMMPS calculates the mixed Lennard-Jones
  coefficients for pairs of atoms having distinct atom types using geometric averages:
  $\epsilon_{ij} = \sqrt{\epsilon_{ii} \epsilon_{jj}}$,
  $\sigma_{ij} = \sqrt{\sigma_{ii} \sigma_{jj}}$.  In the present case,
  $\epsilon_{12} = \sqrt{1.0 \times 0.5} = 0.707$, and
  $\sigma_{12} = \sqrt{1.0 \times 3.0} = 1.732$.  Other
  rules can be selected using the \lmpcmd{pair\_modify} command.
\end{note}

\paragraph{Single-point energy}

\begin{figure}
\centering
\includegraphics[width=0.55\linewidth]{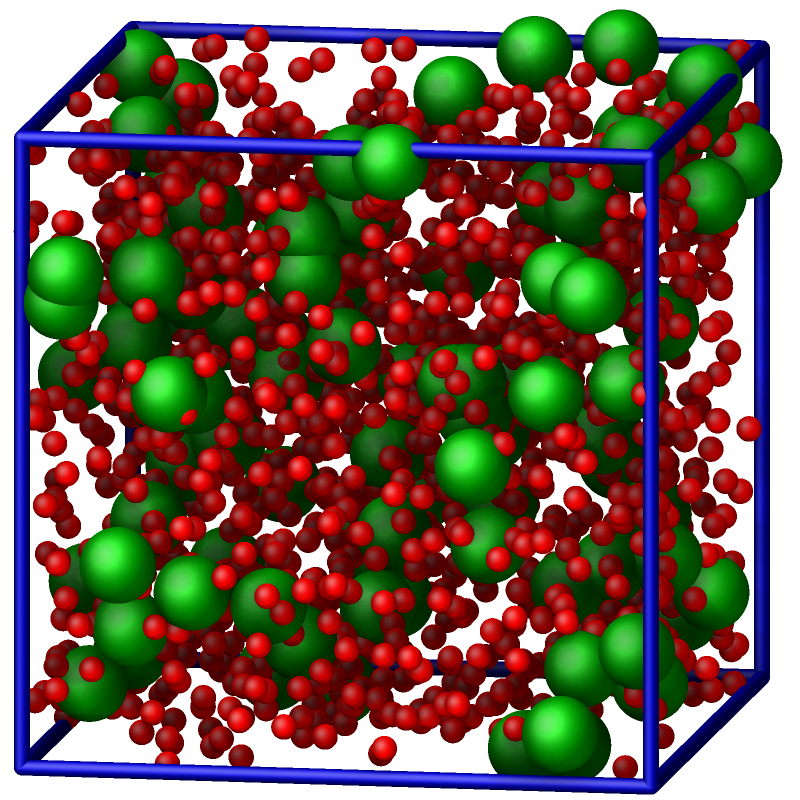}
\caption{The binary mixture simulated in \hyperref[lennard-jones-label]{Tutorial 1}.
  This image was generated directly from the \lammpsgui{}.  Atoms of
  type 1 are represented as small red spheres, atoms of type 2 as large
  green spheres, and the edges of the simulation box are represented as blue sticks.}
\label{fig:LJ}
\end{figure}

The system is now fully parameterized.  Let us complete the two remaining categories,
\lmpcmd{Monitoring} and \lmpcmd{Run}, by adding the following lines
to \flecmd{initial.lmp}:
\begin{lstlisting}
# 4) Monitoring
thermo 10
thermo_style custom step etotal press
# 5) Run
run 0 post no
\end{lstlisting}
The \lmpcmd{thermo 10} command instructs LAMMPS to print thermodynamic
information to the console every specified number of steps, in this case,
every 10 simulation steps.  The \lmpcmd{thermo\_style custom} command
defines the specific outputs, which in this case are the step number
(\lmpcmd{step}), total energy $E$ (\lmpcmd{etotal}), and pressure $p$ (\lmpcmd{press}).
The \lmpcmd{run 0 post no} command instructs LAMMPS to initialize forces and energy
without actually running the simulation.  The \lmpcmd{post no} option disables
the post-run summary and statistics output.

\begin{note}
  The `thermodynamic information' printed by
  LAMMPS using \lmpcmd{thermo\_style custom} keywords refers to
  instantaneous values of the specified thermodynamic properties
  at each output step, not cumulative averages.  However, LAMMPS also
  allows to reference a wide variety of custom data from compute styles, fix
  styles, and variables.  These can be used for on-the-fly analysis,
  including cumulative and sliding-window averages.
\end{note}

You can now run LAMMPS (see subsection \ref{running-lammps-label}
for details on running LAMMPS).  The simulation should finish quickly, and, with the default
settings, \lammpsgui{} will open two windows: one displaying the console
output and another with a chart.  The \guicmd{Output} window will display information from
the executed commands, including the total energy and pressure at step 0,
as specified by the thermodynamic data request.  Since no actual simulation
steps were performed, the \guicmd{Charts} window will be empty.

\paragraph{Snapshot Image}

At this point, you can create a snapshot image of the current system
using the \guicmd{Image Viewer} window, which can be accessed by
clicking the \guicmd{Create Image} button in the \guicmd{Run} menu.  The
image viewer works by instructing LAMMPS to render an image of the
current system using its internal rendering library via the \lmpcmd{dump
  image} command.  The resulting image is then displayed, with various
buttons available to adjust the view and rendering style.  The image
shown in Fig.~\ref{fig:LJ} was created this way.  This will always
capture the current state of the system.  Save the image for future
comparisons by clicking the \guicmd{Save as} button
in the \guicmd{File} menu.

\paragraph{Energy minimization}

Now, replace the \lmpcmd{run 0 post no} command line with the
following \lmpcmd{minimize} command:
\begin{lstlisting}
# 5) Run
minimize 1.0e-6 1.0e-6 1000 10000
\end{lstlisting}
This tells LAMMPS to perform an iterative energy
minimization of the system.  Specifically, LAMMPS will compute the
forces on all atoms and then update their positions according to a
selected algorithm using the computed forces, aiming to reduce the
potential energy.  By default, LAMMPS uses the conjugate gradient (CG)
algorithm~\cite{hestenes1952methods}.  The simulation will stop as soon
as one of the four minimizer criteria is met.  LAMMPS
will then report which stopping criterion was satisfied, along with
selected system properties at both the initial and final steps.
Note that, except for trivial systems, minimization algorithms will find a
local minimum rather than the global minimum.

\begin{figure}
\centering
\includegraphics[width=\linewidth]{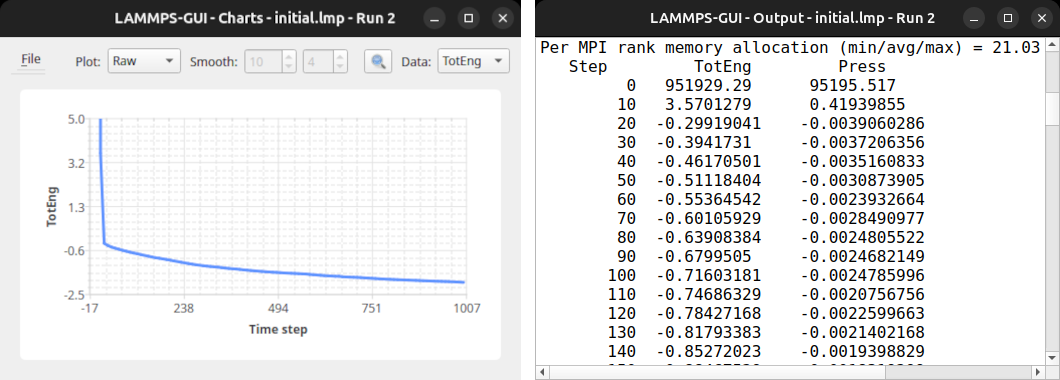}
\caption{\guicmd{Charts} (left) and \guicmd{Output} (right) windows of \lammpsgui{}
  after performing the minimization simulation of \hyperref[lennard-jones-label]{Tutorial 1}.}
\label{fig:chart-log}
\end{figure}

Run the minimization and observe that \lammpsgui{} captures the output
and update the chart in real time (see Fig.~\ref{fig:chart-log}).  This
run executes quickly (depending on your computer's capabilities)
and thus \lammpsgui{} may fail to capture some of the
thermodynamic data.  In that case, use the \guicmd{Preferences} dialog
to reduce the data update interval and switch to single-threaded,
unaccelerated execution in the \guicmd{Accelerators} tab.  You can
repeat the run; each new attempt will start fresh, resetting the system
and re-executing the script from the beginning.

\begin{figure}
\centering
\includegraphics[width=\linewidth]{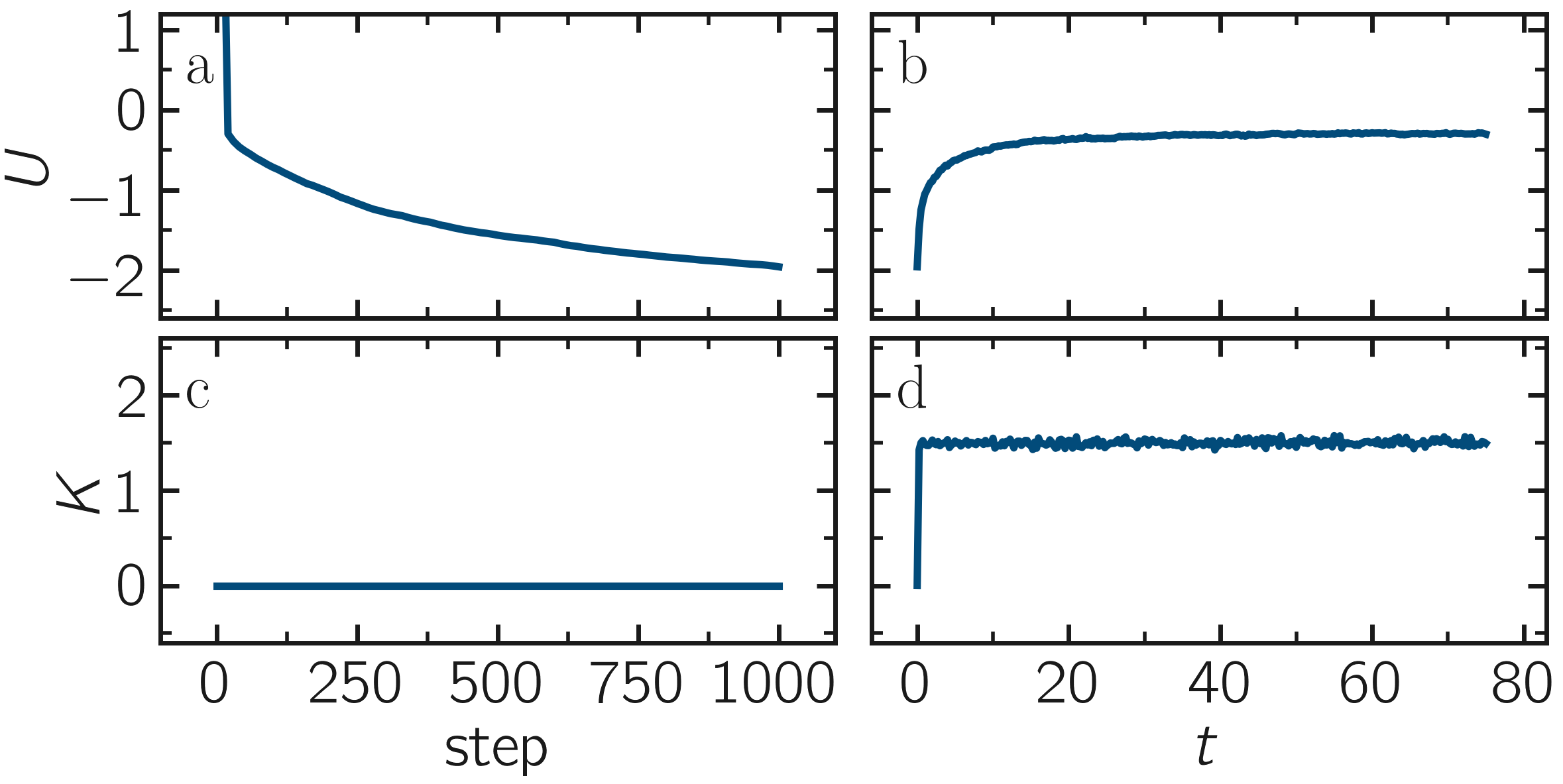}
\caption{(a) Potential energy, $U$, of the binary mixture as a function of the
step during energy minimization in \hyperref[lennard-jones-label]{Tutorial 1}.
(b) Potential energy, $U$, as a function of time during molecular dynamics in
the NVT ensemble.  (c) Kinetic energy, $K$, during energy minimization.
(d) Kinetic energy, $K$, during molecular dynamics.}
\label{fig:evolution-energy}
\end{figure}

The potential energy, $U$, decreases from a positive value to a negative value
(Figs.~\ref{fig:chart-log} and~\ref{fig:evolution-energy}\,a).  Note that
during the energy minimization, the potential energy equals the total energy
of the system, $E = U$, since the kinetic energy, $K$, is zero.  The
initially positive potential energy is expected, as the atoms are
created at random positions within the simulation box, with some in very
close proximity to each other.  This proximity results in a large
initial potential energy due to the repulsive branch of the
Lennard-Jones potential [i.e.,~the term $1/r^{12}$ in
Eq.~\eqref{eq:LJ}].  As the energy minimization progresses, the energy
decreases - first rapidly - then more gradually, before plateauing at a
negative value.  This indicates that the atoms have moved to reasonable
distances from one another.

\begin{note}
  Since the \lmpcmd{thermo\_style} command also includes the \lmpcmd{press}
  keyword, you can switch from plotting the total energy to
  displaying the pressure by selecting \guicmd{Press} in the \guicmd{Select data}
  drop-down menu of the \guicmd{Charts} window.
\end{note}

Create and save a snapshot image of the simulation state after the
minimization, and compare it to the initial image.  You should observe
that the atoms are ``clumping together'' as they move toward positions
of lower potential energy.

\paragraph{Molecular dynamics}

After the energy minimization, any overlapping atoms are displaced, and
the system is ready for a molecular dynamics simulation.  To continue
from the result of the minimization step, append the MD simulation
commands to the same input script, \flecmd{initial.lmp}.  Add the
following lines immediately after the \lmpcmd{minimize} command:
\begin{lstlisting}
# PART B - MOLECULAR DYNAMICS
# 4) Monitoring
thermo 50
thermo_style custom step temp etotal pe ke press
\end{lstlisting}

Since LAMMPS reads inputs from top to bottom, these lines will be
executed \emph{after} the energy minimization.  Therefore, there is no
need to re-initialize or re-define the system.  The \lmpcmd{thermo}
command is called a second time to update the output frequency from 10
to 50 as soon as \lmpcmd{PART B} of the simulation starts.  In addition,
a new \lmpcmd{thermo\_style} command is introduced to specify the
thermodynamic information LAMMPS should print during during \lmpcmd{PART
  B}.  This adjustment is made because, during molecular dynamics, the
system exhibits a non-zero temperature $T$ (and consequently a non-zero
kinetic energy $K$, keyword \lmpcmd{ke}), which are useful to monitor.
The \lmpcmd{pe} keyword represents the potential energy of the system,
$U$, such that $U + K = E$.

Then, add a second \lmpcmd{Run} category by including the following
lines in \lmpcmd{PART B} of \flecmd{initial.lmp}:
\begin{lstlisting}
# 5) Run
fix mynve all nve
timestep 0.005
run 50000
\end{lstlisting}
The \lmpcmd{fix nve} command updates the positions and velocities of the
atoms in the group \lmpcmd{all} at every step.  More
  specifically, this command integrates Newton's equations of motion
  using the velocity-Verlet algorithm.  The group \lmpcmd{all} is a
default group that contains all atoms.  The last two lines specify the
value of the \lmpcmd{timestep} and the number of steps for the
\lmpcmd{run}, respectively, for a total duration of 250 time units.

\begin{note}
  Since the \emph{only} command affecting forces
  and velocities in the present script is \lmpcmd{fix nve}, \emph{and}
  periodic boundary conditions are applied in all directions, the MD
  simulation will be performed in the microcanonical (NVE) 
  statistical mechanical ensemble, which maintains a constant number
  of particles and a fixed box volume. In this ensemble, the system does
  not exchange energy with anything outside the simulation box.
\end{note}

Run the simulation using LAMMPS.  Initially, the system is
not equilibrated, as the potential energy
decreases while the kinetic energy increases.  After approximately
40\,000 steps, the values for both kinetic and potential energy
plateau, indicating that the system has reached equilibrium, with
the total energy fluctuating around a certain constant value.

Now, we change the second \lmpcmd{Run} section to 
(note the smaller number of MD steps):
\begin{lstlisting}
# 5) Run
fix mynve all nve
fix mylgv all langevin 1.0 1.0 0.1 10917
timestep 0.005
run 15000
\end{lstlisting}
The new command adds a Langevin thermostat to the atoms in the group
\lmpcmd{all}, with a target temperature of 1.0 temperature units
throughout the run (the two numbers represent the target temperature at
the beginning and at the end of the run, which results in a temperature
ramp if they differ)~\cite{schneider1978molecular}.  A \lmpcmd{damping}
parameter of 0.1 is used.  It determines how rapidly the temperature is
relaxed to its desired value.  In a Langevin thermostat, the atoms are
subject to friction and random noise (in the form of randomly added
velocities).  Since a constant friction term removes more kinetic energy
from fast atoms and less from slow atoms, the system will eventually
reach a dynamic equilibrium where the kinetic energy removed and added
are about the same.  The number 10917 is a seed used to initialize the
random number generator used inside of \lmpcmd{fix langevin}; you can
change it to perform statistically independent simulations.  In the
presence of a thermostat, the MD simulation will be performed in the
canonical or NVT ensemble.

Run the simulation again using \lammpsgui{}.  From the information
printed in the \guicmd{Output} window, one can see that the temperature
starts from 0 but rapidly reaches the requested value and
stabilizes itself near $T=1$ temperature units.  One can also observe that
the potential energy, $U$, rapidly decreases during energy
minimization (see also Fig.~\ref{fig:evolution-energy}\,a).  After
the molecular dynamics simulation starts, $U$ increases until
it reaches a plateau value of about -0.25.  The kinetic energy,
$K$, is equal to zero during energy minimization and then
increases rapidly during molecular dynamics until it reaches
a plateau value of about 1.5 (Fig.~\ref{fig:evolution-energy}\,d).

\begin{note}
  All simulations presented in these tutorials are deliberately kept
  short so they can be executed on a personal computer.  These runs are not intended
  to produce statistically meaningful results, and should not be considered suitable
  for publication (see for instance Ref.~\citenum{grossfield2009quantifying}).
\end{note}

\paragraph{Trajectory visualization}

So far, the simulation has been mostly monitored through the analysis of
thermodynamic information.  To better follow the evolution of the system
and visualize the trajectories of the atoms, let us use the \lmpcmd{dump
  image} command to create snapshot images during the simulation.  We
have already explored the \guicmd{Image Viewer} window.  Open it again
and adjust the visualization to your liking using the available buttons.
Now you can copy the commands used to create this visualization to the
clipboard by either using the \guicmd{Ctrl-D} keyboard shortcut or
selecting \guicmd{Copy dump image command} from the \guicmd{File} menu.
This text can be pasted into the into the \lmpcmd{Monitoring} section
of \lmpcmd{PART B} of the \flecmd{initial.lmp} file.  This may look like
the following:
\begin{lstlisting}
dump viz all image 100 myimage-*.ppm type type &
  size 800 800 zoom 1.452 shiny 0.7 fsaa yes &
  view 80 10 box yes 0.025 axes no 0.0 0.0 &
  center s 0.483725 0.510373 0.510373
dump_modify viz pad 9 boxcolor royalblue &
  backcolor white adiam 1 1.6 adiam 2 4.8
\end{lstlisting}
The `$\&$' characters at the end are used to extend the
commands across multiple lines. These two commands tell LAMMPS to
generate NetPBM format images every 100
steps.  The two \lmpcmd{type} keywords are for \lmpcmd{color} and
\lmpcmd{diameter}, respectively.  Run the \flecmd{initial.lmp} using
LAMMPS again, and a new window named \guicmd{Slide Show} will pop up.
It will show each image created by the \lmpcmd{dump image} as it is
created. After the simulation is finished (or stopped), the slideshow
viewer allows you to animate the trajectory by cycling through the
images.  The window also allows you to export the animation to a movie
(provided the FFMpeg program is installed) and to bulk delete those
image files.

The rendering of the system can be further adjusted using the many
options of the \lmpcmd{dump image} command.  For instance, the value for the
\lmpcmd{shiny} keyword is used to adjust the shininess of the atoms, the
\lmpcmd{box} keyword adds or removes a representation of the box, and
the \lmpcmd{view} and \lmpcmd{zoom} keywords adjust the camera (and so
on).

\subsubsection{Improving the script}

Let us improve the input script and perform more advanced operations,
such as specifying initial positions for the atoms and restarting the
simulation from a previously saved configuration.

\paragraph{Control the initial atom positions}

Open the \flecmd{improved.min.lmp}, which was downloaded during the
tutorial setup.  This file contains the \lmpcmd{Part A} of the
\flecmd{initial.lmp} file, but \emph{without} any
commands in the \lmpcmd{System definition} section:
\begin{lstlisting}
# 1) Initialization
units lj
dimension 3
atom_style atomic
boundary p p p
# 2) System definition
# 3) Settings
mass 1 1.0
mass 2 10.0
pair_style lj/cut 4.0
pair_coeff 1 1 1.0 1.0
pair_coeff 2 2 0.5 3.0
# 4) Monitoring
thermo 10
thermo_style custom step etotal press
# 5) Run
minimize 1.0e-6 1.0e-6 1000 10000
\end{lstlisting}
We want to create the atoms of types 1 and 2 in two separate
regions.  To achieve this, we need to add two \lmpcmd{region} commands and then
reintroduce the \lmpcmd{create\_atoms} commands, this time using the new
regions instead of the simulation box region to place the atoms:
\begin{lstlisting}
# 2) System definition
region simbox block -20 20 -20 20 -20 20
create_box 2 simbox
# for creating atoms
region cyl_in cylinder z 0 0 10 INF INF side in
region cyl_out cylinder z 0 0 10 INF INF side out
create_atoms 1 random 1000 34134 cyl_out
create_atoms 2 random 150 12756 cyl_in
\end{lstlisting}
The \lmpcmd{side in} and \lmpcmd{side out} keywords are used to define
regions representing the inside and outside of the cylinder of radius
10 length units, respectively.  Then, append a sixth section titled \lmpcmd{Save system} at the end
of the file, ensuring that the \lmpcmd{write\_data} command is placed \emph{after}
the \lmpcmd{minimize} command:
\begin{lstlisting}
# 6) Save system
write_data improved.min.data
\end{lstlisting}

\begin{note}
  A key improvement to the input is the addition of the
  \lmpcmd{write\_data} command.  This command writes the state of the
  system to a text file called \flecmd{improved.min.data}.  This
  \flecmd{.data} file will be used later to restart the simulation from
  the final state of the energy minimization step, eliminating the need
  to repeat the system creation and minimization.
\end{note}
Run the \flecmd{improved.min.lmp} file using \lammpsgui{}.  At the end
of the simulation, a file called \flecmd{improved.min.data} is created.
You can view the contents of this file from \lammpsgui{}, by
right-clicking on the file name in the editor and selecting the entry
\guicmd{View file improved.min.data}.

\begin{figure}
\centering
\includegraphics[width=0.65\linewidth]{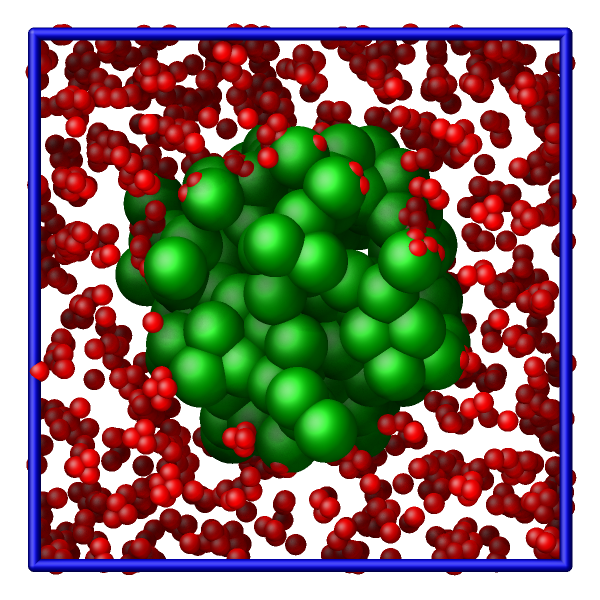}
\caption{Visualization of the improved binary mixture input after minimization
  during \hyperref[lennard-jones-label]{Tutorial 1}.  Colors are the same as in
  Fig.~\ref{fig:LJ}.}
\label{fig:improved-min}
\end{figure}

The created \flecmd{.data} file contains all the information necessary
to restart the simulation, such as the number of atoms, the box size,
the masses, and the pair coefficients.  This \flecmd{.data} file also
contains the final positions of the atoms, along with their IDs and types,
within the \lmpcmd{Atoms} section.  The first five columns of the \lmpcmd{Atoms} section
correspond (from left to right) to the atom indexes (from 1 to the total
number of atoms, 1150), the atom types (1 or 2 here), and the positions
of the atoms $x$, $y$, $z$.  The last three columns are image flags that
keep track of which atoms crossed the periodic boundary.  The exact
format of each line in the \lmpcmd{Atoms} section depends on the choice
of \lmpcmd{atom\_style}, which determines which per-atom data is set and
stored internally in LAMMPS.

\begin{note}
  Instead of the \lmpcmdnote{write\_data} command, you can also use the
  \lmpcmdnote{write\_restart} command to save the state
  of the simulation to a binary restart file.  Binary restart files are
  more compact, faster to write, and contain more information, making them often
  more convenient to use.  For example, the choice of \lmpcmdnote{atom\_style}
  or \lmpcmdnote{pair\_style} is recorded, so those commands do not need to be issued
  before reading the restart.  Note however that restart files are not expected to be
  portable across LAMMPS versions or platforms.  Therefore, in these tutorials,
  and with the exception of \hyperref[all-atom-label]{Tutorial 3}, we primarily
  use \lmpcmdnote{write\_data} to provide you with a reference
  copy of the data file that works regardless of your LAMMPS version and platform.
\end{note}

\paragraph{Restarting from a saved configuration}

To continue a simulation from the saved configuration, open the
\flecmd{improved.md.lmp} file, which was downloaded during the tutorial setup.
This file contains the \textit{Initialization} part from \flecmd{initial.lmp}
and \flecmd{improved.min.lmp}:
\begin{lstlisting}
# 1) Initialization
units lj
dimension 3
atom_style atomic
boundary p p p
# 2) System definition
# 3) Settings
# 4) Monitoring
# 5) Run
\end{lstlisting}
Since we read most of the information from the data file, we don't need
to repeat all the commands from the \lmpcmd{System definition}
and \lmpcmd{Settings} categories.  The exception is the \lmpcmd{pair\_style}
command, which now must come \emph{before} the simulation box is defined,
meaning before the \lmpcmd{read\_data} command.  Add the following
lines to \flecmd{improved.md.lmp}:
\begin{lstlisting}
# 2) System definition
pair_style lj/cut 4.0
read_data improved.min.data
\end{lstlisting}

By visualizing the system (see Fig.~\ref{fig:improved-min}), you may
have noticed that some atoms left their original region during
minimization.  To start the simulation from a clean slate, with only
atoms of type 2 inside the cylinder and atoms of type 1 outside the
cylinder, let us delete the misplaced atoms by adding the following
commands to the \lmpcmd{System definition} section of 
the \flecmd{improved.md.lmp}:

\begin{lstlisting}
region cyl_in cylinder z 0 0 10 INF INF side in
region cyl_out cylinder z 0 0 10 INF INF side out
group grp_t1 type 1
group grp_t2 type 2
group grp_in region cyl_in
group grp_out region cyl_out
group grp_t1_in intersect grp_t1 grp_in
group grp_t2_out intersect grp_t2 grp_out
delete_atoms group grp_t1_in
delete_atoms group grp_t2_out
\end{lstlisting}
The first two \lmpcmd{region} commands recreate the previously defined
regions, which is necessary since regions are not saved by the
\lmpcmd{write\_data} command.  The first two \lmpcmd{group} commands
create groups containing all the atoms of type 1 and all the
atoms of type 2, respectively.  The next two \lmpcmd{group} commands
create atom groups based on their positions at the beginning of the
simulation, i.e.,~when the commands are being read by LAMMPS.  The last
two \lmpcmd{group} commands create atom groups based on the intersection
between the previously defined groups.  Finally, the two
\lmpcmd{delete\_atoms} commands delete the atoms of type 1
located inside the cylinder and the atoms of type 2 located
outside the cylinder, respectively.

Since LAMMPS has a limited number of custom groups (30), it is good practice
to delete groups that are no longer needed.  This can be done by adding the
following four commands to \flecmd{improved.md.lmp}:
\begin{lstlisting}
# delete no longer needed groups
group grp_in delete
group grp_out delete
group grp_t1_in delete
group grp_t2_out delete
\end{lstlisting}

Let us monitor the number of atoms of each type inside the cylinder as a
function of time by creating the following equal-style variables:
\begin{lstlisting}
variable n1_in equal count(grp_t1,cyl_in)
variable n2_in equal count(grp_t2,cyl_in)
\end{lstlisting}
The equal-style \lmpcmd{variables} are expressions evaluated
during the run and return a number.  Here, they are defined to count
the number of atoms of a specific group within the \lmpcmd{cyl\_in} region.

\begin{note}
  The \lmpcmd{n1\_in} and \lmpcmd{n2\_in} defined above are
  equal-style variables, which evaluate a numerical expression using the
  \lmpcmd{count()} function.  Other LAMMPS variable styles include
  atom, index, file, loop, string, and vector.
\end{note}

In addition to counting the atoms in each region, we will also extract
the coordination number of type 2 atoms around type 1 atoms.  The
coordination number measures the number of type 2 atoms near
type 1 atoms, defined by a cutoff distance.  Taking the average provides
as a good indicator of the degree of mixing in a binary mixture.  This
is done using two \lmpcmd{compute} commands:  the first counts the
coordinated atoms, and the second calculates the average over all type 1
atoms.  Add the following lines to \flecmd{improved.md.lmp}:
\begin{lstlisting}
compute coor12 grp_t1 coord/atom cutoff 2 group grp_t2
compute sumcoor12 grp_t1 reduce ave c_coor12
\end{lstlisting}
The \lmpcmd{compute reduce ave} command is used to average the per-atom
coordination number calculated by the \lmpcmd{compute coord/atom}
command.  Compute commands do not print or output
anything by themselves, nor are they automatically executed; they
require a ``consumer'' command that references the compute.  In this case, the
first compute is referenced by the second, and we reference the second
in a \lmpcmd{thermo\_style custom} command (see below).

\begin{note}
  LAMMPS \lmpcmd{compute} commands can produce
  a wide variety of data and one can identify the category from the
  name of the compute style: global data (no suffix), local data
  (/local suffix), per-atom data (/atom suffix), per-chunk data
  (/chunk suffix), per-grid data (/grid suffix).  In the example
  above, the \lmpcmd{compute coord/atom} produces per-atom data, which
  is used as input for \lmpcmd{compute reduce} which returns global
  data.  For global data three kinds of data exists: scalars (single
  values), vectors (one-dimensional arrays), or arrays
  (two-dimensional tables).  When referencing results of a compute,
  you can use indices: for example, \lmpcmd{c\_mycompute} refers to
  the entire scalar, vector, or array, and \lmpcmd{c\_mycompute[1]}
  refers to its first element or column (in case of vector or array).  In some
  cases also wildcards like ``*'' can be used to, for instance, refer to all elements
  of a vector instead of having specify all elements individually.
  In general, ``consumer'' commands (\lmpcmd{fix} styles or \lmpcmd{dump} styles,
  \lmpcmd{variables}, or other \lmpcmd{compute} styles) can only work with certain data
  types or need to have keywords set to select which data to use.
  You need to check the documentation of each command to ensure
  compatibility.
\end{note}

There is no need for a \lmpcmd{Settings} section, as the settings are
taken from the \flecmd{.data} file.
Finally, let us complete the script by adding the following lines to
\flecmd{improved.md.lmp}:
\begin{lstlisting}
# 4) Monitoring
thermo 1000
thermo_style custom step temp pe ke etotal &
  press v_n1_in v_n2_in c_sumcoor12
dump viz all image 1000 myimage-*.ppm type type &
  shiny 0.1 box no 0.01 view 0 0 zoom 1.8 fsaa yes size 800 800
dump_modify viz adiam 1 1 adiam 2 3 acolor 1 &
  turquoise acolor 2 royalblue backcolor white
\end{lstlisting}
The two variables \lmpcmd{n1\_in}, \lmpcmd{n2\_in}, along with the compute
\lmpcmd{sumcoor12}, were added to the list of information printed during
the simulation.  Additionally, images of the system will be created with
slightly less saturated colors than the default ones.

Finally, add the following lines to \flecmd{improved.md.lmp}:
\begin{lstlisting}
# 5) Run
velocity all create 1.0 49284 mom yes dist gaussian
fix mynve all nve
fix mylgv all langevin 1.0 1.0 0.1 10917 zero yes
timestep 0.005
run 300000
\end{lstlisting}
Here, there are a few more differences from the previous simulation.
First, the \lmpcmd{velocity create} command assigns an initial velocity
to each atom.  The initial velocity is chosen so that the average
initial temperature is equal to 1.0 temperature units.  The additional
keywords ensure that no linear momentum (\lmpcmd{mom yes}) is given to
the system and that the generated velocities are distributed according
to a Gaussian distribution.  Another improvement is the \lmpcmd{zero
  yes} keyword in the Langevin thermostat, which ensures that the total
random force applied to the atoms is equal to zero. These steps are
important to prevent the system from starting to drift or move as a
whole.
\begin{figure}
\centering
\includegraphics[width=\linewidth]{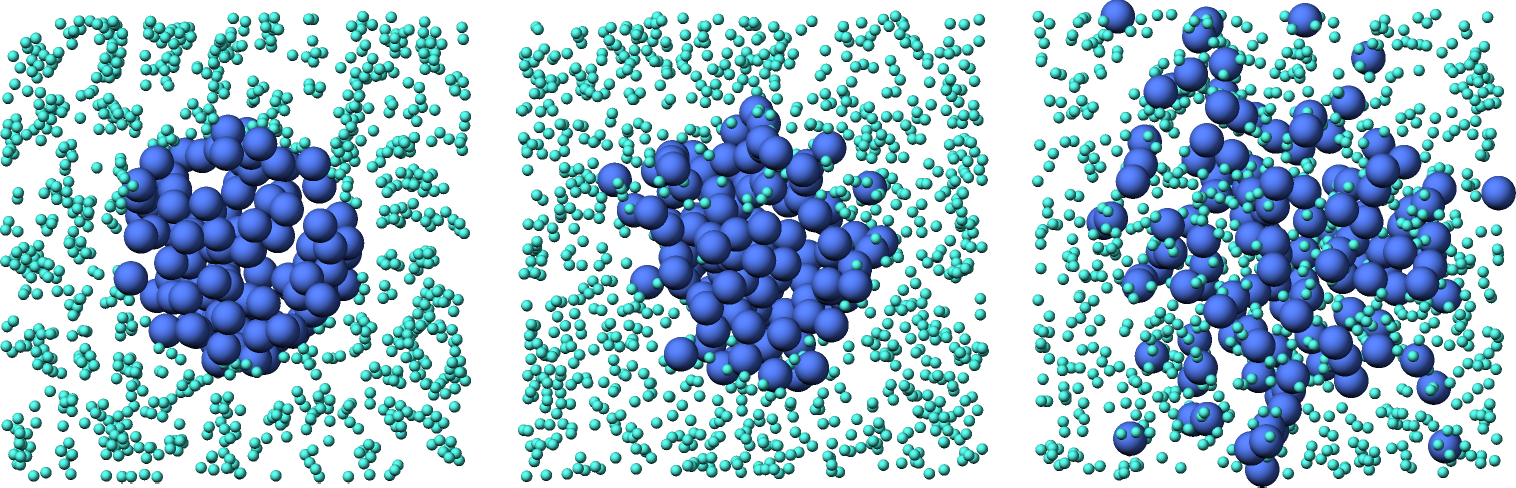}
\caption{Evolution of the system from \hyperref[lennard-jones-label]{Tutorial 1}
during mixing.  The three snapshots show respectively the system
at $t=0$ (left panel), $t=75$ (middle panel), and $t=1500$ (right panel).  The
atoms of type 1 are represented as small turquoise spheres and the atoms of type 2
as large blue spheres.}
\label{fig:evolution-population}
\end{figure}

\begin{figure}
\centering
\includegraphics[width=\linewidth]{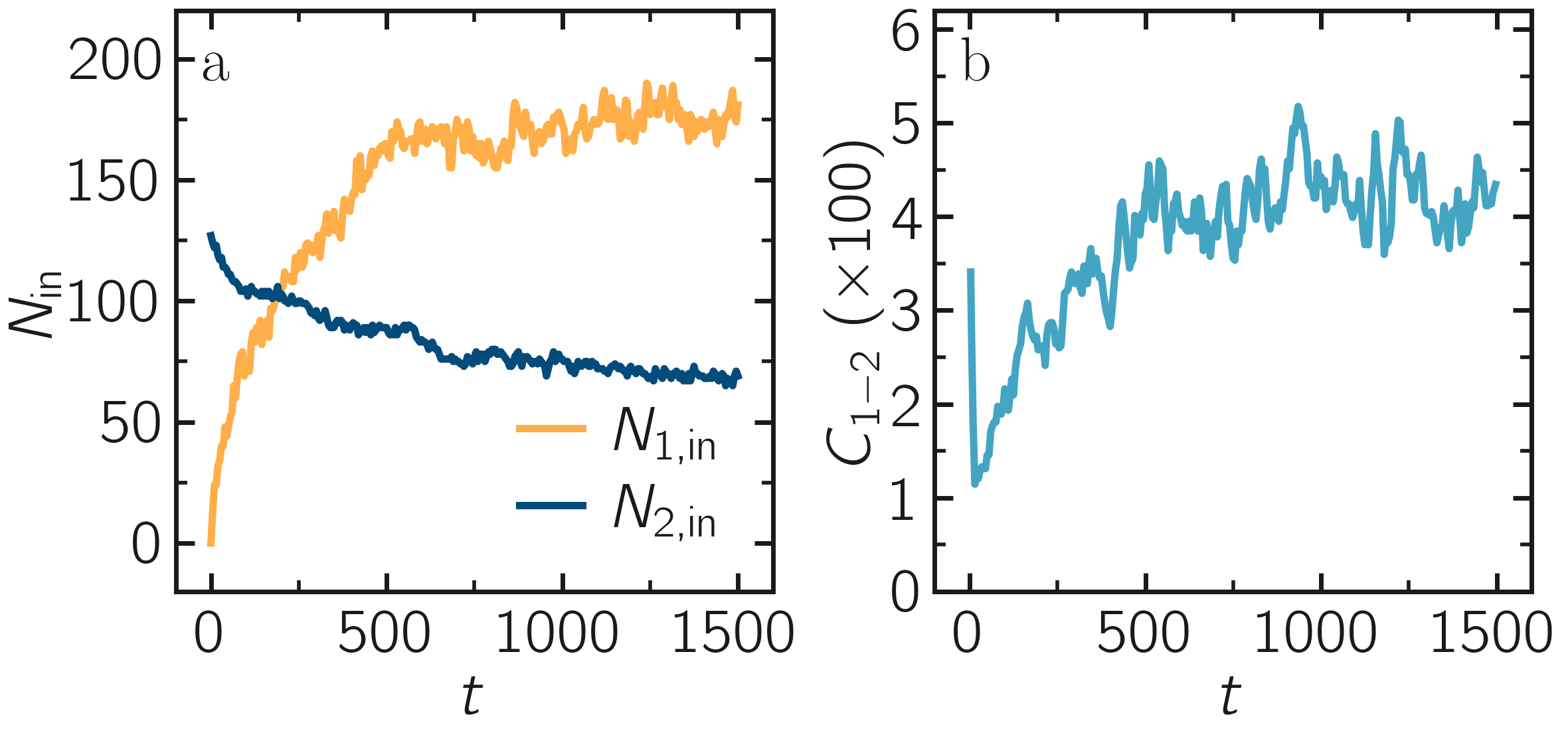}\\[-2ex]
\caption{a)~Evolution of the numbers $N_\text{1, in}$ and $N_\text{2, in}$ of atoms
of types 1 and 2, respectively, within the \lmpcmd{cyl\_in} region as functions
of time $t$.  b)~Evolution of the coordination number $C_{1-2}$ (compute \lmpcmd{sumcoor12})
between atoms of types 1 and 2.}
\label{fig:mixing}
\end{figure}

\begin{note}
  A bulk system with periodic boundary conditions is expected to remain
  in place.  Accordingly, when computing the temperature from the
  kinetic energy, we use $3N-3$ degrees of freedom since there is no
  global translation.  In a drifting system, some of the kinetic energy
  is due to the drift, which means the system itself cools down.  In
  extreme cases, the system can freeze while its center of mass drifts
  very quickly.  This phenomenon is sometimes referred to as the
  ``flying ice cube syndrome''~\cite{wong2016good}.
\end{note}
Run \flecmd{improved.md.lmp} and observe the mixing of the two populations
over time (see also Fig.~\ref{fig:evolution-population}).  From the
variables \lmpcmd{n1\_in} and \lmpcmd{n2\_in}, you can track the number
of atoms in each region as a function of time
(Fig.~\ref{fig:mixing}\,a).  To view their evolution, select the entries
\guicmd{v\_n1\_in} or \guicmd{v\_n2\_in} in the \guicmd{Data} drop-down
menu in the \guicmd{Charts} window of \lammpsgui{}.

In addition, as the mixing progresses, the average coordination number
between atoms of types 1 and 2 increases from about $0.01$ to $0.04$
(Fig.~\ref{fig:mixing}\,b).  This indicates that, over time, more and
more particles of type 1 come into contact with particles of type 2, as
expected during mixing.  This can be observed using the entry
\guicmd{c\_sumcoor12} in the \guicmd{Charts} drop-down menu.

\paragraph{Experiments}

Here are some suggestions for further experiments with this system that
may lead to additional insights into how different systems are configured
and how various features function:
\begin{itemize}
\item Use a Nos\'e-Hoover thermostat (\lmpcmd{fix nvt}) instead of a Langevin thermostat
  (\lmpcmd{fix nve} + \lmpcmd{fix langevin}).
\item Omit the energy minimization step before starting the MD simulation using either
the Nos\'e-Hoover or the Langevin thermostat.
\item Apply a thermostat to only one type of atoms each and observe the
  temperature for each type separately.
\item Append an NVE run (i.e.~without any thermostat) and observe the energy levels.
\end{itemize}

\begin{figure}
\centering
\includegraphics[width=0.50\linewidth]{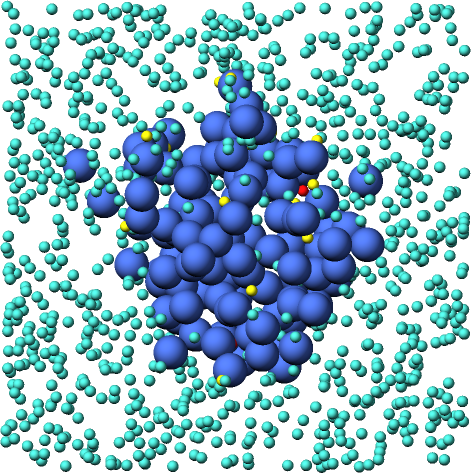}
\caption{Snapshot of the binary mixture simulated
  during \hyperref[lennard-jones-label]{Tutorial 1} with atoms of type 1
  colored according to their computed $1-2$ coordination
  number from the compute \lmpcmd{coor12}, ranging from turquoise,\lmpcmd{c\_coor12 = 0},
  to yellow, \lmpcmd{c\_coor12 = 1}, and red, \lmpcmd{c\_coor12 = 2}.}
\label{fig:coords-viz}
\end{figure}

\begin{note}
  In contrast to the \lmpcmd{fix nve} command, which integrates Newton's equations
  of motion without any thermostatting, the \lmpcmd{fix nvt} command adds a Nosé-Hoover
  thermostat to control the system temperature.
\end{note}

Another useful experiment is coloring the atoms in the \guicmd{Slide Show}
according to an observable, such as their respective coordination
numbers.  To do this, replace the
\lmpcmd{dump} and \lmpcmd{dump\_modify} commands with the following lines:
\begin{lstlisting}
variable coor12 atom (type==1)*(c_coor12)+(type==2)*-1
dump viz all image 1000 myimage-*.ppm v_coor12 type &
  shiny 0.1 box no 0.01 view 0 0 zoom 1.8 fsaa yes size 800 800
dump_modify viz adiam 1 1 adiam 2 3 backcolor white &
  amap -1 2 ca 0.0 4 min royalblue 0 turquoise 1 yellow max red
\end{lstlisting}
Run LAMMPS again.  Atoms of type 1 are now colored based on the value
of \lmpcmd{c\_coor12}, which is mapped continuously from turquoise to yellow
and red for atoms with the highest coordination (Fig.~\ref{fig:coords-viz}).
In the definition of the variable \lmpcmd{v\_coor12}, atoms of type 2 are
all assigned a value of -1, and will therefore always be colored their default blue.

\subsection{Tutorial 2: Pulling on a carbon nanotube}
\label{carbon-nanotube-label}

\begin{figure}
\centering
\includegraphics[width=0.55\linewidth]{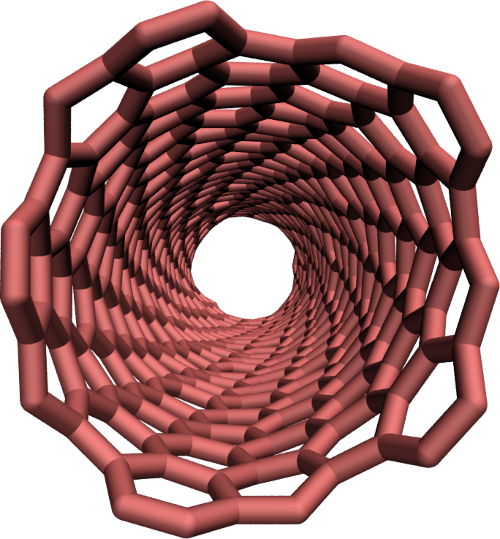}
\caption{The carbon nanotube (CNT) simulated during
\hyperref[carbon-nanotube-label]{Tutorial 2}.}
\label{fig:CNT}
\end{figure}

In this tutorial, the system of interest is a small, single-walled
carbon nanotube (CNT) in an empty box (Fig.~\ref{fig:CNT}).  The CNT is
strained by imposing a constant velocity on the edge atoms.  To
illustrate the difference between conventional and reactive force
fields, this tutorial is divided into two parts: in the first part, a
conventional molecular force field (called
OPLS-AA~\cite{jorgensenDevelopmentTestingOPLS1996}) is used and the
functional form of the bonded potential ensures that the
bonds between the atoms of the CNT are unbreakable.  In the second part,
a reactive, many-body force field (called
AIREBO~\cite{stuart2000reactive}) is used, which allows chemical bonds
to break under large strain.

To set up this tutorial, select \guicmd{Start Tutorial 2} from the
\guicmd{Tutorials} menu of \lammpsgui{} and follow the instructions.
This will select a folder, create one if necessary, and place several
files into it.  The initial input file, set up for a single-point energy
calculation, will also be loaded into the editor under the name
\flecmd{unbreakable.lmp}.  Additional files are a data file containing
the CNT topology and geometry, named \flecmd{unbreakable.data}, a
parameters file named \flecmd{unbreakable.inc}, as well as the scripts
required for the second part of the tutorial.

\subsubsection{Unbreakable bonds}

With most conventional molecular force fields, the chemical bonds
between atoms are defined at the start of the simulation and remain
fixed, regardless of the forces applied to the atoms.  In
this tutorial, these bonds are explicitly specified in the
\lmpcmd{.data} file, which is read using the \lmpcmd{read\_data}
command (see below).
Bonds are typically modeled as springs following Hooke's
law with equilibrium distances $r_0$, force constants $k_\text{b}$,
and bond potential energy
$U_\text{b} = k_\text{b} \left( r - r_0 \right)^2$.  Additionally,
angular and dihedral interactions are often imposed to
preserve the molecular structure by maintaining the relative
orientations of neighboring atoms.

\paragraph{The LAMMPS input}

After completing the setup, the editor should display the following content:
\begin{lstlisting}
units real
atom_style molecular
boundary f f f

pair_style lj/cut 14.0
bond_style harmonic
angle_style harmonic
dihedral_style opls
improper_style harmonic
special_bonds lj 0.0 0.0 0.5

read_data unbreakable.data
include unbreakable.inc

run 0 post no
\end{lstlisting}
The chosen unit system is \lmpcmd{real} (therefore distances are in
Ångströms (Å), times in femtoseconds (fs), and energies in (kcal/mol)), the
\lmpcmd{atom\_style} is \lmpcmd{molecular} (therefore atoms are point
particles that can form bonds with each other), and the boundary
conditions are fixed.  The boundary conditions do not matter here, as
the box boundaries were placed far from the CNT.  Just like in the
previous tutorial, \hyperref[lennard-jones-label]{Lennard-Jones fluid},
the pair style is \lmpcmd{lj/cut} (i.e.~a Lennard-Jones potential with
cutoff) and its cutoff is set to 14~Å, which means that only the
atoms closer than this distance interact through the Lennard-Jones
potential.

The \lmpcmd{bond\_style}, \lmpcmd{angle\_style},
\lmpcmd{dihedral\_style}, and \lmpcmd{improper\_style} commands specify
the different potentials used to constrain the relative positions of the
atoms.  The \lmpcmd{special\_bonds} command sets the weighting factors
for the Lennard-Jones interactions between atoms sitting one,
two, or three bonds away from each other, respectively.  This is done for
convenience when parameterizing the force constants for bonds, angles, and
so on.  By excluding the non-bonded (Lennard-Jones) interactions for
these pairs, those interactions do not need to be considered when determining
the force constants.

The \lmpcmd{read\_data} command imports the
\href{\filepath tutorial2/unbreakable.data}{\dwlcmd{unbreakable.data}}
file that should have been downloaded next
to \lmpcmd{unbreakable.lmp} during the tutorial setup. This file
contains information about the box size, atom positions, as well as the
identity of the atoms that are
linked by \lmpcmd{bonds}, \lmpcmd{angles}, \lmpcmd{dihedrals}, and
\lmpcmd{impropers} interactions. It was created using VMD and TopoTools
\cite{kohlmeyer2017topotools}.

\begin{note}
  Bonds, angles, dihedrals, and impropers in LAMMPS are assigned types and IDs, just like atoms.
  The ID uniquely identifies each interaction instance, while the type determines which parameters
  (from the \lmpcmd{bond\_coeff}, \lmpcmd{angle\_coeff}, etc. commands) are applied.
  In this tutorial, these types and IDs are specified in the \lmpcmd{.data} file and
  read by the \lmpcmd{read\_data} command.
\end{note}

\begin{note}
  The format details of the
  different sections in a data file change with different settings.  In
  particular, the \lmpcmd{Atoms} section may have a different number of
  columns, or the columns may represent different properties when the
  \lmpcmd{atom\_style} is changed.  To help users, LAMMPS and tools like
  VMD and TopoTools will add a comment (here \lmpcmd{\# molecular}) to the
  \lmpcmd{Atoms} header line in the data files that indicates the intended
  \lmpcmd{atom\_style}.  LAMMPS will print a warning when the chosen atom
  style does not match what is written in that comment.
\end{note}

The \flecmd{.data} file does not contain any sections with potential parameters; thus,
we need to specify the parameters of both the bonded and
non-bonded potentials.  The parameters we use are taken
from the OPLS-AA (Optimized Potentials for Liquid Simulations-All-Atom)
force field~\cite{jorgensenDevelopmentTestingOPLS1996}, and are given
in a separate \lmpcmd{unbreakable.inc} file (also downloaded during
the tutorial setup).  This file - that must be placed
next to \flecmd{unbreakable.lmp} - contains the following lines:
\begin{lstlisting}
pair_coeff 1 1 0.066 3.4
bond_coeff 1 469 1.4
angle_coeff 1 63 120
dihedral_coeff 1 0 7.25 0 0
improper_coeff 1 5 180
\end{lstlisting}
The \lmpcmd{pair\_coeff} command sets the parameters for non-bonded
Lennard-Jones interactions between atoms type 1 to
$\epsilon_{11} = 0.066 \, \text{kcal/mol}$ and
$\sigma_{11} = 3.4 \, \text{\AA{}}$.  The \lmpcmd{bond\_coeff} provides
the equilibrium distance $r_0= 1.4 \, \text{\AA{}}$ and the
spring constant $k_\text{b} = 469 \, \text{kcal/mol/\AA{}}^2$ for the
harmonic potential imposed between two bonded carbon atoms.  The potential
is given by $U_\text{b} = k_\text{b} ( r - r_0)^2$.  The
\lmpcmd{angle\_coeff} gives the equilibrium angle $\theta_0$ and
constant for the potential between atoms forming an angle:
$U_\theta = k_\theta ( \theta - \theta_0)^2$.  The
\lmpcmd{dihedral\_coeff} and \lmpcmd{improper\_coeff} define the potentials
for the constraints between 4 atoms.

\begin{note}
  Rather than copying the contents of the file into the input, we
  incorporate it using the \lmpcmd{include} command.  Using \lmpcmd{include} allows
  us to conveniently reuse the parameter settings
  in other inputs or switch them with others.  This will become more general
  when using type labels~\cite{gissinger2024type}, which is shown in the next
  tutorial.
\end{note}

\paragraph{Prepare the initial state}

In this tutorial, a deformation will be applied to the CNT by displacing
the atoms located at its edges.  To achieve this, we will first isolate the
atoms at the two edges and place them into groups named \lmpcmd{rtop} and
\lmpcmd{rbot}.  Add the following lines to \flecmd{unbreakable.lmp},
just before the \lmpcmd{run 0} command:
\begin{lstlisting}
group carbon_atoms type 1
variable xmax equal bound(carbon_atoms,xmax)-0.5
variable xmin equal bound(carbon_atoms,xmin)+0.5
region rtop block ${xmax} INF INF INF INF INF
region rbot block INF ${xmin} INF INF INF INF
region rmid block ${xmin} ${xmax} INF INF INF INF
\end{lstlisting}
The first command includes all the atoms of type 1 (i.e.~all the atoms here)
in a group named \lmpcmd{carbon\_atoms}.
The variable $x_\text{max}$ corresponds to the coordinate of the
last atoms along $x$ minus 0.5~Å, and $x_\text{min}$ to the coordinate
of the first atoms along $x$ plus 0.5~Å.  Then, three regions are defined,
corresponding to the following: $x < x_\text{min}$, (\lmpcmd{rbot}, for region
bottom), $x_\text{min} > x > x_\text{max}$ (\lmpcmd{rmid}, for region middle),
and $x > x_\text{max}$ (\lmpcmd{rtop}, for region top).

\begin{note}
  So far, variables have been referenced
  dynamically during the run using the \lmpcmd{v\_} prefix, which
  evaluates the variable as it evolves over time.  Here, a dollar sign
  (\$) is used to expand the variable immediately at the time the input
  script is read.
\end{note}

Finally, let us define 3 groups of atoms corresponding to the atoms
in each of the 3 regions by adding to \flecmd{unbreakable.lmp}
just before the \lmpcmd{run 0} command:
\begin{lstlisting}
group cnt_top region rtop
group cnt_bot region rbot
group cnt_mid region rmid
set group cnt_top mol 1
set group cnt_bot mol 2
set group cnt_mid mol 3
\end{lstlisting}
With the three \lmpcmd{set} commands, we assign unique, otherwise unused molecule
IDs to atoms in those three groups.  A molecule ID is an
integer that groups atoms into a `molecule' for bookkeeping purposes, and can be
useful for tracking and post-processing.  We will use these IDs later to assign
different colors to these groups of atoms.

\begin{figure}
\centering
\includegraphics[width=\linewidth]{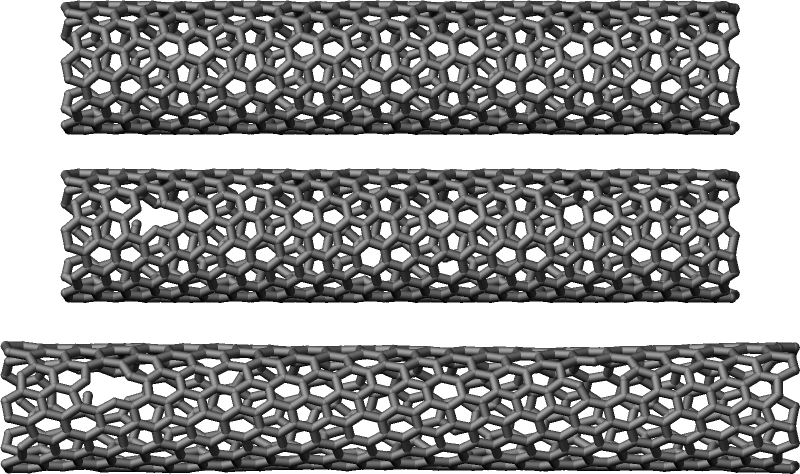}
\caption{The unbreakable CNT simulated during \hyperref[carbon-nanotube-label]{Tutorial 2}
before the removal of atoms (top), after the removal of 10 atoms from the \lmpcmd{rmid}
region (middle), and after deformation (bottom).}
\label{fig:CNT-unbreakable}
\end{figure}

Run the simulation using LAMMPS.  The number of atoms in each group is given in
the \guicmd{Output} window.  It is an important check to make sure that the number
of atoms in each group corresponds to what is expected, as shown here:
\begin{lstlisting}
700 atoms in group carbon_atoms
10 atoms in group cnt_top
10 atoms in group cnt_bot
680 atoms in group cnt_mid
\end{lstlisting}

Finally, to start from a less ideal state and create a system with some defects,
let us randomly delete a small fraction of the carbon atoms.  To avoid deleting
atoms that are too close to the edges, let us define a new region named \lmpcmd{rdel}
that starts at $2\,\text{\AA{}}$ from the CNT edges:
\begin{lstlisting}
variable xmax_del equal ${xmax}-2
variable xmin_del equal ${xmin}+2
region rdel block ${xmin_del} ${xmax_del} INF INF INF INF
group rdel region rdel
delete_atoms random fraction 0.02 no rdel NULL 2793 bond yes
\end{lstlisting}
The \lmpcmd{delete\_atoms} command randomly deletes $2\,\%$ of the atoms from
the \lmpcmd{rdel} group, here about 10 atoms (compare the top
and the middle panels in Fig.~\ref{fig:CNT-unbreakable}).

\paragraph{The molecular dynamics}

Let us give an initial temperature to the atoms of the group \lmpcmd{cnt\_mid}
by adding the following commands to \flecmd{unbreakable.lmp}:
\begin{lstlisting}
reset_atoms id sort yes
velocity cnt_mid create 300 48455 mom yes rot yes
\end{lstlisting}
Re-setting the atom IDs is necessary before using the \lmpcmd{velocity} command
when atoms were deleted, which is done here with the \lmpcmd{reset\_atoms} command.
The \lmpcmd{velocity} command assigns random initial velocities to the atoms of the middle
group \lmpcmd{cnt\_mid} from a uniform distribution, ensuring an initial temperature of $T = 300\,\text{K}$
for these atoms.

Let us specify the thermalization and the dynamics of the system.  Add the following
lines into \flecmd{unbreakable.lmp}:
\begin{lstlisting}
fix mynve1 cnt_top nve
fix mynve2 cnt_bot nve
fix mynvt cnt_mid nvt temp 300 300 100
\end{lstlisting}
The \lmpcmd{fix nve} commands are applied to the atoms of \lmpcmd{cnt\_top} and
\lmpcmd{cnt\_bot}, respectively, and will ensure that the positions of the atoms
from these groups are recalculated at every step.  The \lmpcmd{fix nvt} does the
same for the \lmpcmd{cnt\_mid} group, while also applying a Nos\'e-Hoover thermostat
with desired temperature of 300\,K~\cite{nose1984unified, hoover1985canonical}.

\begin{note}
  The Nosé-Hoover thermostat only controls the temperature of
  the atoms belonging to the specified \lmpcmd{cnt\_mid} group. Atoms outside
  this group are not affected.
\end{note}

To immobilize the atoms at the edges, let us add the following
commands to \flecmd{unbreakable.lmp}:
\begin{lstlisting}
fix mysf1 cnt_top setforce 0 0 0
fix mysf2 cnt_bot setforce 0 0 0
velocity cnt_top set 0 0 0
velocity cnt_bot set 0 0 0
\end{lstlisting}
The two \lmpcmd{setforce} commands cancel the forces applied on the atoms of the
two edges, respectively.  The cancellation of the forces is done at every step,
and along all 3 directions of space, $x$, $y$, and $z$, due to the use of
\lmpcmd{0 0 0}.  Although the force on these atoms is set to zero,
the \lmpcmd{fix} stores the force vector acting on the group \emph{before}
cancellation, which can later be extracted for analysis (see below).
The two \lmpcmd{velocity} commands set the initial velocities
along $x$, $y$, and $z$ to 0 for the atoms of \lmpcmd{cnt\_top} and
\lmpcmd{cnt\_bot}, respectively.  As a consequence of these last four commands,
the atoms of the edges will remain immobile during the simulation (or at least
they would if no other command was applied to them).

\begin{note}
  The \lmpcmdnote{velocity set} command adjusts the velocities of
  a group of atoms immediately but has no effect
  \emph{during} the simulation.  When \lmpcmdnote{velocity set} is used
  in combination with \lmpcmdnote{setforce 0 0 0}, as is the case here, the
  initial velocity will persist during the entire simulation, thus producing
  a constant velocity motion or no motion at all.
\end{note}

\paragraph{Outputs}

\begin{figure}
\centering
\includegraphics[width=\linewidth]{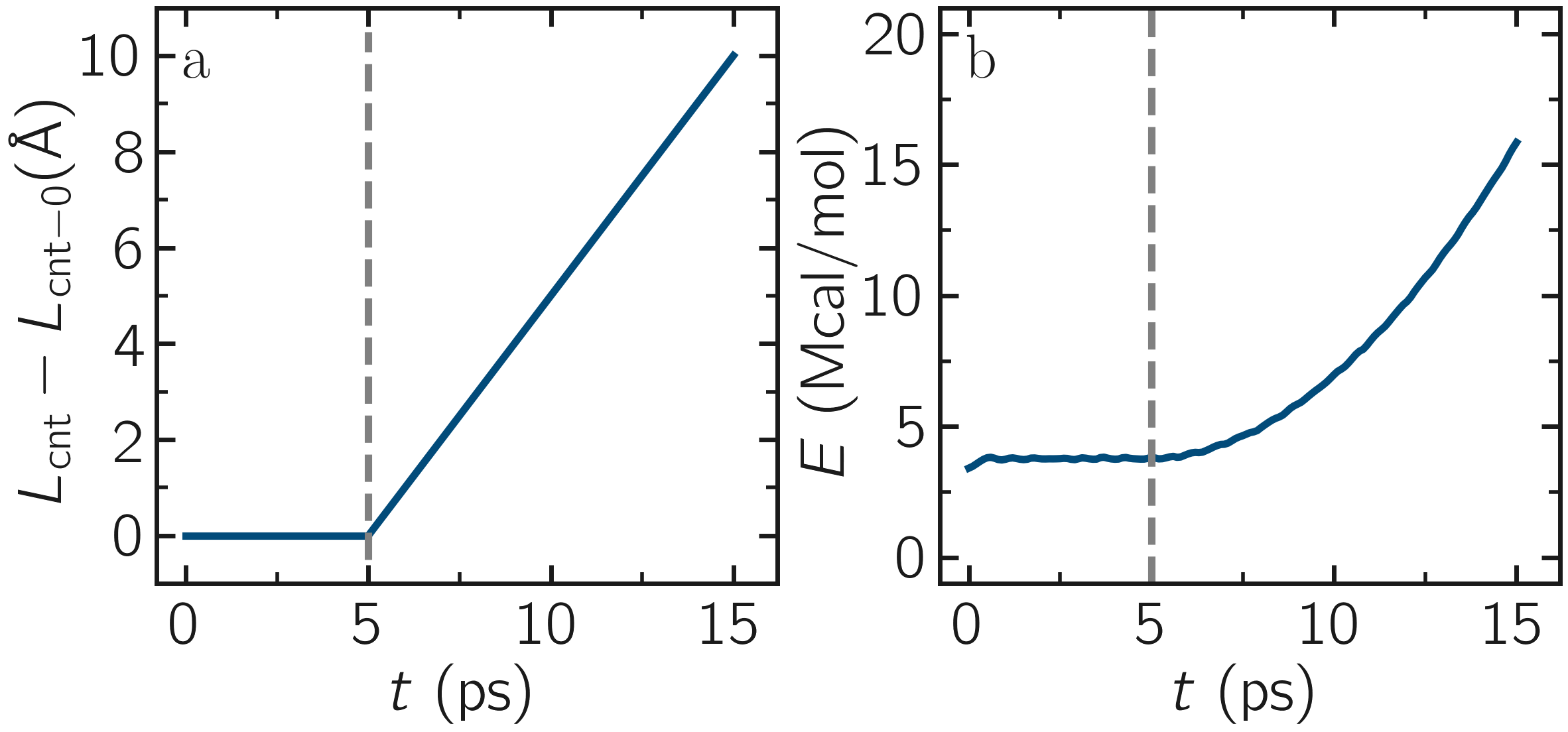}\\[-2ex]
\caption{a) Evolution of the length $L_\text{cnt}$ of the CNT with time,
as simulated during \hyperref[carbon-nanotube-label]{Tutorial 2}.
The CNT starts deforming at $t = 5\,\text{ps}$, and $L_\text{cnt-0}$ is the
CNT initial length.  b) Evolution of the total energy $E$ of the system
with time $t$.  Here, the potential is OPLS-AA, and the CNT is unbreakable.}
\label{fig:CNT-unbreakable-LE}
\end{figure}

Next, to measure the strain and stress applied to the CNT, let us create a
variable for the distance $L_\text{cnt}$ between the two edges,
as well as a variable $F_\text{cnt}$ for the force applied on the edges:
\begin{lstlisting}
variable Lcnt equal xcm(cnt_top,x)-xcm(cnt_bot,x)
variable Fcnt equal f_mysf1[1]-f_mysf2[1]
\end{lstlisting}
Here, the force is extracted from the fixes \lmpcmd{mysf1} and \lmpcmd{mysf2}
using \lmpcmd{f\_}, similarly to the use of \lmpcmd{v\_} to call a variable,
and \lmpcmd{c\_} to call a compute, as seen in \hyperref[lennard-jones-label]{Tutorial 1}.

Let us also add a \lmpcmd{dump image} command to visualize the system
every 500 steps:
\begin{lstlisting}
dump viz all image 500 myimage-*.ppm element type size &
  1000 400 zoom 6 shiny 0.3 fsaa yes bond atom 0.8 &
  view 0 90 box no 0.0 axes no 0.0 0.0
dump_modify viz pad 9 backcolor white adiam 1 0.85 bdiam 1 1.0
\end{lstlisting}
Let us run a small equilibration step to bring the system to the required
temperature before applying any deformation.  Replace the \lmpcmd{run 0 post no}
command in \flecmd{unbreakable.lmp} with the following lines:
\begin{lstlisting}
compute Tmid cnt_mid temp
thermo 100
thermo_style custom step temp etotal v_Lcnt v_Fcnt
thermo_modify temp Tmid line yaml

timestep 1.0
run 5000
\end{lstlisting}
With the \lmpcmd{thermo\_modify} command, we specify to LAMMPS that the
temperature $T_\mathrm{mid}$ of the middle group, \lmpcmd{cnt\_mid},
must be outputted, instead of the temperature of the entire system.
This choice is motivated by the presence of
frozen parts with an effective temperature of 0\,K, which makes the average
temperature of the entire system less relevant.  The \lmpcmd{thermo\_modify}
command also imposes the use of the YAML format that can easily be read by
Python (see below).
\begin{figure}
\centering
\includegraphics[width=0.55\linewidth]{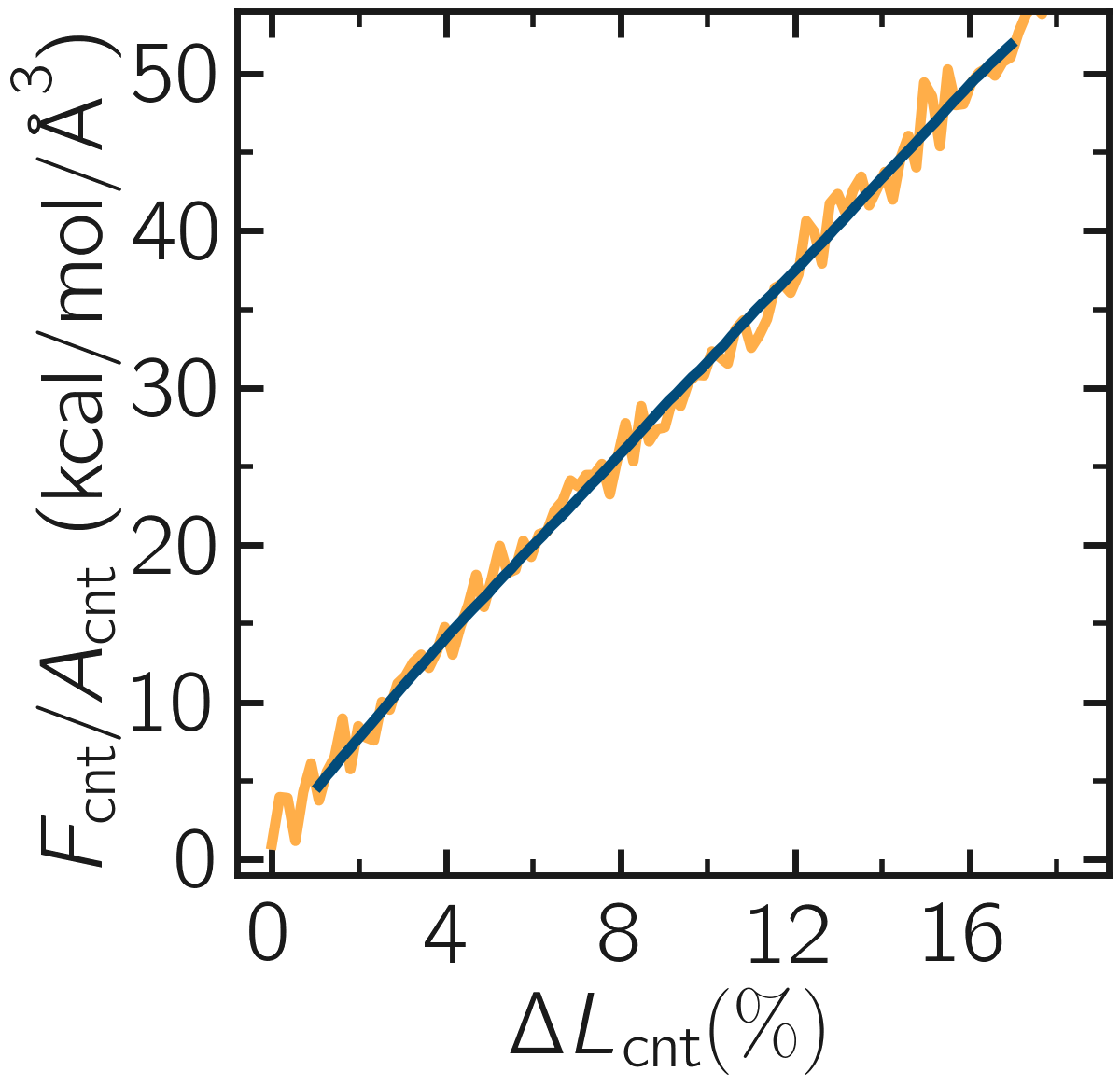}\\[-2ex]
\caption{Stress applied on the CNT during deformation, $F_\text{cnt}/A_\text{cnt}$,
where $F_\text{cnt}$ is the force and $A_\text{cnt}$ the CNT surface area,
as a function of the strain, $\Delta L_\text{cnt} = (L_\text{cnt}-L_\text{cnt-0}/L_\text{cnt-0})$, where
$L_\text{cnt}$ is the CNT length and $L_\text{cnt-0}$ the CNT initial length,
as simulated during \hyperref[carbon-nanotube-label]{Tutorial 2}.
Here, the potential is OPLS-AA, and the CNT is unbreakable.  The orange line
shows the raw data, and the blue line represents a time-averaged curve.
}
\label{fig:CNT-stress-strain-unbreakable}
\end{figure}

Let us impose a constant velocity deformation on the CNT
by combining the \lmpcmd{velocity set} command with previously defined
\lmpcmd{fix setforce}.  Add the following lines in the \lmpcmd{unbreakable.lmp}
file, right after the last \lmpcmd{run 5000} command:
\begin{lstlisting}
velocity cnt_top set 0.0005 0 0
velocity cnt_bot set -0.0005 0 0

run 10000
\end{lstlisting}
The chosen velocity for the deformation is $100\,\text{m/s}$, or
$0.001\,\text{\AA{}/fs}$.

Run the simulation using LAMMPS.  As can be seen from the variable $L_\text{cnt}$, the length
of the CNT increases linearly over time for $t > 5\,\text{ps}$ (Fig.~\ref{fig:CNT-unbreakable-LE}\,a),
as expected from the imposed constant velocity.  What you observe in the \guicmd{Slide Show}
windows should resembles Fig.~\ref{fig:CNT-unbreakable}.  The total energy of the system
shows a non-linear increase with $t$ once the deformation starts, which is expected
from the typical dependency of bond energy with bond distance,
$U_\text{b} = k_\text{b} \left( r - r_0 \right)^2$ (Fig.~\ref{fig:CNT-unbreakable-LE}\,b).

\paragraph{Importing YAML log file into Python}

Let us import the simulation data into Python, and generate a stress-strain curve.
Here, the stress is defined as $F_\text{cnt}/A_\text{cnt}$,
where $A_\text{cnt} = \pi r_\text{cnt}^2$ is the surface area of the
CNT, and $r_\text{cnt}=5.2$\,\AA{} the CNT radius.  The strain is defined
as $(L_\text{cnt}-L_\text{cnt-0})/L_\text{cnt-0}$, where $L_\text{cnt-0}$ is the initial CNT length.

Right-click inside the \guicmd{Output} window, and select
\guicmd{Export YAML data to file}.  Call the output \flecmd{unbreakable.yaml}, and save
it within the same folder as the input files, where a Python script named
\href{\filepath tutorial2/unbreakable-yaml-reader.py}{\dwlcmd{unbreakable-yaml-reader.py}} should also
be located.  When executed using Python, this .py file first imports
the \flecmd{unbreakable.yaml} file.  Then, a certain pattern is
identified and stored as a string character named `docs'.  The string is
then converted into a list, and $F_\text{cnt}$ and $L_\text{cnt}$
are extracted.  The stress and strain are then calculated, and the result
is saved in a data file named \flecmd{unbreakable.dat} using
the NumPy `savetxt' function.  `thermo[0]' can be used to access the
information from the first minimization run, and `thermo[1]' to access the
information from the second MD run.  The data extracted from
the \flecmd{unbreakable.yaml} file can then be used to plot the stress-strain
curve, see Fig.~\ref{fig:CNT-stress-strain-unbreakable}.

\subsubsection{Breakable bonds}

\begin{figure}
\centering
\includegraphics[width=\linewidth]{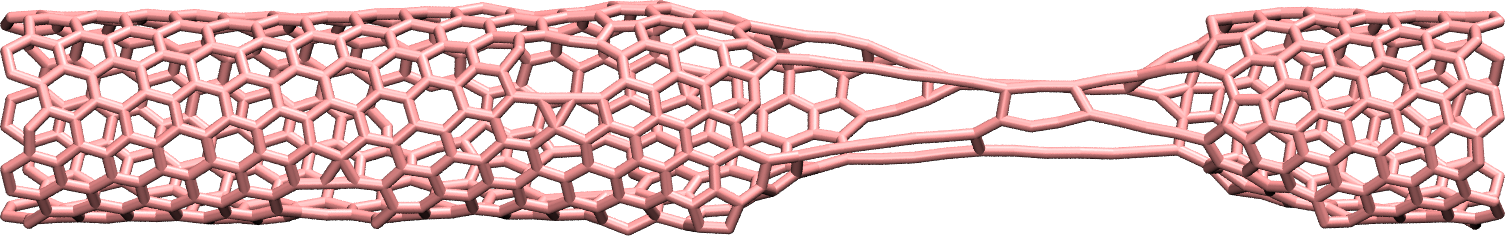}
\caption{CNT with broken bonds.  This image was generated using
VMD~\cite{vmd_home,humphrey1996vmd} with the \guicmd{DynamicBonds} representation.}
\label{fig:CNT-deformed-breakable}
\end{figure}

When using a conventional molecular force field, as we have just done,
the bonds between the atoms are non-breakable.  Let us perform a similar
simulation and deform a small CNT again, but this time with a reactive
force field that allows bonds to break if the applied deformation is
large enough.

\paragraph{Input file initialization}

Open the input named \href{\filepath tutorial2/breakable.lmp}{\dwlcmd{breakable.lmp}}
that should have been downloaded next to \lmpcmd{unbreakable.lmp} during
the tutorial setup.  There are only a few differences with the previous
input.  First, the AIREBO force field requires the \lmpcmd{metal} units
setting instead of \lmpcmd{real} for OPLS-AA.  A second difference is
the use of \lmpcmd{atom\_style atomic} instead of
\lmpcmd{molecular}, since no explicit bond information is required with
AIREBO.  The following commands are setting up the AIREBO force field:
\begin{lstlisting}
pair_style airebo 3.0
pair_coeff * * CH.airebo C
\end{lstlisting}
\begin{note}
  The AIREBO force field is a many-body
  potential, where interactions are not only between pairs of atoms,
  but also triples and quadruples representing angle and dihedral
  interactions.  This means that there are different rules for the
  \lmpcmd{pair\_coeff} command: there must be only one command that
  covers all permutations of atom types by using two '*' wildcards.
  After the potential file follows a list of elements.  These element
  names are used to look up the parameter sets in the potential file.
  There must be a list with as many elements as atom types following
  the filename.  In our system, there is only one atom type (1), which is
  mapped to the element 'C' in the \lmpcmd{pair\_coeff} command.
  Which elements are supported is determined by the contents of the
  potential file.
\end{note}
Here, \href{\filepath tutorial2/CH.airebo}{\dwlcmd{CH.airebo}} is the
file containing the parameters for AIREBO, and must be placed next to
\lmpcmd{breakable.lmp}.

\begin{note}
  With \lmpcmdnote{metal} units, time values are in units of picoseconds
  ($10^{-12}$\,s) instead of femtoseconds ($10^{-15}$\,s) in the case of
  \lmpcmdnote{real} units.  It is important to keep this in mind when
  setting parameters that are expressed units containing time, such as
  the timestep or the time constant of a thermostat, or velocities.
\end{note}

Since bonds, angles, and dihedrals do not need to be explicitly set when
using AIREBO, some simplification must be made to the \flecmd{.data}
file.  The new \flecmd{.data} file is named
\href{\filepath tutorial2/breakable.data}{\dwlcmd{breakable.data}},
and must be placed within the same folder as the input file.  Just like
\flecmd{unbreakable.data}, the \flecmd{breakable.data} contains the
information required for placing the atoms in the box, but no
bond/angle/dihedral information.  Another difference between the
\flecmd{unbreakable.data} and \flecmd{breakable.data} files is that,
here, a larger distance of 120~Å was used for the box size along
the $x$-axis, to allow for larger deformation of the CNT.

\paragraph{Start the simulation}

Here, let us perform a similar deformation as the previous one.
In \lmpcmd{breakable.lmp}, replace the \lmpcmd{run 0 post no} line with:
\begin{lstlisting}
fix mysf1 cnt_bot setforce 0 0 0
fix mysf2 cnt_top setforce 0 0 0
velocity cnt_bot set 0 0 0
velocity cnt_top set 0 0 0

variable Lcnt equal xcm(cnt_top,x)-xcm(cnt_bot,x)
variable Fcnt equal f_mysf1[1]-f_mysf2[1]

dump viz all image 500 myimage.*.ppm type type size 1000 400 &
  zoom 4 shiny 0.3 adiam 1.5 box no 0.01 view 0 90 &
  shiny 0.1 fsaa yes
dump_modify viz pad 5 backcolor white acolor 1 gray

compute Tmid cnt_mid temp
thermo 100
thermo_style custom step temp etotal v_Lcnt v_Fcnt
thermo_modify temp Tmid line yaml

timestep 0.0005
run 10000
\end{lstlisting}
Note the relatively small timestep of $0.0005$\,ps ($= 0.5$\,fs) used.  Reactive force
fields like AIREBO usually require a smaller timestep than conventional ones.  When running
\flecmd{breakable.lmp} with LAMMPS, you can see that the temperature deviates
from the target temperature of $300\,\text{K}$ at the start of the equilibration,
but that after a few steps, it reaches the target value.

\begin{note}
  Bonds cannot be displayed by the \lmpcmdnote{dump image} when using
  the \lmpcmdnote{atom\_style atomic}, as it contains no bonds.  A
  \hyperref[tip-dynamic-bonds]{tip for displaying bonds} with the
  present system using LAMMPS is provided at the end of the tutorial.
  You can also use external tools like VMD or OVITO (see the
  \hyperref[tip-external-viz]{tip for tutorial 3}).
\end{note}

\paragraph{Launch the deformation}

\begin{figure}
\centering
\includegraphics[width=\linewidth]{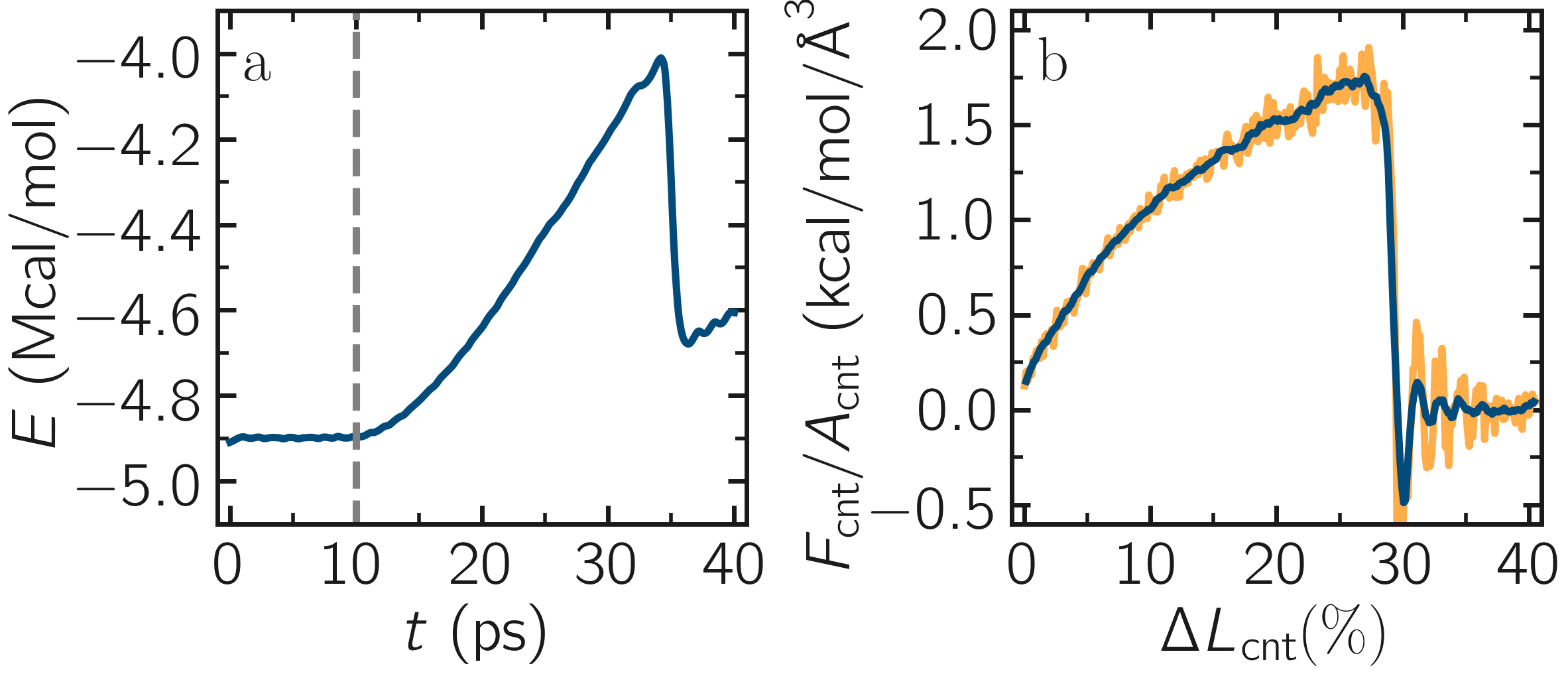}\\[-2ex]
\caption{a) Evolution of the total energy $E$ of the CNT with time $t$.
b) Stress applied on the CNT during deformation, $F_\text{cnt}/A_\text{cnt}$,
where $F_\text{cnt}$ is the force and $A_\text{cnt}$ the CNT surface area,
as a function of the strain, $\Delta L_\text{cnt} = (L_\text{cnt}-L_\text{cnt-0}/L_\text{cnt-0})$, where
$L_\text{cnt}$ is the CNT length and $L_\text{cnt-0}$ the CNT initial length,
as simulated during \hyperref[carbon-nanotube-label]{Tutorial 2}.
Here, the potential is AIREBO, and the CNT is breakable.  The orange line
shows the raw data, and the blue line represents a time-averaged curve.}
\label{fig:CNT-breakable-energy-stress}
\end{figure}

After equilibration, let us set the velocity of the edges equal to
$75~\text{m/s}$ (or $0.75~\text{\AA{}/ps}$) and run for a longer duration than
previously.  Add the following lines into \flecmd{breakable.lmp}:
\begin{lstlisting}
velocity cnt_top set 0.75 0 0
velocity cnt_bot set -0.75 0 0

run 30000
\end{lstlisting}
Run the simulation.  Some bonds are expected to break before the end of the
simulation (Fig.~\ref{fig:CNT-deformed-breakable}).

Looking at the evolution of the energy, one can see that the total
energy $E$ is initially increasing with the deformation.  When bonds
break, the energy relaxes abruptly, as can be seen near $t=32~\text{ps}$
in Fig.~\ref{fig:CNT-breakable-energy-stress}\,a.  Using a similar
script as previously,
i.e.,~\href{\filepath tutorial2/unbreakable-yaml-reader.py}{\dwlcmd{unbreakable-yaml-reader.py}},
import the data into Python and generate the stress-strain curve
(Fig.~\ref{fig:CNT-breakable-energy-stress}\,b).  The stress-strain
curve reveals a linear (elastic) regime where
$F_\text{cnt} \propto \Delta L_\text{cnt}$ for
$\Delta L_\text{cnt} < 5\,\%$, and a non-linear (plastic) regime for
$5\,\% < \Delta L_\text{cnt} < 25\,\%$.

\paragraph{Tip: bonds representation with AIREBO}
\label{tip-dynamic-bonds}

In the input file
\href{\filepath tutorial2/solution/breakable-with-tip.lmp}{\dwlcmd{solution/breakable-with-tip.lmp}},
which is an alternate solution for \flecmd{breakable.lmp}, a trick is
used to represent bonds while using AIREBO.  A detailed explanation of
the script is beyond the scope of the present tutorial.  In short, the
trick is to use AIREBO with the \lmpcmd{molecular} atom style, and use
the \lmpcmd{fix bond/break} and \lmpcmd{fix bond/create/angle} commands
to update the status of the bonds during the simulation:
\begin{lstlisting}
fix break all bond/break 1000 1 2.5
fix form all bond/create/angle 1000 1 1 2.0 1 aconstrain 90.0 180
\end{lstlisting}
This ``hack'' works because AIREBO does not pay any attention to bonded
interactions and computes the bond topology dynamically inside the pair
style.  Thus adding bonds of bond style \lmpcmd{zero} does not add any
interactions but allows the visualization of them with \lmpcmd{dump
  image}.  It is required to change the \lmpcmd{special\_bonds}
setting to disable any neighbor list exclusions as they are common for
force fields with explicit bonds.
\begin{lstlisting}
bond_style zero
bond_coeff 1 1.4
special_bonds lj/coul 1.0 1.0 1.0
\end{lstlisting}

\subsection{Tutorial 3: Polymer in water}
\label{all-atom-label}

\begin{figure}
\centering
\includegraphics[width=0.55\linewidth]{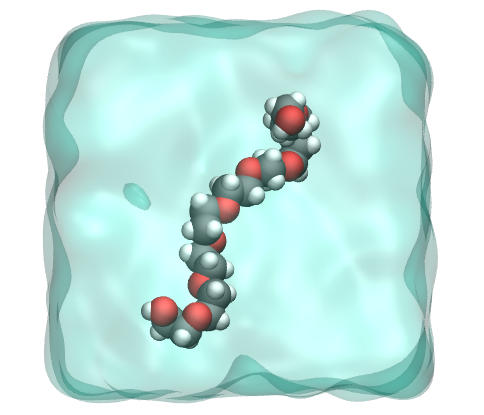}
\caption{The polymer molecule (PEG - polyethylene glycol) solvated in water as
simulated during \hyperref[all-atom-label]{Tutorial 3}.  Water molecules are
represented as a transparent continuum field for clarity.}
\label{fig:PEG}
\end{figure}

The goal of this tutorial is to use LAMMPS to solvate a small
hydrophilic polymer molecule (PEG - polyethylene glycol) in a reservoir of water
(Fig.~\ref{fig:PEG}).  Once the water reservoir is properly equilibrated
at the desired temperature and pressure, the polymer molecule is added
and a constant stretching force is applied to both ends of the polymer.
The evolution of the polymer length is measured as a function of time.
The {GROMOS} 54A7 force field~\cite{schmid2011definition} is used for the
PEG, the SPC/Fw model~\cite{wu2006flexible} is used for the water, and
the long-range Coulomb interactions are solved using the PPPM
solver~\cite{luty1996calculating}.  This tutorial was inspired by a
publication by Liese and coworkers, in which molecular dynamics
simulations are compared with force spectroscopy experiments, see
Ref.\,~\citenum{liese2017hydration}.

\begin{note}
  When mixing different force fields, as is done here with GROMOS
  and SPC/Fw, users should exercise caution.  The choices made in these tutorials
  prioritize progressive learning of LAMMPS functionality
  over strict physical accuracy.  While GROMOS is commonly used with water
  models from the SPC family~\cite{oostenbrink2004biomolecular},
  the inter-compatibility of force fields is not generally guaranteed.
\end{note}

\subsubsection{Preparing the water reservoir}

In this tutorial, the water reservoir is first prepared in the absence of the polymer.
A rectangular box of water is created and equilibrated at ambient temperature and
pressure.  The SPC/Fw water model is used~\cite{wu2006flexible}, which is
a flexible variant of the rigid SPC (simple point charge) model~\cite{berendsen1981interaction}.
To set up this tutorial, select \guicmd{Start Tutorial 3} from the
\guicmd{Tutorials} menu of \lammpsgui{} and follow the instructions.
The editor should display the following content corresponding to \flecmd{water.lmp}:
\begin{lstlisting}
units real
atom_style full
bond_style harmonic
angle_style harmonic
dihedral_style harmonic
pair_style lj/cut/coul/long 10
kspace_style ewald 1e-5
special_bonds lj 0.0 0.0 0.5 coul 0.0 0.0 1.0 angle yes
\end{lstlisting}
With the unit style \lmpcmd{real}, masses are in g/mol, distances in Å,
time in fs, and energies in kcal/mol.  With the \lmpcmd{atom\_style
  full}, each atom is a dot with a mass and a charge that can be linked
by bonds, angles, dihedrals, and/or impropers.  The
\lmpcmd{bond\_style}, \lmpcmd{angle\_style}, and
\lmpcmd{dihedral\_style} commands define the potentials for the bonds,
angles, and dihedrals used in the simulation, here \lmpcmd{harmonic}.
With the \lmpcmd{pair\_style} named \lmpcmd{lj/cut/coul/long}, atoms
interact through both a Lennard-Jones (LJ) potential and Coulomb
interactions.  The value of $10\,\text{\AA{}}$ is the cutoff, and the
\lmpcmd{kspace\_style} command defines the long-range solver for the Coulomb
interactions~\cite{ewald1921berechnung}.  Finally, the
\lmpcmd{special\_bonds} command, which was already seen in
\hyperref[carbon-nanotube-label]{Tutorial 2}, sets the LJ and Coulomb
weighting factors for the interaction between neighboring atoms.

\begin{note}
  With Coulomb interactions, additional rules
  apply to the \lmpcmd{pair\_coeff} command: (a) atom type values
  only matter for assignment of LJ potential parameters; (b) for Coulomb interactions,
  there are no parameters outside the cutoff, and when using a
  \lmpcmd{coul/long} pair style, that cutoff can only be set globally
  for all atoms with the \lmpcmd{pair\_style}  command;  (c) for
  Coulomb interactions, only the per-atom charge and any
  \lmpcmd{special\_bonds} exclusions are relevant.
\end{note}

\noindent Let us create a 3D simulation box of dimensions $6 \times 3 \times 3 \; \text{nm}^3$,
and make space for 8 atom types (2 for the water, 6 for the polymer), 7 bond types
(1 for the water, 6 for the polymer), 8 angle types (1 for the water, 7 for the polymer),
and 4 dihedral types (only for the polymer).  Copy the following lines into \flecmd{water.lmp}:
\begin{lstlisting}
 region box block -30 30 -15 15 -15 15
 create_box 8 box &
 bond/types 7 &
 angle/types 8 &
 dihedral/types 4 &
 extra/bond/per/atom 3 &
 extra/angle/per/atom 6 &
 extra/dihedral/per/atom 10 &
 extra/special/per/atom 14
\end{lstlisting}
The \lmpcmd{extra/x/per/atom} commands reserve memory for adding bond topology
data later. We use the file \href{\filepath tutorial3/parameters.inc}{\dwlcmd{parameters.inc}}
to set all the parameters (masses, interaction energies, bond equilibrium
distances, etc).  Thus add to \flecmd{water.lmp} the line:
\begin{lstlisting}
include parameters.inc
\end{lstlisting}

\begin{note}
  This tutorial uses type labels~\cite{gissinger2024type} to map each
  numeric atom type to a string (see the \flecmd{parameters.inc} file):
  \lmpcmdnote{labelmap atom 1 OE 2 C 3 HC 4 H 5 CPos 6 OAlc 7 OW 8 HW}
  Therefore, the oxygen and hydrogen atoms of water (respectively types
  7 and 8) can be referred to as `OW' and `HW', respectively.  Similar
  maps are used for the bond types, angle types, and dihedral types.
\end{note}

\noindent Let us create water molecules.  To do so, let us import a molecule template called
\flecmd{water.mol} and then randomly create 700 molecules.  Add the following
lines into \flecmd{water.lmp}:
\begin{lstlisting}
molecule h2omol water.mol
create_atoms 0 random 700 87910 NULL mol h2omol 454756 &
  overlap 1.0 maxtry 50
\end{lstlisting}
The first parameter is 0, meaning that the atom types from
the \flecmd{water.mol} file will be used.
The \lmpcmd{overlap 1.0} option of the \lmpcmd{create\_atoms} command ensures
that no atoms are placed exactly in the same position, as this would cause the
simulation to crash.  The \lmpcmd{maxtry 50} asks LAMMPS to try at most 50 times
to insert the molecules, which is useful in case some insertion attempts are
rejected due to overlap.  In some cases, depending on the system and the values
of \lmpcmd{overlap} and \lmpcmd{maxtry}, LAMMPS may not create the desired number
of molecules.  Always check the number of created atoms in the \lmpcmd{log} file
(or in the \guicmd{Output} window), where you should see:
\begin{lstlisting}
Created 2100 atoms
\end{lstlisting}
When LAMMPS fails to create the desired number of molecules, a WARNING
appears.  The molecule template called
\href{\filepath tutorial3/water.mol}{\dwlcmd{water.mol}}
must be downloaded and saved
next to \flecmd{water.lmp}.  This template contains the necessary
structural information of a water molecule, such as the number of atoms,
or the IDs of the atoms that are connected by bonds and angles.

\begin{figure}
\centering
\includegraphics[width=\linewidth]{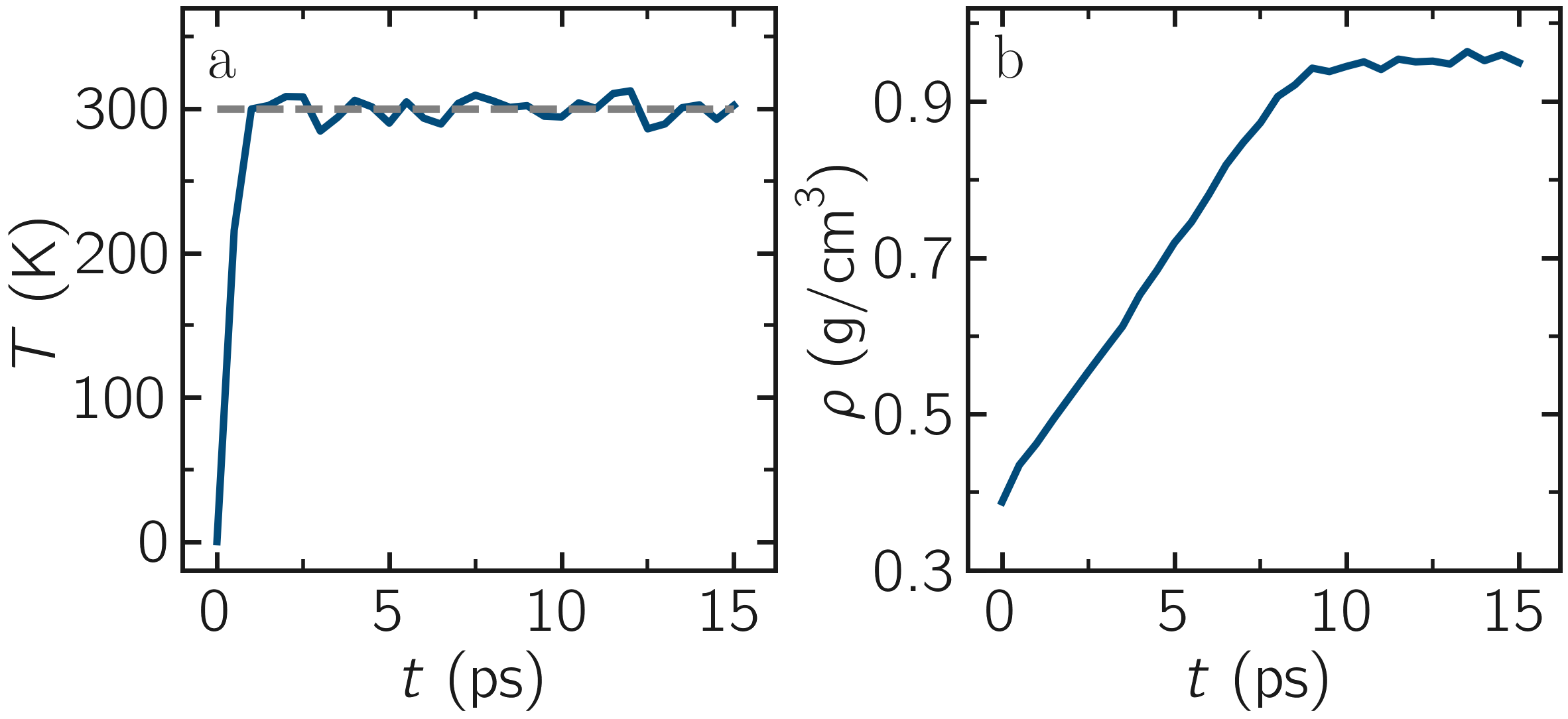}\\[-2ex]
\caption{a) Temperature, $T$, of the water reservoir from \hyperref[all-atom-label]{Tutorial 3}
as a function of the time, $t$.  The horizontal dashed line is the target temperature of 300\,K.
b) Evolution of the system density, $\rho$, with $t$.}
\label{fig:PEG-density}
\end{figure}

Then, let us organize the atoms of types OW and HW of the water
molecules in a group named \lmpcmd{H2O} and perform a small energy
minimization.  The energy minimization is mandatory here because of the
small \lmpcmd{overlap} value of 1~Å chosen in the \lmpcmd{create\_atoms}
command.  Add the following lines into \flecmd{water.lmp}:
\begin{lstlisting}
group H2O type OW HW
minimize 1.0e-4 1.0e-6 100 1000
reset_timestep 0
\end{lstlisting}
Resetting the step of the simulation to 0 using the
\lmpcmd{reset\_timestep} command is optional.
It is used here because the number of iterations performed by the \lmpcmd{minimize}
command is usually not a round number, since the minimization stops when one of
four criteria is reached, which can disrupt the intended frequency
of outputs such as \lmpcmd{dump} commands that depend on the timestep count.
We will use \lmpcmd{fix npt} to control the temperature
and pressure of the molecules with a Nosé-Hoover thermostat and barostat,
respectively~\cite{nose1984unified, hoover1985canonical, martyna1994constant}.
Add the following line into \flecmd{water.lmp}:
\begin{lstlisting}
fix mynpt all npt temp 300 300 100 iso 1 1 1000
\end{lstlisting}
The \lmpcmd{fix npt} allows us to impose both a temperature of $300\,\text{K}$
(with a damping constant of $100\,\text{fs}$), and a pressure of 1 atmosphere
(with a damping constant of $1000\,\text{fs}$).  With the \lmpcmd{iso} keyword,
the three dimensions of the box will be re-scaled isotropically,
maintaining the same proportion in all directions.

\begin{figure}
\centering
\includegraphics[width=\linewidth]{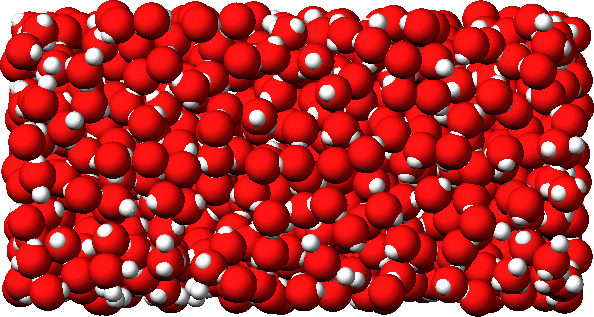}
\caption{The water reservoir from \hyperref[all-atom-label]{Tutorial 3}
after equilibration.  Oxygen atoms are in red, and hydrogen atoms are in white.}
\label{fig:PEG-water}
\end{figure}

Let us output the system into images by adding the following commands to \flecmd{water.lmp}:
\begin{lstlisting}
dump viz all image 250 myimage-*.ppm type type &
  shiny 0.1 box no 0.01 view 0 90 zoom 3 size 1000 600
dump_modify viz backcolor white &
  acolor OW red acolor HW white &
  adiam OW 3 adiam HW 1.5
\end{lstlisting}
Let us also extract the volume and density, among others, every 500 steps:
\begin{lstlisting}
thermo 500
thermo_style custom step temp etotal vol density
\end{lstlisting}
With the real units system, the volume is in Å$^3$, and
the density is in g/cm$^3$.

Finally, let us set the timestep to 1.0 fs, and run the simulation for 15~ps by
adding the following lines into \flecmd{water.lmp}:
\begin{lstlisting}
timestep 1.0
run 15000

write_restart water.restart
\end{lstlisting}
The final state is saved in a binary file named \flecmd{water.restart}.
Run the input using LAMMPS.  The system reaches its equilibrium temperature
after just a few picoseconds, and its equilibrium density after approximately
10~picoseconds (Fig.~\ref{fig:PEG-density}).  A snapshot of the equilibrated
system can also be seen in Fig.~\ref{fig:PEG-water}.

\begin{note}
  The binary file created by the \lmpcmdnote{write\_restart} command contains the
  complete state of the simulation, including atomic positions, velocities, and
  box dimensions (similar to \lmpcmdnote{write\_data}), but also the groups,
  the compute, or the \lmpcmdnote{atom\_style}.  Use the \guicmd{Inspect Restart}
  option of the \lammpsgui{} to vizualize the content saved in \flecmd{water.restart}.
\end{note}

\subsubsection{Solvating the PEG in water}

\begin{figure}
\centering
\includegraphics[width=0.8\linewidth]{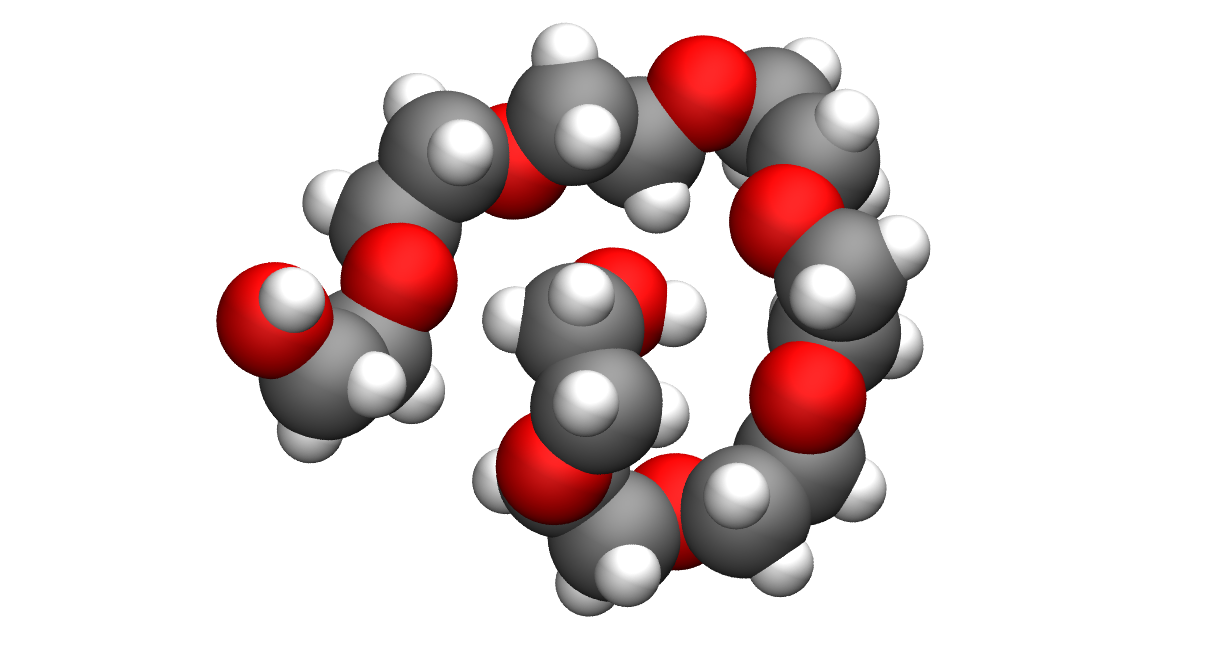}
\caption{The PEG molecule from \hyperref[all-atom-label]{Tutorial 3}.
The carbon atoms are in gray, the oxygen atoms in red, and the hydrogen atoms in white.}
\label{fig:PEG-in-vacuum}
\end{figure}

Now that the water reservoir is equilibrated, we can safely add the PEG polymer
to the water.  The PEG molecule topology was downloaded from the ATB repository
\cite{malde2011automated, oostenbrink2004biomolecular}.  It has a formula
$\text{C}_{16}\text{H}_{34}\text{O}_{9}$, and the parameters are taken from
the {GROMOS} 54A7 force field~\cite{schmid2011definition} (Fig.~\ref{fig:PEG-in-vacuum}).

Open the file named \flecmd{merge.lmp} that was downloaded
alongside \flecmd{water.lmp} during the tutorial setup.  It only contain one line:
\begin{lstlisting}
read_restart water.restart
\end{lstlisting}
Most of the commands that were initially present in \flecmd{water.lmp}, such as
the \lmpcmd{units} of the \lmpcmd{atom\_style} commands do not need to be repeated,
as they were saved within the \flecmd{.restart} file.  There is also no need to
re-include the parameters from the \flecmd{.inc} file.  The \lmpcmd{kspace\_style}
command, however, is not saved by the \lmpcmd{write\_restart} command and must be
repeated.  Since Ewald summation is not the most efficient choice for such dense
system, let us use PPPM (for particle-particle particle-mesh) for the rest
of the tutorial.  Add the following command to \flecmd{merge.lmp}:
\begin{lstlisting}
kspace_style pppm 1e-5
\end{lstlisting}
Using the molecule template for the polymer called
\href{\filepath tutorial3/peg.mol}{\dwlcmd{peg.mol}},
let us create a single molecule in the middle of the box by adding the following
commands to \flecmd{merge.lmp}:
\begin{lstlisting}
molecule pegmol peg.mol
create_atoms 0 single 0 0 0 mol pegmol 454756
\end{lstlisting}
Let us create a group for the atoms of the PEG (the previously created
group H2O was saved by the restart and can be omitted):
\begin{lstlisting}
group PEG type C CPos H HC OAlc OE
\end{lstlisting}
Water molecules that are overlapping with the PEG must be deleted to avoid future
crashing.  Add the following line into \flecmd{merge.lmp}:
\begin{lstlisting}
delete_atoms overlap 2.0 H2O PEG mol yes
\end{lstlisting}
Here the value of 2.0~Å for the overlap cutoff was fixed arbitrarily and can
be chosen through trial and error.  If the cutoff is too small, the simulation will
crash because atoms that are too close to each other undergo forces
that can be extremely large.  If the cutoff is too large, too many water
molecules will unnecessarily be deleted.

Let us use the \lmpcmd{fix npt} to control the temperature, as
well as the pressure by allowing the box size to be rescaled along the $x$-axis:
\begin{lstlisting}
fix mynpt all npt temp 300 300 100 x 1 1 1000
\end{lstlisting}
Let us also use the \lmpcmd{recenter} command to always keep the PEG at
the position $(0,~0,~0)$:
\begin{lstlisting}
fix myrct PEG recenter 0 0 0 shift all
\end{lstlisting}

\begin{note}
  Note that the \lmpcmd{recenter} command has no impact on the dynamics,
  it simply repositions the frame of reference so that any drift of the
  system is ignored, which can be convenient for visualizing and analyzing
  the system. However, be aware that using \lmpcmd{fix recenter} can sometimes
  mask underlying issues in the simulation, such as a net momentum or the so-called
  ``flying ice cube syndrome''~\cite{wong2016good}.
\end{note}

\noindent Let us create images of the systems:
\begin{lstlisting}
dump viz all image 250 myimage-*.ppm type type size 1100 600 &
  box no 0.1 shiny 0.1 view 0 90 zoom 3.3 fsaa yes bond atom 0.8
dump_modify viz backcolor white acolor OW red adiam OW 0.2 &
  acolor OE darkred adiam OE 2.6 acolor HC white adiam HC 1.4 &
  acolor H white adiam H 1.4 acolor CPos gray adiam CPos 2.8 &
  acolor HW white adiam HW 0.2 acolor C gray  adiam C 2.8 &
  acolor OAlc darkred adiam OAlc 2.6
thermo 500
\end{lstlisting}
\begin{figure}
\centering
\includegraphics[width=\linewidth]{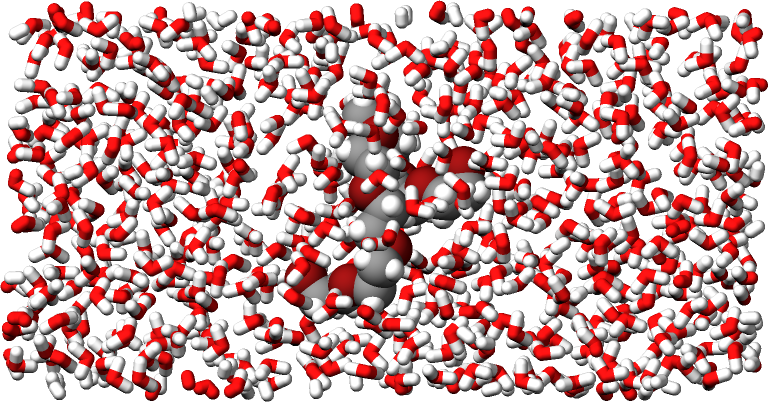}
\caption{The PEG molecule solvated in water during
\hyperref[all-atom-label]{Tutorial 3}.}
\label{fig:PEG-solvated}
\end{figure}
Finally, to perform a short equilibration and save the final state to
a \lmpcmd{.restart} file, add the following lines to the input:
\begin{lstlisting}
timestep 1.0
run 10000

write_restart merge.restart
\end{lstlisting}
Run the simulation using LAMMPS.  From the outputs, you can make
sure that the temperature remains close to the
target value of $300~\text{K}$ throughout the entire simulation, and that
the volume and total energy are almost constant, indicating
that the system was in a reasonable configuration from the start.
See a snapshot of the system in Fig.~\ref{fig:PEG-solvated}.

\subsubsection{Stretching the PEG molecule}

Here, a constant force is applied to both ends of the PEG molecule until it
stretches.  Open the file named \flecmd{pull.lmp}, which
only contains two lines:
\begin{lstlisting}
kspace_style pppm 1e-5
read_restart merge.restart
\end{lstlisting}
Next, we'll create new atom groups, each containing a single oxygen atom.  The atoms of type OAlc
correspond to the hydroxyl (alcohol) group oxygen atoms located at the ends
of the PEG molecule, which we will use to apply the force.  Add the
following lines to \flecmd{pull.lmp}:
\begin{lstlisting}
group ends type OAlc
variable xcm equal xcm(ends,x)
variable oxies atom type==label2type(atom,OAlc)
variable end1 atom v_oxies*(x>v_xcm)
variable end2 atom v_oxies*(x<v_xcm)
group topull1 variable end1
group topull2 variable end2
\end{lstlisting}
These lines identify the oxygen atoms (type OAlc) at the ends of the PEG
molecule and calculates their center of mass along the $x$-axis.  It then
divides these atoms into two groups, \lmpcmd{end1} (i.e.,~the OAlc atom to
the right of the center) and \lmpcmd{end2} (i.e.,~the OAlc atom to the right
of the center), for applying force during the stretching process.

\begin{figure}
\centering
\includegraphics[width=\linewidth]{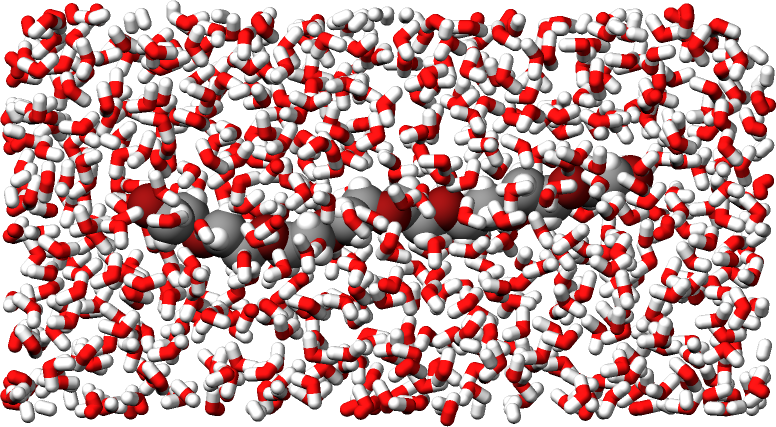}
\caption{PEG molecule stretched along the $x$ direction in water
as simulated during \hyperref[all-atom-label]{Tutorial 3}.}
\label{fig:PEG-in-water}
\end{figure}

Add the following \lmpcmd{dump} command to create images of the system:
\begin{lstlisting}
dump viz all image 250 myimage-*.ppm type &
  type shiny 0.1 box no 0.01 &
  view 0 90 zoom 3.3 fsaa yes bond atom 0.8 size 1100 600
dump_modify viz backcolor white &
  acolor OW red acolor HW white &
  acolor OE darkred acolor OAlc darkred &
  acolor C gray acolor CPos gray &
  acolor H white acolor HC white &
  adiam OW 0.2 adiam HW 0.2 &
  adiam C 2.8 adiam CPos 2.8 adiam OAlc 2.6 &
  adiam H 1.4 adiam HC 1.4 adiam OE 2.6
\end{lstlisting}
Let us use a single Nosé-Hoover thermostat applied to all the atoms,
and let us keep the PEG in the center of the box, by adding
the following lines to \flecmd{pull.lmp}:
\begin{lstlisting}
timestep 1.0
fix mynvt all nvt temp 300 300 100
fix myrct PEG recenter 0 0 0 shift all
\end{lstlisting}
To investigate the stretching of the PEG molecule, let us compute its radius of
gyration~\cite{fixmanRadiusGyrationPolymer1962a} and the angles of its dihedral
constraints using the following commands:
\begin{lstlisting}
compute rgyr PEG gyration
compute dphi PEG dihedral/local phi
\end{lstlisting}
The radius of gyration can be directly printed with the \lmpcmd{thermo\_style} command:
\begin{lstlisting}
thermo_style custom step temp etotal c_rgyr
thermo 250
dump mydmp all local 100 pull.dat index c_dphi
\end{lstlisting}
By contrast with the radius of gyration (compute \lmpcmd{rgyr}), the dihedral angle
$\phi$ (compute \lmpcmd{dphi}) is returned as a vector by the \lmpcmd{compute dihedral/local}
command and must be written to a file using the \lmpcmd{dump local} command.

\begin{figure}
\centering
\includegraphics[width=\linewidth]{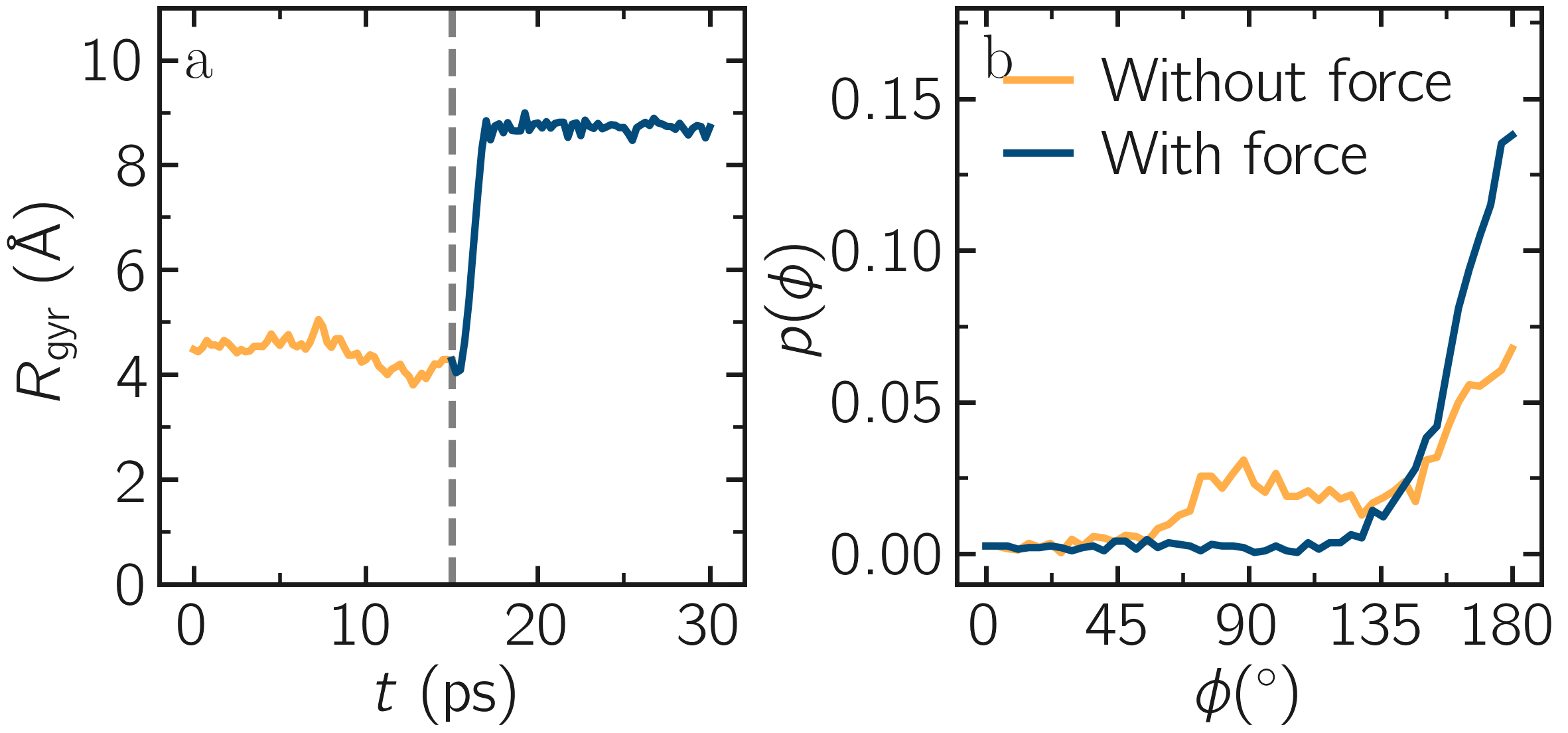}\\[-2ex]
\caption{a) Evolution of
the radius of gyration $R_\text{gyr}$ of the PEG molecule
from \hyperref[all-atom-label]{Tutorial 3}, with the force
applied starting at $t = 15\,\text{ps}$.  b) Histograms of the dihedral angles of type 1
in the absence (orange) and in the presence (blue) of the applied force.}
\label{fig:PEG-distance}
\end{figure}

Finally, let us simulate 15 picoseconds without any external force:
\begin{lstlisting}
run 15000
\end{lstlisting}
This initial run will serve as a benchmark to quantify the changes caused by
the applied force in later steps.  Next, let us apply a force to the two selected
oxygen atoms using two \lmpcmd{addforce} commands, and then run the simulation
for an extra 15~ps:
\begin{lstlisting}
fix myaf1 topull1 addforce 10 0 0
fix myaf2 topull2 addforce -10 0 0
run 15000
\end{lstlisting}
Each applied force has a magnitude of $10\,\text{kcal/mol/\AA{}}$, corresponding to $0.67\,\text{nN}$.
This value was chosen to be sufficiently large to overcome both the thermal agitation and
the entropic contributions from the molecules.

Run the \flecmd{pull.lmp} file using LAMMPS.  From the generated images of the system,
you should observe that the PEG molecule eventually aligns
in the direction of the applied force (as seen in Fig.~\ref{fig:PEG-in-water}).
The evolutions of the radius of gyration over
time indicates that the PEG quickly adjusts to the external force
(Fig.~\ref{fig:PEG-distance}\,a).  Additionally, from the values of the dihedral angles
printed in the \flecmd{pull.dat} file, you can create a histogram
of dihedral angles for a specific type.  For example, the angle $\phi$ for dihedrals
of type 1 (C-C-OE-C) is shown in Fig.~\ref{fig:PEG-distance}\,b.

\paragraph{Tip: using external visualization tools}
\label{tip-external-viz}

Trajectories can be visualized using external tools such as VMD or
OVITO~\cite{humphrey1996vmd, stukowski2009visualization}.  To do so, the IDs and
positions of the atoms must be regularly written to a file during the
simulation.  This can be accomplished by adding a \lmpcmd{dump} command
to the input file.  For instance, create a duplicate of
\flecmd{pull.lmp} and name it
\href{\filepath tutorial3/solution/pull-with-tip.lmp}{\dwlcmd{pull-with-tip.lmp}}.
Then, replace the existing \lmpcmd{dump} and \lmpcmd{dump\_modify} commands with:
\begin{lstlisting}
dump mydmp all atom 1000 pull.lammpstrj
\end{lstlisting}
Running the \flecmd{pull-with-tip.lmp} file using LAMMPS will generate a trajectory file
named \flecmd{pull.lammpstrj}, which can be opened in OVITO or VMD.

\begin{note}
  Since the default trajectory dump file does not contain information about
  topology and elements, it is usually preferred to first write out a
  data file and import it directly (in the case of OVITO) or convert it
  to a PSF file (for VMD).  This allows the topology to be loaded before
  \emph{adding} the trajectory file to it.  When using \lammpsgui{},
  this process can be automated through the \guicmd{View in OVITO} or
  \guicmd{View in VMD} options in the \guicmd{Run} menu.  Afterwards
  only the trajectory dump needs to be added.  Alternatively, the
  \lmpcmd{dump custom} command can be combined with \lmpcmd{dump} command to
  include element names in the dump file and simplify visualization.
\end{note}

\begin{note}
  Microstates collected during a simulation in the form of a trajectory
  can be analyzed within LAMMPS using the \lmpcmd{rerun} command.  This is
  particularly useful, for example, for computing properties not set up in
  the original simulation without having to run it again.  A possible use of
  the \lmpcmd{rerun} command is estimating the self-diffusion coefficient
  by using the \lmpcmd{compute msd} command~\cite{frenkel2023understanding}.
\end{note}

\subsection{Tutorial 4: Nanosheared electrolyte}
\label{sheared-confined-label}

\begin{figure}
\centering
\includegraphics[width=0.55\linewidth]{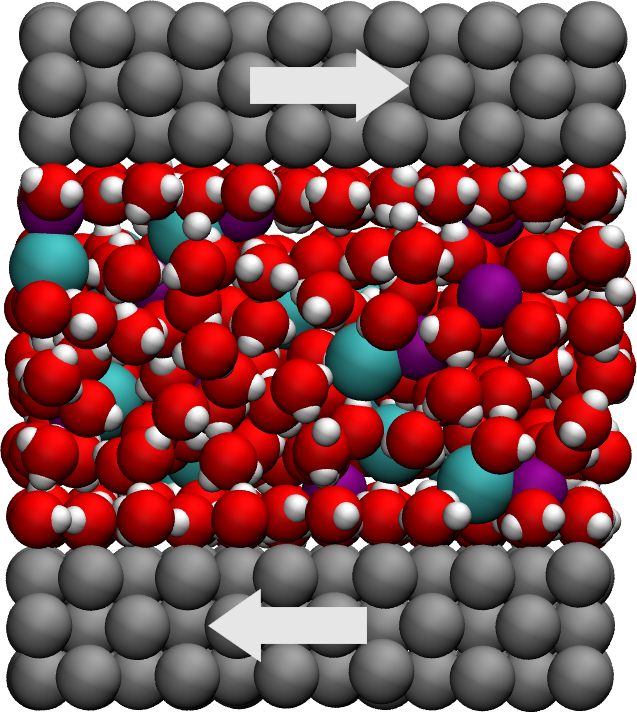}
\caption{The electrolyte confined in a nanometer slit pore as simulated during
\hyperref[sheared-confined-label]{Tutorial 4}.  $\text{Na}^+$ ions are represented
as purple spheres, $\text{Cl}^-$ ions as cyan spheres, water molecules are colored
in red and white, and the walls are colored in gray.  The arrows indicate the
imposed lateral motion of the walls.}
\label{fig:NANOSHEAR}
\end{figure}

The objective of this tutorial is to simulate an electrolyte
nanoconfined and sheared between two walls (Fig.~\ref{fig:NANOSHEAR}).  The density
and velocity profiles of the fluid in the direction normal to the walls are
extracted to highlight the effect of confining a fluid on its local properties.
This tutorial demonstrates key concepts of combining a fluid and a solid in
the same simulation.  A major difference from the previous tutorial,
\hyperref[all-atom-label]{Polymer in water}, is that here a rigid four-point
water model named TIP4P/2005 is used~\cite{abascal2005general}.

\begin{note}
  Four-point water models such as TIP4P/2005 are widely used as they offer a
  good compromise between accuracy and computational cost~\cite{kadaoluwa2021systematic}.
\end{note}

\subsubsection{System preparation}

The fluid and walls must first be generated, followed by equilibration at the
desired temperature and pressure.

\paragraph{System generation}

To set up this tutorial, select \guicmd{Start Tutorial 4} from the
\guicmd{Tutorials} menu of \lammpsgui{} and follow the instructions.
The editor should display the following content corresponding to \flecmd{create.lmp}:
\begin{lstlisting}
boundary p p f
units real
atom_style full
bond_style harmonic
angle_style harmonic
pair_style lj/cut/tip4p/long O H O-H H-O-H 0.1546 12.0
kspace_style pppm/tip4p 1.0e-5
kspace_modify slab 3.0
\end{lstlisting}
These lines are used to define the most basic parameters, including the 
atom style, the forms of the non-bonded, bond, and angle potentials, 
as well as other specifics of the non-bonded interactions.  Here, \lmpcmd{lj/cut/tip4p/long}
imposes a Lennard-Jones potential with a cut-off at $12\,\text{\AA{}}$ and a long-range
Coulomb potential.  The parameters \lmpcmd{O}, \lmpcmd{H}, \lmpcmd{O-H},
and \lmpcmd{H-O-H} correspond respectively to the oxygens, hydrogens, O-H bonds, and
H-O-H angle constraints of the water molecules; their definitions, provided by the
\lmpcmd{labelmap} commands, will be clarified below.

So far, the commands are relatively similar to those in the previous tutorial,
\hyperref[all-atom-label]{Polymer in water}, with two major differences: the use
of \lmpcmd{lj/cut/tip4p/long} instead of \lmpcmd{lj/cut/coul/long}, and \lmpcmd{pppm/tip4p}
instead of \lmpcmd{pppm}.  When using \lmpcmd{lj/cut/tip4p/long} and \lmpcmd{pppm/tip4p},
the interactions resemble the conventional Lennard-Jones and Coulomb interactions,
except that they are specifically designed for the four-point water model.  As a result,
LAMMPS automatically adds the fourth point to the water molecules, assigning type O
atoms as oxygen and type H atoms as hydrogen.  The fourth massless atom (M) of the
TIP4P water molecule does not have to be defined explicitly, and the value of
$0.1546\,\text{\AA{}}$ corresponds to the O-M distance of the
TIP4P-2005 water model~\cite{abascal2005general}.  All other atoms in the simulation
are treated as usual, with long-range Coulomb interactions.  Another novelty, here, is
the use of \lmpcmd{kspace\_modify slab 3.0} that is combined with the non-periodic
boundaries along the $z$ coordinate: \lmpcmd{boundary p p f}.  With the \lmpcmd{slab}
option, the system is treated as periodical along $z$, but with an empty volume inserted
between the periodic images of the slab, and the interactions along $z$ effectively turned off.

Let us create the box and the label maps by adding the following lines to \flecmd{create.lmp}:
\begin{lstlisting}
lattice fcc 4.04
region box block -3 3 -3 3 -5 5
create_box 5 box bond/types 1 angle/types 1 &
  extra/bond/per/atom 2 extra/angle/per/atom 1 &
  extra/special/per/atom 2
labelmap atom 1 O 2 H 3 Na+ 4 Cl- 5 WALL
labelmap bond 1 O-H
labelmap angle 1 H-O-H
\end{lstlisting}
The \lmpcmd{lattice} command defines the unit cell.  Here, the face-centered cubic (fcc) lattice
with a scale factor of 4.04 has been chosen for the future positioning of the atoms
of the walls.  The \lmpcmd{region} command defines a geometric region of space.  By choosing
$\text{xlo}=-3$ and $\text{xlo}=3$, and because we have previously chosen a lattice with a scale
factor of 4.04, the region box extends from $-12.12~\text{\AA{}}$ to $12.12~\text{\AA{}}$
along the $x$ direction.  The \lmpcmd{create\_box} command creates a simulation box with
5 types of atoms: the oxygen and hydrogen of the water molecules, the two ions ($\text{Na}^+$,
$\text{Cl}^-$), and the atoms from the walls.  The simulation contains 1 type of bond
and 1 type of angle (both required by the water molecules).
The parameters for these bond and angle constraints will be given later.  The \lmpcmd{extra/ (...)}
keywords are for memory allocation.  Finally, the \lmpcmd{labelmap} commands assign
alphanumeric type labels to each numeric atom type, bond type, and angle type,
concepts already introduced in previous tutorials.

Now, we can add atoms to the system.  First, let us create two sub-regions corresponding
respectively to the two solid walls, and create a larger region from the union of the
two regions.  Then, let us create atoms of type WALL within the two regions.  Add the
following lines to \flecmd{create.lmp}:
\begin{lstlisting}
region rbotwall block -3 3 -3 3 -4 -3
region rtopwall block -3 3 -3 3 3 4
region rwall union 2 rbotwall rtopwall
create_atoms WALL region rwall
\end{lstlisting}
Atoms will be placed in the positions of the previously defined lattice, thus
forming fcc solids.

To add the water molecules, the molecule
template called \href{\filepath tutorial4/water.mol}{\dwlcmd{water.mol}}
must be located next to \flecmd{create.lmp}.  The template contains all the
necessary information concerning the water molecule, such as atom positions,
bonds, and angles.  Add the following lines to \flecmd{create.lmp}:
\begin{lstlisting}
region rliquid block INF INF INF INF -2 2
molecule h2omol water.mol
create_atoms 0 region rliquid mol h2omol 482793
\end{lstlisting}
Within the last three lines, a region named \lmpcmd{rliquid} is
created based on the last defined lattice, \lmpcmd{fcc 4.04}.  \lmpcmd{rliquid}
will be used for introducing the water molecules in the simulation domain.  The \lmpcmd{molecule} command
opens up the molecule template called \flecmd{water.mol}, and names the
associated molecule \lmpcmd{h2omol}.  The new molecules are placed on the
\lmpcmd{fcc 4.04} lattice by the \lmpcmd{create\_atoms} command.  The first
parameter is 0, meaning that the atom types from the \flecmd{water.mol} file
will be used.  The number \lmpcmd{482793} is a seed that is required by LAMMPS,
it can be any positive integer.

Finally, let us create 30 ions (15 $\text{Na}^+$ and 15 $\text{Cl}^-$) in between
the water molecules, by adding the following commands to \flecmd{create.lmp}:
\begin{lstlisting}
create_atoms Na+ random 15 5802 rliquid overlap 0.3 maxtry 500
create_atoms Cl- random 15 9012 rliquid overlap 0.3 maxtry 500
set type Na+ charge 1
set type Cl- charge -1
\end{lstlisting}
Each \lmpcmd{create\_atoms} command will add 15 ions at random positions
within the \lmpcmd{rliquid} region, ensuring that there is no \lmpcmd{overlap}
with existing molecules.  Feel free to increase or decrease the salt concentration
by changing the number of desired ions.  To keep the system charge neutral,
always insert the same number of $\text{Na}^+$ and $\text{Cl}^-$, unless there
are other charges in the system.  The charges of the newly added ions are specified
by the two \lmpcmd{set} commands.

Before starting the simulation, we need to define the parameters of the
simulation: the mass of the 5 atom types (O, H, $\text{Na}^+$, $\text{Cl}^-$,
and wall), the pairwise interaction parameters (in this case, for the
Lennard-Jones potential), and the bond and angle parameters.  Copy the following
lines into \flecmd{create.lmp}:
\begin{lstlisting}
include parameters.inc
include groups.inc
\end{lstlisting}
Both \href{\filepath tutorial4/parameters.inc}{\dwlcmd{parameters.inc}}
and \href{\filepath tutorial4/groups.inc}{\dwlcmd{groups.inc}} files
must be located next to \flecmd{create.lmp}.

The \flecmd{parameters.inc} file contains the masses, as follows:
\begin{lstlisting}
mass O 15.9994
mass H 1.008
mass Na+ 22.990
mass Cl- 35.453
mass WALL 26.9815
\end{lstlisting}
Each \lmpcmd{mass} command assigns a mass in g/mol to an atom type.
The \flecmd{parameters.inc} file also contains the pair coefficients:
\begin{lstlisting}
pair_coeff O O 0.185199 3.1589
pair_coeff H H 0.0 1.0
pair_coeff Na+ Na+ 0.04690 2.4299
pair_coeff Cl- Cl- 0.1500 4.04470
pair_coeff WALL WALL 11.697 2.574
pair_coeff O WALL 0.4 2.86645
\end{lstlisting}
Each \lmpcmd{pair\_coeff} assigns the depth of the LJ potential (in
kcal/mol), and the distance (in Ångströms) at which the
particle-particle potential energy is 0.  As noted in previous
tutorials, with the important exception of \lmpcmd{pair\_coeff O WALL},
pairwise interactions were only assigned between atoms of identical
types.  By default, LAMMPS calculates the pair coefficients for the
interactions between atoms of different types (i and j) by using
geometric average: $\epsilon_{ij} = \sqrt{\epsilon_{ii} \epsilon_{jj}}$,
$\sigma_{ij} = \sqrt{\sigma_{ii} \sigma_{jj}}$.  However, if the default
value of $1.472\,\text{kcal/mol}$ was used for $\epsilon_\text{O-WALL}$,
the solid walls would be extremely hydrophilic, causing the water
molecules to form dense layers.  As a comparison, the water-water energy
$\epsilon_\text{O-O}$ is only $0.185199\,\text{kcal/mol}$.  Therefore,
to make the walls less hydrophilic, the value of
$\epsilon_\text{O-WALL}$ was reduced.

Finally, the \flecmd{parameters.inc} file contains the following two lines:
\begin{lstlisting}
bond_coeff O-H 0 0.9572
angle_coeff H-O-H 0 104.52
\end{lstlisting}
The \lmpcmd{bond\_coeff} command, used here for the O-H bond of the
water molecule, sets both the spring constant of the harmonic potential
and the equilibrium bond distance of $0.9572~\text{\AA{}}$.  The force
constant can be 0 for a rigid water molecule because the SHAKE
algorithm, which will be used in the input at a later
step, will constrain the intramolecular structure of the water molecules (see
below)~\cite{ryckaert1977numerical, andersen1983rattle}.  Similarly, the
\lmpcmd{angle\_coeff} command for the H-O-H angle of the water molecule
sets the force constant of the angular harmonic potential to 0 and the
equilibrium angle to $104.52^\circ$.

Alongside \flecmd{parameters.inc}, the \flecmd{groups.inc} file contains
several \lmpcmd{group} commands to define groups of atoms based
on their types:
\begin{lstlisting}
group H2O type O H
group Na type Na+
group Cl type Cl-
group ions union Na Cl
group fluid union H2O ions
\end{lstlisting}
The \flecmd{groups.inc} file also defines the \lmpcmd{walltop} and \lmpcmd{wallbot}
groups, which contain the WALL atoms located in the $z > 0$ and $z < 0$ regions, respectively:
\begin{lstlisting}
group wall type WALL
region rtop block INF INF INF INF 0 INF
region rbot block INF INF INF INF INF 0
group top region rtop
group bot region rbot
group walltop intersect wall top
group wallbot intersect wall bot
\end{lstlisting}

Currently, the fluid density between the two walls is slightly too high.  To avoid
excessive pressure, let us add the following lines into \flecmd{create.lmp}
to delete about $15~\%$ of the water molecules:
\begin{lstlisting}
delete_atoms random fraction 0.15 yes H2O NULL 482793 mol yes
\end{lstlisting}

\begin{figure}
\centering
\includegraphics[width=\linewidth]{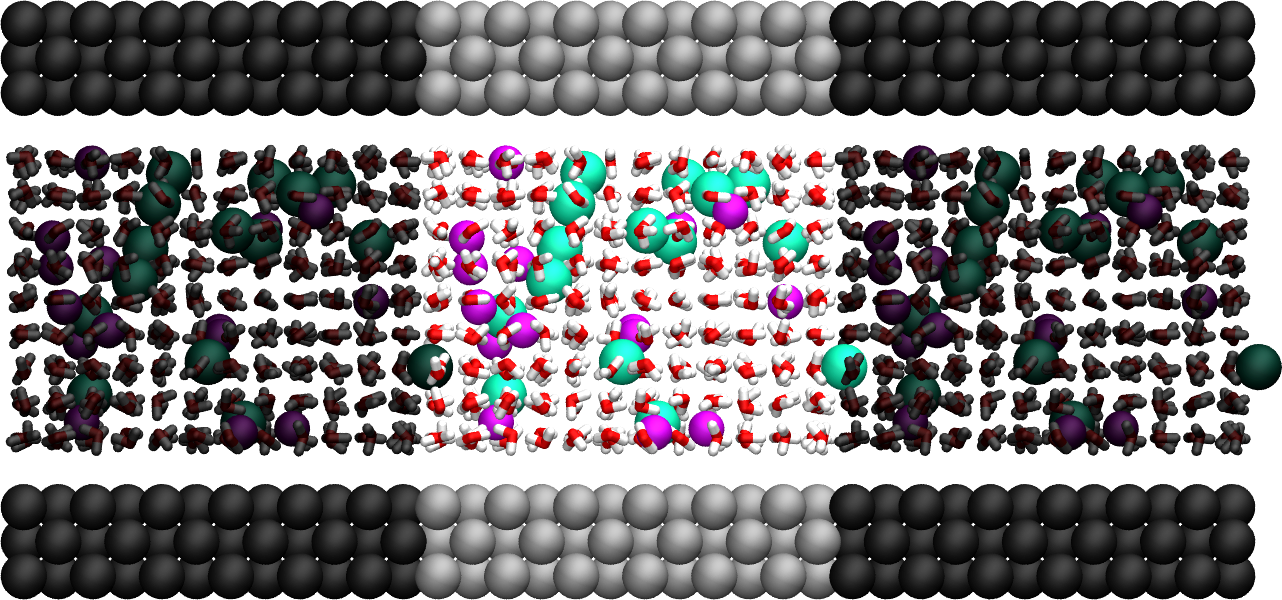}
\caption{Side view of the system.  Periodic images are represented in darker colors.
Water molecules are in red and white, $\text{Na}^+$ ions in purple, $\text{Cl}^-$
ions in lime, and wall atoms in gray.  Note the absence of atomic defect at the
cell boundaries.}
\label{fig:NANOSHEAR-system}
\end{figure}

To create an image of the system, add the following \lmpcmd{dump} image
into \flecmd{create.lmp} (see also Fig.~\ref{fig:NANOSHEAR-system}):
\begin{lstlisting}
dump mydmp all image 200 myimage-*.ppm type type &
  shiny 0.1 box no 0.01 view 90 0 zoom 1.8
dump_modify mydmp backcolor white &
  acolor O red adiam O 2 &
  acolor H white adiam H 1 &
  acolor Na+ blue adiam Na+ 2.5 &
  acolor Cl- cyan adiam Cl- 3 &
  acolor WALL gray adiam WALL 3
\end{lstlisting}

Finally, add the following lines into \flecmd{create.lmp}:
\begin{lstlisting}
run 0

write_data create.data nocoeff
\end{lstlisting}
The \lmpcmd{run 0} command initializes the simulation, which is required for
cleanly saving the state, but it
does not advance positions or velocities.  The \lmpcmd{write\_data} command
generates a file called \lmpcmd{system.data} containing the information required
to restart the simulation from the final configuration produced by this input
file.  With the \lmpcmd{nocoeff} option, the parameters from the force field are
not included in the \flecmd{.data} file.  Run the \flecmd{create.lmp} file using LAMMPS,
and a file named \flecmd{create.data} will be created alongside \flecmd{create.lmp}.

\paragraph{Energy minimization}

Let us move the atoms and place them in more energetically favorable positions
before starting the actual molecular dynamics simulation.
Open the \flecmd{equilibrate.lmp} file that was downloaded alongside
\flecmd{create.lmp} during the tutorial setup.  Same as
before, it contains the following lines:
\begin{lstlisting}
boundary p p f
units real
atom_style full
bond_style harmonic
angle_style harmonic
pair_style lj/cut/tip4p/long O H O-H H-O-H 0.1546 12.0
kspace_style pppm/tip4p 1.0e-5
kspace_modify slab 3.0

read_data create.data

include parameters.inc
include groups.inc
\end{lstlisting}
The only difference from the previous input is that, instead of creating a new
box and new atoms, we open the previously created \flecmd{create.data} file.

Now, let us use the SHAKE algorithm to maintain the shape of the
water molecules~\cite{ryckaert1977numerical, andersen1983rattle}
by adding the following line to the script.
\begin{lstlisting}
fix myshk H2O shake 1.0e-5 200 0 b O-H a H-O-H kbond 2000
\end{lstlisting}
Here the SHAKE algorithm applies to the \lmpcmd{O-H} bond and the \lmpcmd{H-O-H} angle
of the water molecules.  The \lmpcmd{kbond} keyword specifies the force constant that will be
used to apply a restraint force when used during minimization.  This last keyword is important
here, because the spring constants of the rigid water molecules were set
to 0 (see the \flecmd{parameters.inc} file).

\begin{note}
  LAMMPS provides several ways to keep molecules rigid during a simulation.
  The \lmpcmd{fix shake} command is appropriate and efficient for constraining
  individual bonds or bonds and angles within small molecules like water while using
  a per-atom time integration fix command like \lmpcmd{fix nve} or \lmpcmd{fix nvt}.
  However, it fails for linear molecules like CO$_2$ or constraining larger or more
  complex objects.
  In such cases, the \lmpcmd{fix rigid} family of commands can be used to perform
  time integration for translation and rotation of groups of atoms as rigid bodies.
\end{note}

Let us also create images of the system and control
the printing of thermodynamic outputs by adding the following lines
to \flecmd{equilibrate.lmp}:
\begin{lstlisting}
  dump mydmp all image 1 myimage-*.ppm type type &
  shiny 0.1 box no 0.01 view 90 0 zoom 1.8
dump_modify mydmp backcolor white &
  acolor O red adiam O 2 &
  acolor H white adiam H 1 &
  acolor Na+ blue adiam Na+ 2.5 &
  acolor Cl- cyan adiam Cl- 3 &
  acolor WALL gray adiam WALL 3

thermo 1
thermo_style custom step temp etotal press
\end{lstlisting}

Let us perform an energy minimization by adding the following lines to \flecmd{equilibrate.lmp}:
\begin{lstlisting}
minimize 1.0e-6 1.0e-6 1000 1000
reset_timestep 0
\end{lstlisting}
When running the \flecmd{equilibrate.lmp} file with LAMMPS, you should observe that the
total energy of the system is initially very high but rapidly decreases.  From the generated
images of the system, you will notice that the atoms and molecules are moving to adopt more favorable positions.








\paragraph{System equilibration}

Let us equilibrate further the entire system by letting both fluid and wall
relax at ambient temperature.  Here, the commands are written within the same
\flecmd{equilibrate.lmp} file, right after the \lmpcmd{reset\_timestep} command.

Let us do a molecular dynamics simulation using the Nosé-Hoover
thermostat.  Add the following lines to \flecmd{equilibrate.lmp}:
\begin{lstlisting}
fix mynvt all nvt temp 300 300 100
fix myshk H2O shake 1.0e-5 200 0 b O-H a H-O-H
fix myrct all recenter NULL NULL 0
timestep 1.0
\end{lstlisting}
As mentioned previously, the \lmpcmd{fix recenter} does not influence the dynamics,
but will keep the system in the center of the box, which makes the
visualization easier.  Then, add the following lines into \flecmd{equilibrate.lmp}
for the trajectory visualization:
\begin{lstlisting}
undump mydmp
dump mydmp all image 250 myimage-*.ppm type type &
  shiny 0.1 box no 0.01 view 90 0 zoom 1.8
dump_modify mydmp backcolor white &
  acolor O red adiam O 2 &
  acolor H white adiam H 1 &
  acolor Na+ blue adiam Na+ 2.5 &
  acolor Cl- cyan adiam Cl- 3 &
  acolor WALL gray adiam WALL 3
\end{lstlisting}
The \lmpcmd{undump} command is used to cancel the previous \lmpcmd{dump} command.
Then, a new \lmpcmd{dump} command with a larger dumping period is used.

\begin{note}
  Just like the \lmpcmd{undump} command can cancel an active \lmpcmd{dump}, other
  objects defined in a LAMMPS input script can be cancelled when no longer needed.
  For example, you can use \lmpcmd{unfix} to remove a previously defined \lmpcmd{fix}, and
  \lmpcmd{uncompute} to delete a \lmpcmd{compute}.
\end{note}

To monitor the system equilibration, let us print the distance between
the two walls.  Add the following lines to \flecmd{equilibrate.lmp}:
\begin{lstlisting}
variable walltopz equal xcm(walltop,z)
variable wallbotz equal xcm(wallbot,z)
variable deltaz equal v_walltopz-v_wallbotz

thermo 250
thermo_style custom step temp etotal press v_deltaz
\end{lstlisting}
The first two variables extract the z coordinate of the centers of mass of the two walls.  The
\lmpcmd{deltaz} variable is then used to calculate the difference between the two
variables \lmpcmd{walltopz} and \lmpcmd{wallbotz}, i.e.~the distance in the z direction between the
two centers of mass of the walls.

\begin{figure}
\centering
\includegraphics[width=\linewidth]{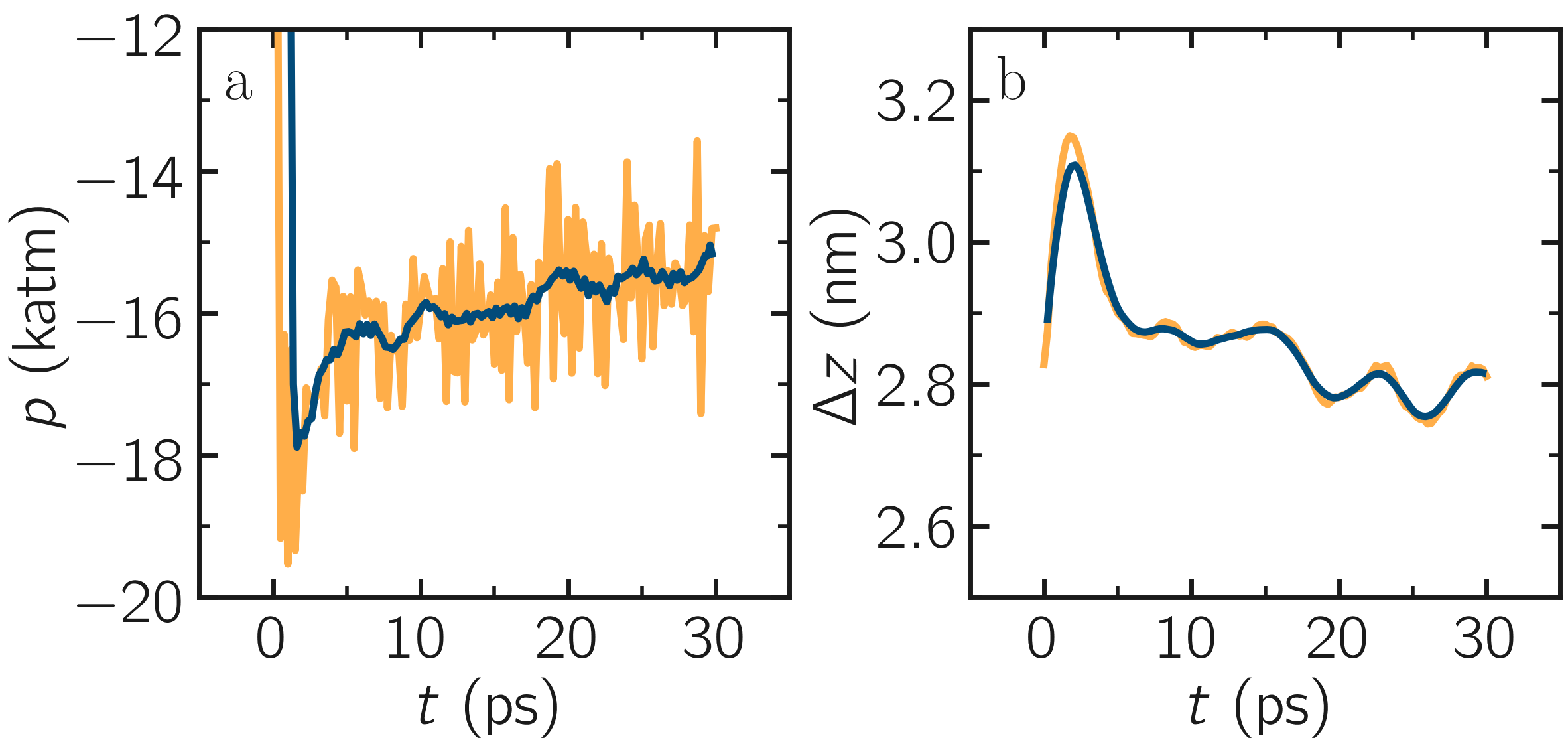}\\[-2ex]
\caption{a)~Pressure, $p$, of the nanosheared electrolyte system
simulated in \hyperref[sheared-confined-label]{Tutorial 4} as a function of the
time, $t$.  b)~Distance between the walls, $\Delta z$, as a function of $t$.
The orange line shows the raw data, and the blue line represents a time-averaged curve.}
\label{fig:NANOSHEAR-equilibration}
\end{figure}

Finally, let us run the simulation for 30~ps by adding a \lmpcmd{run} command
to \flecmd{equilibrate.lmp}:
\begin{lstlisting}
run 30000

write_data equilibrate.data nocoeff
\end{lstlisting}
Run the \flecmd{equilibrate.lmp} file using LAMMPS.  Both the pressure and the distance
between the two walls show oscillations at the start of the simulation
but eventually stabilize at their equilibrium values toward
the end of the simulation (Fig.~\ref{fig:NANOSHEAR-equilibration}).

\begin{note}
  Note that it is generally recommended to run a longer equilibration.  In this case,
  the slowest process in the system is likely ionic diffusion.
  Therefore, the equilibration period should, in principle, exceed the time required
  for the ions to diffuse across the size of the pore, i.e.~$H_\text{pore}^2/D_\text{ions}$.
  Using $H_\text{pore} \approx 1.2~\text{nm}$ as the final pore size
  and $D_\text{ions} \approx 1.5 \cdot 10^{-9}~\text{m}^2/\text{s}$
  as the typical diffusion coefficient for sodium chloride in water at room
  temperature~\cite{mills1955remeasurement}, one finds that the equilibration
  should be on the order of one nanosecond.
\end{note}

\subsubsection{Imposed shearing}

From the equilibrated configuration, let us impose a lateral motion on the two
walls and shear the electrolyte.  Open the last input file named \flecmd{shearing.lmp}.
It starts as follows:
\begin{lstlisting}
boundary p p f
units real
atom_style full
bond_style harmonic
angle_style harmonic
pair_style lj/cut/tip4p/long O H O-H H-O-H 0.1546 12.0
kspace_style pppm/tip4p 1.0e-5
kspace_modify slab 3.0

read_data equilibrate.data

include parameters.inc
include groups.inc
\end{lstlisting}

To address the dynamics of the system, add the following lines to
\flecmd{shearing.lmp}:
\begin{lstlisting}
compute Tfluid fluid temp/partial 0 1 1
fix mynvt1 fluid nvt temp 300 300 100
fix_modify mynvt1 temp Tfluid

compute Twall wall temp/partial 0 1 1
fix mynvt2 wall nvt temp 300 300 100
fix_modify mynvt2 temp Twall

fix myshk H2O shake 1.0e-5 200 0 b O-H a H-O-H
fix myrct all recenter NULL NULL 0
timestep 1.0
\end{lstlisting}

One key difference with the previous input is that, here, two thermostats are used,
one for the fluid (\lmpcmd{mynvt1}) and one for the solid (\lmpcmd{mynvt2}).
The combination of \lmpcmd{fix\_modify} with \lmpcmd{compute temp} ensures
that the correct temperature values are used by the thermostats.  Using
\lmpcmd{compute} commands for the temperature with \lmpcmd{temp/partial 0 1 1} is
intended to exclude the $x$ coordinate from the thermalization, which is important since a
large velocity will be imposed along the $x$ direction.

Then, let us impose the velocity of the two walls by adding the following
commands to \flecmd{shearing.lmp}:
\begin{lstlisting}
fix mysf1 walltop setforce 0 NULL NULL
fix mysf2 wallbot setforce 0 NULL NULL
velocity wallbot set -2e-4 NULL NULL
velocity walltop set 2e-4 NULL NULL
\end{lstlisting}
The \lmpcmd{setforce} commands cancel the forces on \lmpcmd{walltop} and
\lmpcmd{wallbot} in the $x$ direction.  As a result, the atoms in these two groups will not
experience any forces along $x$ from their interaction with rest of the system.  Consequently, in the absence of
external forces, these atoms will conserve the initial velocities imposed by the
two \lmpcmd{velocity} commands.  As seen previously, although the
forces on these atoms are set to zero, the \lmpcmd{fix setforce} still stores the
forces acting on the group before cancellation, which can later be extracted
for analysis (see below).

\begin{figure}
\centering
\includegraphics[width=\linewidth]{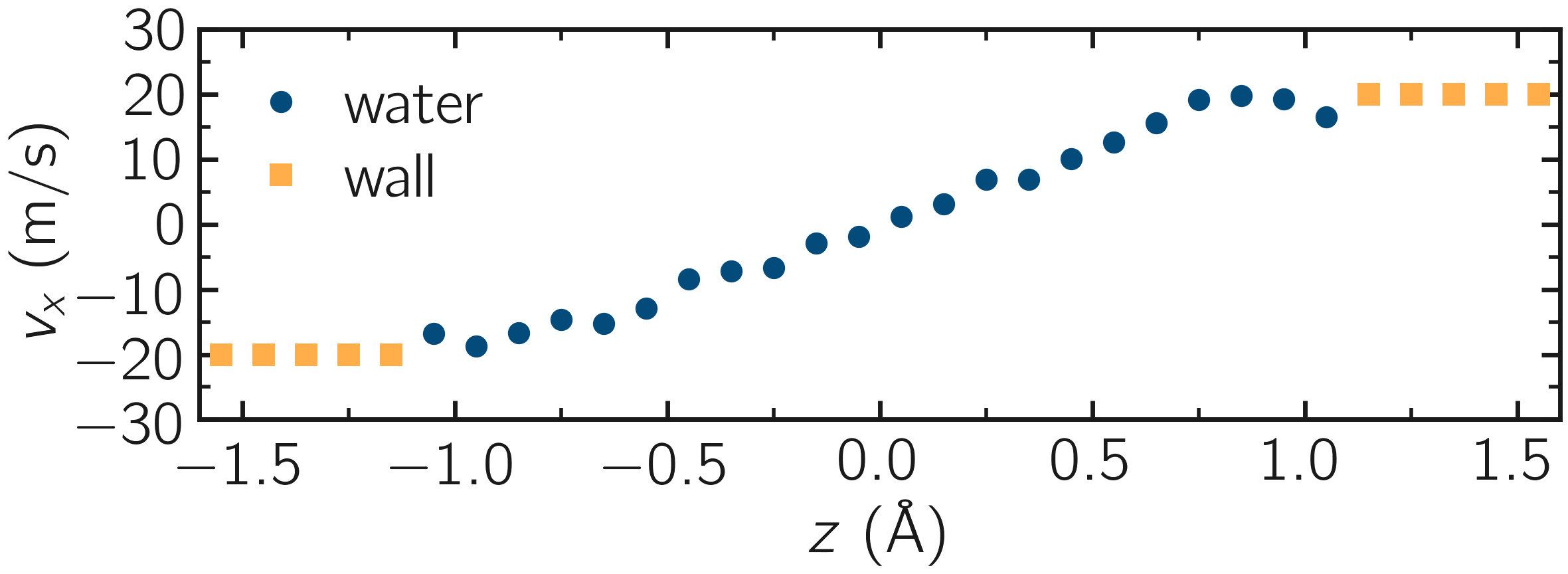}\\[-2ex]
\caption{Velocity profiles for water (blue) and walls (orange) along the $z$-axis as
simulated in \hyperref[sheared-confined-label]{Tutorial 4}.}
\label{fig:NANOSHEAR-profiles}
\end{figure}

Finally, let us generate images of the systems and print the values of the
forces exerted by the fluid on the walls, as given by \lmpcmd{f\_mysf1[1]}
and \lmpcmd{f\_mysf2[1]}.  Add these lines to \flecmd{shearing.lmp}:
\begin{lstlisting}
dump mydmp all image 250 myimage-*.ppm type type &
  shiny 0.1 box no 0.01 view 90 0 zoom 1.8
dump_modify mydmp backcolor white &
  acolor O red adiam O 2 &
  acolor H white adiam H 1 &
  acolor Na+ blue adiam Na+ 2.5 &
  acolor Cl- cyan adiam Cl- 3 &
  acolor WALL gray adiam WALL 3

thermo 250
thermo_modify temp Tfluid
thermo_style custom step temp etotal f_mysf1[1] f_mysf2[1]
\end{lstlisting}
Let us also extract the density and velocity profiles using
the \lmpcmd{chunk/atom} and \lmpcmd{ave/chunk} commands.  When deployed as below, these
commands discretize the simulation domain into spatial bins and compute and output
average properties of the atoms belonging to each bin, here the velocity
along $x$ (\lmpcmd{vx}) within the bins.  Add the following lines to \flecmd{shearing.lmp}:
\begin{lstlisting}
compute cc1 H2O chunk/atom bin/1d z 0.0 0.25
compute cc2 wall chunk/atom bin/1d z 0.0 0.25
compute cc3 ions chunk/atom bin/1d z 0.0 0.25

fix myac1 H2O ave/chunk 10 15000 200000 &
  cc1 density/mass vx file shearing-water.dat
fix myac2 wall ave/chunk 10 15000 200000 &
  cc2 density/mass vx file shearing-wall.dat
fix myac3 ions ave/chunk 10 15000 200000 &
  cc3 density/mass vx file shearing-ions.dat

run 200000
\end{lstlisting}
Here, a bin size of $0.25\,\text{\AA{}}$ is used for the density
profiles generated by the \lmpcmd{ave/chunk} commands, and three
\flecmd{.dat} files are created for the water, the walls, and the ions,
respectively.  With values of \lmpcmd{10 15000 200000}, the velocity
\lmpcmd{vx} will be evaluated every 10 steps during the final 150,000
steps of the simulations.  The result will be averaged and printed only
once at the 200,000\textsuperscript{th} step.

Run the simulation using LAMMPS.  The averaged velocity
profile for the fluid is plotted in Fig.~\ref{fig:NANOSHEAR-profiles}.
As expected for such a Couette flow geometry, the fluid velocity increases
linearly along $z$, and is equal to the walls velocities at the fluid-solid
interfaces (no-slip boundary conditions).

From the force applied by the fluid on the solid, one can extract the stress
within the fluid, which enables the measurement of its viscosity $\eta$
according to
\begin{equation}
\eta = \tau / \dot{\gamma}
\label{eq:eta}
\end{equation}
where $\tau$ is the stress applied by
the fluid on the shearing wall, and $\dot{\gamma}$ the shear rate
\cite{gravelle2021violations}.  Here, the shear rate is
approximately $\dot{\gamma} = 20 \cdot 10^9\,\text{s}^{-1}$ (Fig.~\ref{fig:NANOSHEAR-profiles}),
the average force on each wall is given by \lmpcmd{f\_mysf1[1]} and \lmpcmd{f\_mysf2[1]}
and is approximately $2.7\,\mathrm{kcal/mol/\AA{}}$. 
Using a surface area
for the walls of $A = 6 \cdot 10^{-18}\,\text{m}^2$, one obtains an estimate for
the shear viscosity for the confined fluid of $\eta = 3.1\,\mathrm{mPa\cdot s}$ using Eq.~\eqref{eq:eta}.

\begin{note}
  The viscosity calculated at such a high shear rate may differ from the expected
  \emph{bulk} value.  In general, it is recommended to use a lower value for the
  shear rate.  Note that for lower shear rates, the signal-to-noise ratio is
  smaller, and longer simulations are needed.  Another point to consider
  is that the viscosity of a fluid next to a solid surface is typically larger
  than in bulk due to interaction with the walls~\cite{wolde-kidanInterplayInterfacialViscosity2021}.
  Therefore, one expects the present simulation to yield a viscosity that is slightly
  higher than what would be measured in the absence of walls.
\end{note}

\subsection{Tutorial 5: Reactive silicon dioxide}
\label{reactive-silicon-dioxide-label}

\begin{figure}
\centering
\includegraphics[width=0.55\linewidth]{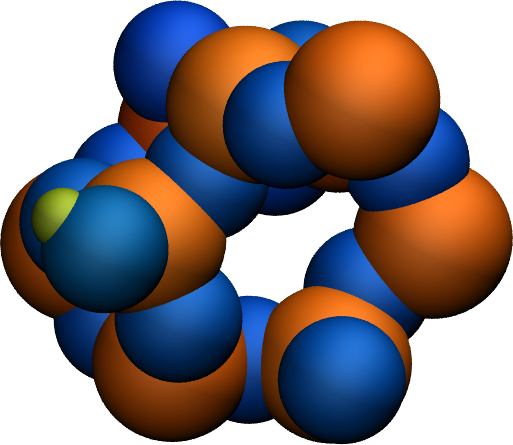}
\caption{A portion of the silicon dioxide structure as simulated during
\hyperref[reactive-silicon-dioxide-label]{Tutorial 5}.  Atoms are colored
by their charges: the hydrogen atoms appear as small greenish spheres, silicon
atoms as large orange spheres, and oxygen atoms as blue spheres of intermediate
size.}
\label{fig:SIO}
\end{figure}

The objective of this tutorial is to demonstrate how the reactive force field ReaxFF
can be used to calculate the partial charges of a system undergoing deformation, as well as
the formation and breaking of chemical bonds~\cite{van2001reaxff, zou2012investigation}.
The system simulated in this tutorial is a block of silicon dioxide $\text{SiO}_2$ (Fig.~\ref{fig:SIO})
which is deformed until it ruptures.  Particular attention is given to the evolution
of atomic charges during deformation, with a focus on tracking chemical reactions
resulting from the deformation over time.

\subsubsection{Prepare and relax}

The first step is to relax the structure with ReaxFF, which which will be achieved using
molecular dynamics.  To ensure the system equilibrates properly, we will monitor certain
parameters over time, such as the system volume.  To set up this
tutorial, select \guicmd{Start Tutorial 5} from the
\guicmd{Tutorials} menu of \lammpsgui{} and follow the instructions.
The editor should display the following content corresponding to \flecmd{relax.lmp}:
\begin{lstlisting}
units real
atom_style full

read_data silica.data

\end{lstlisting}
So far, the input is very similar to what was seen in the previous tutorials.
Some basic parameters are defined (\lmpcmd{units} and \lmpcmd{atom\_style}),
and a \flecmd{.data} file is imported by the \lmpcmd{read\_data} command.

The initial topology given by \href{\filepath tutorial5/silica.data}{\dwlcmd{silica.data}}
corresponds to a small amorphous silica structure.
This structure was generated in a prior
simulation using the Vashishta force field~\cite{vashishta1990interaction}.
If you open the \flecmd{silica.data} file, you will find in the \lmpcmd{Atoms}
section that all silicon atoms have a charge of $q = 1.1\,\text{e}$, and all oxygen
atoms have a charge of $q = -0.55\,\text{e}$.

\begin{note}
  Assigning the same charge to all atoms of the same type is common with many
  force fields, including the force fields used in the previous tutorials.  This
  changes once ReaxFF is used: the charge of each atom will adjust to its local
  environment through charge equilibration.
\end{note}

\begin{figure}
\includegraphics[width=\linewidth]{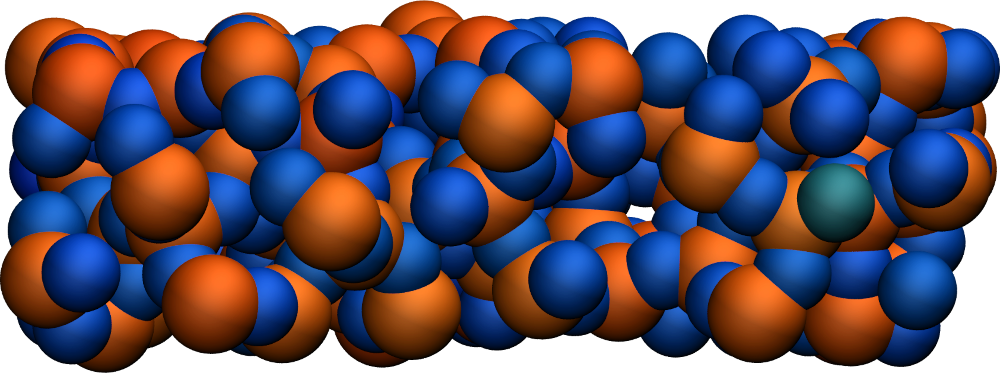}
\caption{A slice of the amorphous silica simulated during
\hyperref[reactive-silicon-dioxide-label]{Tutorial 5}, where atoms are colored by their charges.
Dangling oxygen groups appear in greenish, bulk Si atoms with a charge of about
$1.8~\text{e}$  appear in red/orange, and bulk O atoms with a charge of about
$-0.9~\text{e}$ appear in blue.}
\label{fig:SIO-slice}
\end{figure}

Next, copy the following three crucial lines into the \flecmd{relax.lmp} file:
\begin{lstlisting}
pair_style reaxff NULL safezone 3.0 mincap 150
pair_coeff * * ffield.reax.CHOFe Si O
fix myqeq all qeq/reaxff 1 0.0 10.0 1.0e-6 reaxff maxiter 400
\end{lstlisting}
In this case, the \lmpcmd{pair\_style reaxff} is used without a control file
(see note below).  The
\lmpcmd{safezone} and \lmpcmd{mincap} keywords are added to prevent
allocation issues, which sometimes can trigger segmentation faults and
\lmpcmd{bondchk} errors.  The \lmpcmd{pair\_coeff} command uses the
\href{\filepath tutorial5/ffield.reax.CHOFe}{\dwlcmd{ffield.reax.CHOFe}}
file, which should have been downloaded during the tutorial set up.  Finally, the
\lmpcmd{fix qeq/reaxff} is used to perform charge equilibration~\cite{rappe1991charge},
which occurs at every step.  The values 0.0 and 10.0 represent the
low and the high cutoffs, respectively, and $1.0 \text{e} -6$ is the
tolerance, i.e., the precision to which the atomic charges are
equilibrated during the charge equilibration process.
The \lmpcmd{maxiter} sets an upper limit to the number of attempts to
equilibrate the charge.

\begin{note}
  The \lmpcmd{pair\_style reaxff} command optionally accepts a control file,
  which defines control variables such as
  global parameters of the ReaxFF potential, as well as performance and output settings.
  If no control file is provided, as in this tutorial, LAMMPS uses its default values,
  which correspond to those in Adri van Duin's original stand-alone ReaxFF code~\cite{van2001reaxff}.
\end{note}
Next, add the following commands to the \flecmd{relax.lmp} file to track the
evolution of the charges during the simulation:
\begin{lstlisting}
group grpSi type Si
group grpO type O
variable qSi equal charge(grpSi)/count(grpSi)
variable qO equal charge(grpO)/count(grpO)
variable vq atom q
\end{lstlisting}
The definition of the equal style variables qSi and qO
make use of functions pre-defined within LAMMPS that allow calculating
the total charge of atoms belonging to a group (charge()) and the total
number of atoms in the group (count()).
To print the averaged charges \lmpcmd{qSi} and \lmpcmd{qO} using the
\lmpcmd{thermo\_style} command, and create images of the system.  Add the
following lines to \flecmd{relax.lmp}:
\begin{lstlisting}
thermo 100
thermo_style custom step temp etotal press vol v_qSi v_qO
dump viz all image 100 myimage-*.ppm q &
  type shiny 0.1 box no 0.01 view 180 90 zoom 2.3 size 1200 500
dump_modify viz adiam Si 2.6 adiam O 2.3 backcolor white &
  amap -1 2 ca 0.0 3 min royalblue 0 green max orangered
\end{lstlisting}
Here, the atoms are colored by their charges \lmpcmd{q}, ranging from royal blue
(when $q=-1\,\text{e}$) to orange-red (when $q=2\,\text{e}$).

We can generate histograms of the charges for each atom type using
\lmpcmd{fix ave/histo} commands:
\begin{lstlisting}
fix myhis1 grpSi ave/histo 10 500 5000 -1.5 2.5 1000 v_vq &
  file relax-Si.histo mode vector
fix myhis2 grpO ave/histo 10 500 5000 -1.5 2.5 1000 v_vq &
  file relax-O.histo mode vector
\end{lstlisting}
The \lmpcmd{fix ave/histo} command samples values
over a group of atoms and builds a histogram over a specified range divided into
bins.  In this tutorial, it is used to monitor the charge distributions
of silicon and oxygen atoms.  The parameters \lmpcmd{10 500 5000} specify how often
the histogram is updated and averaged, \lmpcmd{-1.5 2.5} set the value range,
\lmpcmd{1000} is the number of bins, and \lmpcmd{v\_vq} is the variable being histogrammed.

\begin{figure}
\centering
\includegraphics[width=\linewidth]{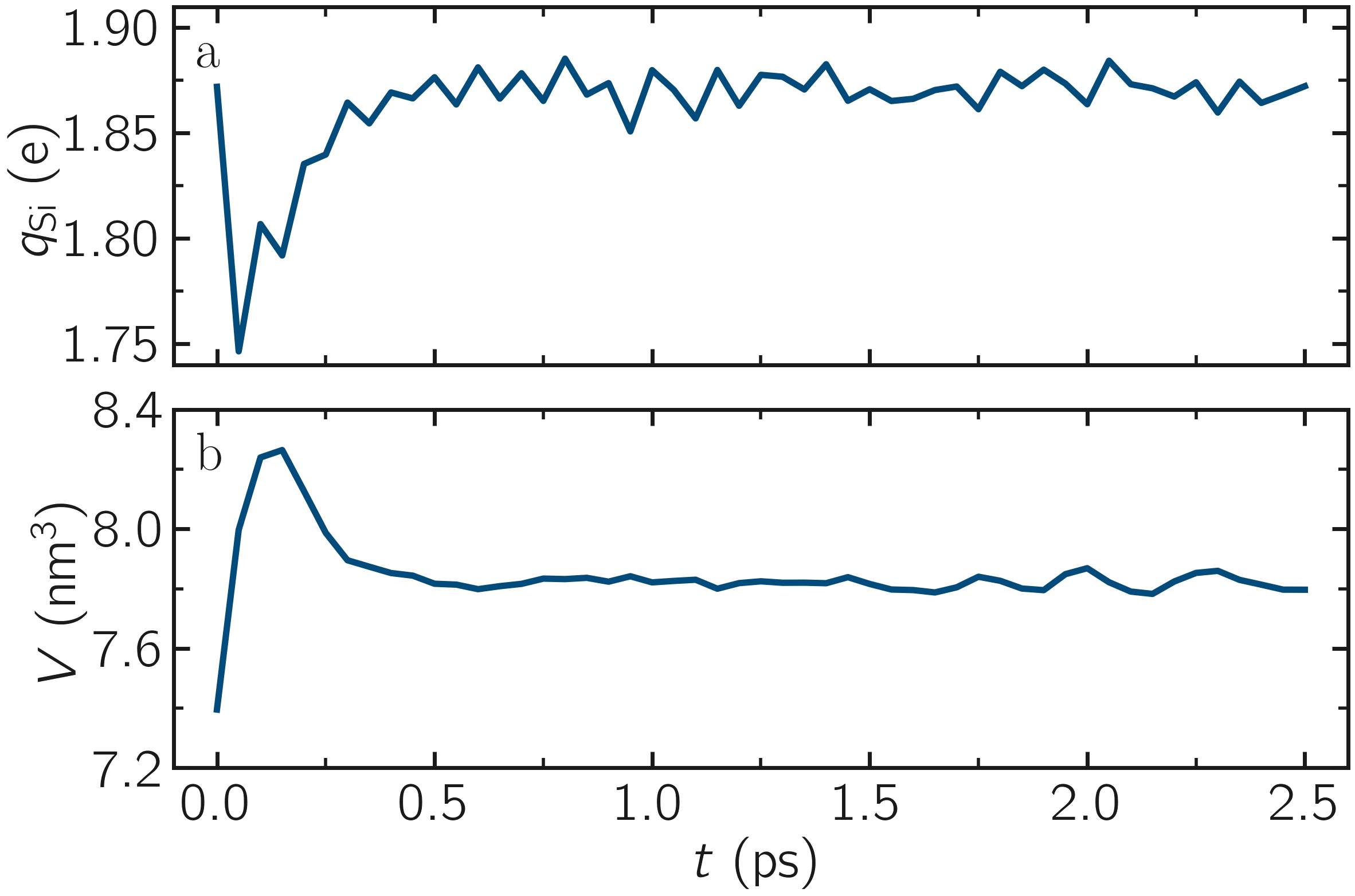}\\[-2ex]
\caption{a) Average charge per atom of the silicon, $q_\text{Si}$, atoms as
a function of time, $t$, during equilibration of the $\text{SiO}_2$ system
from \hyperref[reactive-silicon-dioxide-label]{Tutorial 5}.  b) Volume of the
system, $V$, as a function of $t$.}
\label{fig:SIO-charge}
\end{figure}

We can also use \lmpcmd{fix reaxff/species} to evaluate what species are
present within the simulation.  It will be useful later when the system is deformed,
and bonds are broken:
\begin{lstlisting}
fix myspec all reaxff/species 5 1 5 relax.species element Si O
\end{lstlisting}
Here, the information will be printed every 5 steps in a file called \flecmd{relax.species}.
Let us perform a very short run using the anisotropic NPT command and relax the
density of the system:
\begin{lstlisting}
velocity all create 300.0 32028
fix mynpt all npt temp 300.0 300.0 100 aniso 1.0 1.0 1000
timestep 0.5

run 5000

write_data relax.data nofix
\end{lstlisting}
The \lmpcmd{write\_data} command is used with the \lmpcmd{nofix} keyword to
print a data file without extra sections from the \lmpcmd{reaxff/species} command.
Run the \flecmd{relax.lmp} file using LAMMPS.  As seen from \flecmd{relax.species},
only one species is detected, called \lmpcmd{O384Si192}, representing the entire system.

\begin{note}
  With the \lmpcmd{aniso} keyword, the three dimensions of the simulation
  box can change independently.  This is particularly relevant for solids and other
  systems where anisotropic stresses may develop.
\end{note}

\begin{figure}
\includegraphics[width=\linewidth]{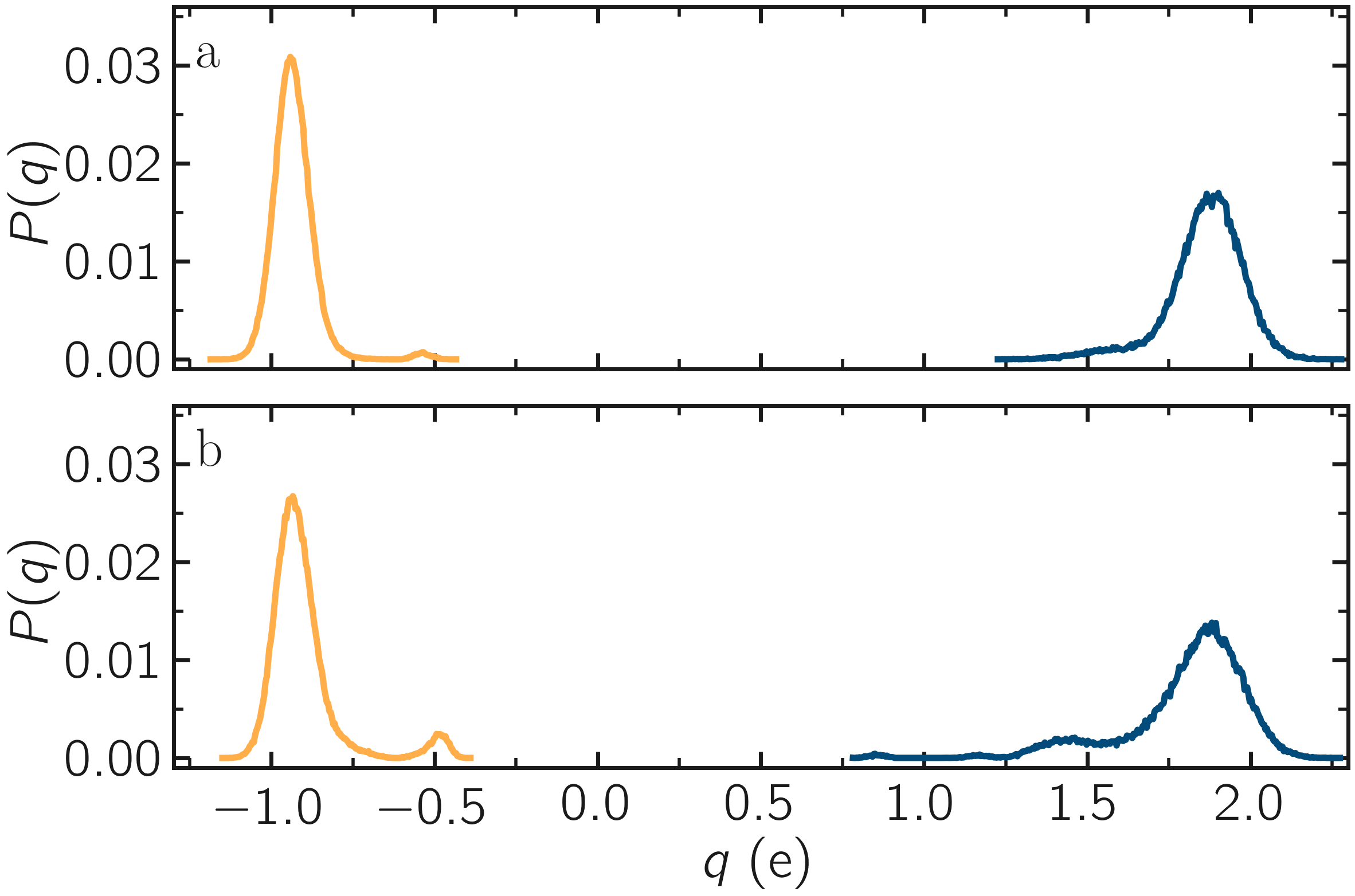}\\[-4ex]
\caption{a) Probability distributions of charge of silicon (positive, blue) and oxygen
(negative, orange) atoms during the equilibration of the $\text{SiO}_2$ system
from \hyperref[reactive-silicon-dioxide-label]{Tutorial 5}.  b) Same probability distributions
as in panel (a) after the deformation.}
\label{fig:SIO-distribution}
\end{figure}

As the simulation progresses, the charge of every atom fluctuates
because it is adjusting to the local environment of the atom (Fig.~\ref{fig:SIO-charge}\,a).
It is also observed that the averaged charges for silicon and oxygen
atoms fluctuate significantly at the beginning of the simulation, corresponding
to a rapid change in the system volume, which causes interatomic distances to
shift quickly (Fig.~\ref{fig:SIO-charge}\,b).  The atoms with the
most extreme charges are located at structural defects,
such as dangling oxygen groups (Fig.~\ref{fig:SIO-slice}).
Finally, the generated \flecmd{.histo} files can be used to
plot the probability distributions, $P(q)$ (see Fig.~\ref{fig:SIO-distribution}\,a).

\subsubsection{Deform the structure}

Let us apply a deformation to the structure to force some $\text{Si}-\text{O}$
bonds to break (and eventually re-assemble).  Open the \flecmd{deform.lmp}
file, which must contain the following lines:
\begin{lstlisting}
units real
atom_style full

read_data relax.data

pair_style reaxff NULL safezone 3.0 mincap 150
pair_coeff * * ffield.reax.CHOFe Si O
fix myqeq all qeq/reaxff 1 0.0 10.0 1.0e-6 reaxff maxiter 400

group grpSi type Si
group grpO type O
variable qSi equal charge(grpSi)/count(grpSi)
variable qO equal charge(grpO)/count(grpO)
variable vq atom q

thermo 200
thermo_style custom step temp etotal press vol v_qSi v_qO
dump viz all image 100 myimage-*.ppm q &
  type shiny 0.1 box no 0.01 view 180 90 zoom 2.3 size 1200 500
dump_modify viz adiam Si 2.6 adiam O 2.3 backcolor white &
  amap -1 2 ca 0.0 3 min royalblue 0 green max orangered

fix myhis1 grpSi ave/histo 10 500 5000 -1.5 2.5 1000 v_vq &
  file deform-Si.histo mode vector
fix myhis2 grpO ave/histo 10 500 5000 -1.5 2.5 1000 v_vq &
  file deform-O.histo mode vector
fix myspec all reaxff/species 5 1 5 deform.species element Si O
\end{lstlisting}
The only difference with the previous \flecmd{relax.lmp} file is the path to
the \flecmd{relax.data} file.

Next, let us use \lmpcmd{fix nvt} instead of \lmpcmd{fix npt} to apply a
Nosé-Hoover thermostat without a barostat:
\begin{lstlisting}
fix mynvt all nvt temp 300.0 300.0 100
timestep 0.5
\end{lstlisting}
Here, no barostat is used because the change in the box volume will be imposed
by the \lmpcmd{fix deform}, see below.

\begin{figure}
\includegraphics[width=\linewidth]{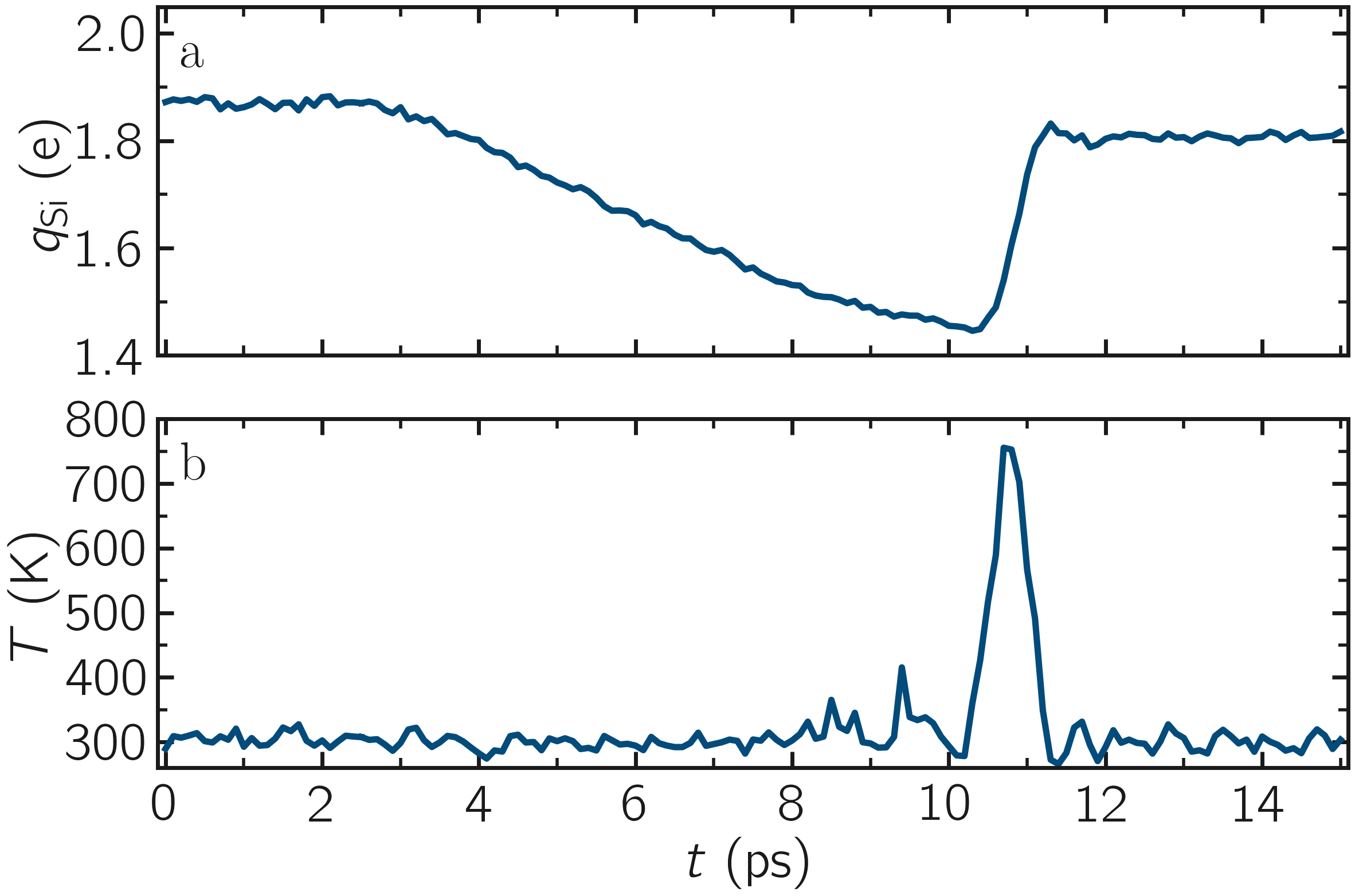}\\[-4ex]
\caption{a) Average charge per atom of the silicon, $q_\text{Si}$, atoms as
a function of time, $t$, during deformation of the $\text{SiO}_2$ system
from \hyperref[reactive-silicon-dioxide-label]{Tutorial 5}. The break down of the
silica structure occurs near $t = 11$\,ps.  b) Temperature, $T$, of the
system as a function of $t$.}
\label{fig:SIO-deformed-charge}
\end{figure}

Let us run for 5000 steps without deformation, then apply the \lmpcmd{fix deform}
to progressively elongate the box along the $x$-axis during 25000 steps.  Add
the following line to \flecmd{deform.lmp}:
\begin{lstlisting}
run 5000

fix mydef all deform 1 x erate 5e-5

run 25000

write_data deform.data nofix
\end{lstlisting}
The \lmpcmd{fix deform} command applies a continuous deformation
by elongating the simulation box along the x-axis at a constant engineering
shear strain rate, specified by \lmpcmd{erate}, of $5 \times 10^{-5}~\text{fs}^{-1}$.

\begin{figure}
\includegraphics[width=\linewidth]{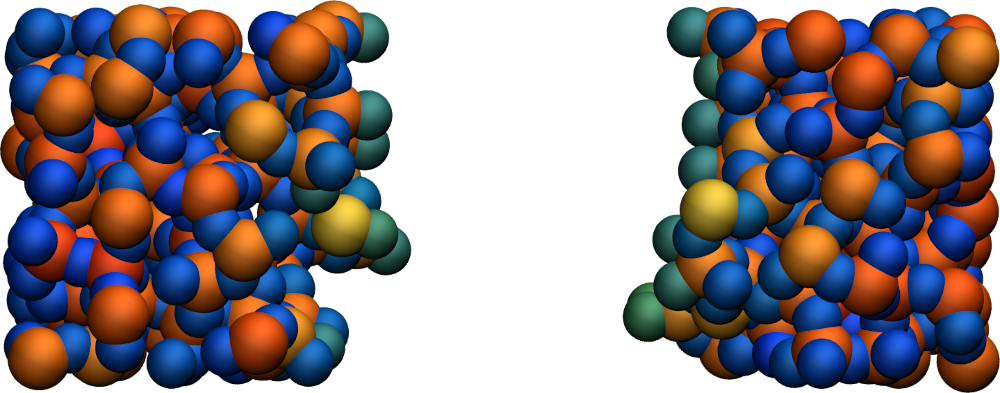}
\caption{Amorphous silicon oxide after deformation during
\hyperref[reactive-silicon-dioxide-label]{Tutorial 5}.  The atoms are colored by their
charges.  Dangling oxygen groups appear in greenish, bulk Si atoms with a charge of
about $1.8~\text{e}$  appear in red/orange, and bulk O atoms with a charge of
about $-0.9 ~ \text{e}$ appear in blue.}
\label{fig:SIO-deformed}
\end{figure}

Run the \lmpcmd{deform.lmp} file using LAMMPS.  During the deformation, the charge
values progressively evolve until the structure eventually breaks down.  After the
structure breaks down, the charges equilibrate near new average values that differ
from the initial averages (Fig.~\ref{fig:SIO-deformed-charge}\,a).  The difference
between the initial and the final charges can be explained by the presence of
defects, as well as new solid/vacuum interfaces, and the fact that surface atoms
typically have different charges compared to bulk atoms (Fig.~\ref{fig:SIO-deformed}).
You can also see a sharp increase in temperature during the rupture of
the material (Fig.~\ref{fig:SIO-deformed-charge}\,b).

You can examine the charge distribution after deformation, as well as during
deformation (Fig.~\ref{fig:SIO-distribution}\,b).  As expected, the final
charge distribution slightly differs from the previously calculated one.  If
no new species were formed during the simulation, the \flecmd{deform.species} file
should look like this:
\begin{lstlisting}
#  Timestep   No_Moles   No_Specs  O384Si192
        5            1          1          1
(...)
#  Timestep   No_Moles   No_Specs  O384Si192
    30000            1          1          1
\end{lstlisting}
Sometimes, $\text{O}_2$ molecules are formed during the deformation.  If this occurs,
a new column \lmpcmd{O2} appears in the \flecmd{deform.species} file.

\subsubsection{Decorate the surface}

Under ambient conditions, some of the surface $\text{SiO}_2$ atoms become chemically
passivated by forming covalent bonds with hydrogen (H) atoms~\cite{sulpizi2012silica}.
We will add hydrogen atoms randomly to the cracked silica and observe how the
system evolves.  To do so, we first need to modify the previously generated data
file \flecmd{deform.data} and make space for a third atom type.
Copy \flecmd{deform.data}, name the copy \flecmd{deform-mod.data}, and modify the
first lines of \flecmd{deform-mod.data} as follows:
\begin{lstlisting}
576 atoms
3 atom types

(...)

Atom Type Labels

1 Si
2 O
3 H

Masses

Si 28.0855
O 15.999
H 1.008

(...)
\end{lstlisting}

Open the \flecmd{decorate.lmp} file, which must contain the following lines:
\begin{lstlisting}
units real
atom_style full

read_data deform-mod.data
displace_atoms all move -12 0 0 # optional

pair_style reaxff NULL safezone 3.0 mincap 150
pair_coeff * * ffield.reax.CHOFe Si O H
fix myqeq all qeq/reaxff 1 0.0 10.0 1.0e-6 reaxff maxiter 400
\end{lstlisting}
The \lmpcmd{displace\_atoms} command is used to move the center of the
crack near the center of the box.  This step is optional but makes for a nicer
visualization.  A different value for the shift may be needed in
your case, depending on the location of the crack.  A difference with the previous
input is that three atom types are specified in the \lmpcmd{pair\_coeff} command, i.e.
\lmpcmd{Si O H}.

\begin{figure}
\includegraphics[width=\linewidth]{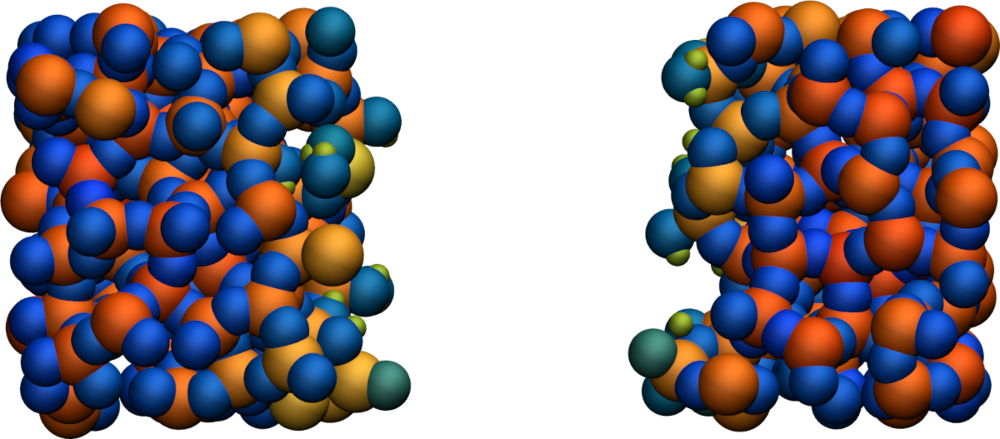}
\caption{Cracked silicon oxide after the addition of hydrogen atoms
during \hyperref[reactive-silicon-dioxide-label]{Tutorial 5}.  The atoms
are colored by their charges, with the newly added hydrogen atoms appearing as small
greenish spheres.}
\label{fig:SIO-decorated}
\end{figure}

Then, let us adapt some familiar commands to measure the charges of all three
types of atoms, and output the charge values into log files:
\begin{lstlisting}
group grpSi type Si
group grpO type O
group grpH type H
variable qSi equal charge(grpSi)/count(grpSi)
variable qO equal charge(grpO)/count(grpO)
variable qH equal charge(grpH)/(count(grpH)+1e-10)

thermo 5
thermo_style custom step temp etotal press v_qSi v_qO v_qH

dump viz all image 100 myimage-*.ppm q &
  type shiny 0.1 box no 0.01 view 180 90 zoom 2.3 size 1200 500
dump_modify viz adiam Si 2.6 adiam O 2.3 adiam H 1.0 &
  backcolor white amap -1 2 ca 0.0 3 min royalblue &
  0 green max orangered

fix myspec all reaxff/species 5 1 5 decorate.species &
  element Si O H
\end{lstlisting}
The commands above are, once again, similar to the ones of the previous script.
Here, the $+1 \mathrm{e}{-10}$ was added to the denominator of the \lmpcmd{variable qH}
to avoid dividing by 0 at the beginning of the simulation, as no hydrogen
atoms exists in the simulation domain yet.  Finally, let us
create a loop with 10 steps, and create two hydrogen atoms at random locations at
every step:
\begin{lstlisting}
fix mynvt all nvt temp 300.0 300.0 100
timestep 0.5

label loop
variable a loop 10

variable seed equal 35672+${a}
create_atoms 3 random 2 ${seed} NULL overlap 2.6 maxtry 50

run 2000

next a
jump SELF loop
\end{lstlisting}
Run the simulation with LAMMPS.  When the simulation is over,
it can be seen from the \flecmd{decorate.species} file that
all the created hydrogen atoms reacted with the $\text{SiO}_{2}$ structure to
form surface groups (such as hydroxyl (-OH) groups).
\begin{lstlisting}
(...)
# Timestep   No_Moles No_Specs H20O384Si192
  20000      1        1        1
\end{lstlisting}
At the end of the simulation, hydroxyl (-OH) groups can be seen at the interfaces
(Fig.~\ref{fig:SIO-decorated}).

\subsection{Tutorial 6: Water adsorption in silica}
\label{gcmc-silica-label}

\begin{figure}
\centering
\includegraphics[width=0.6\linewidth]{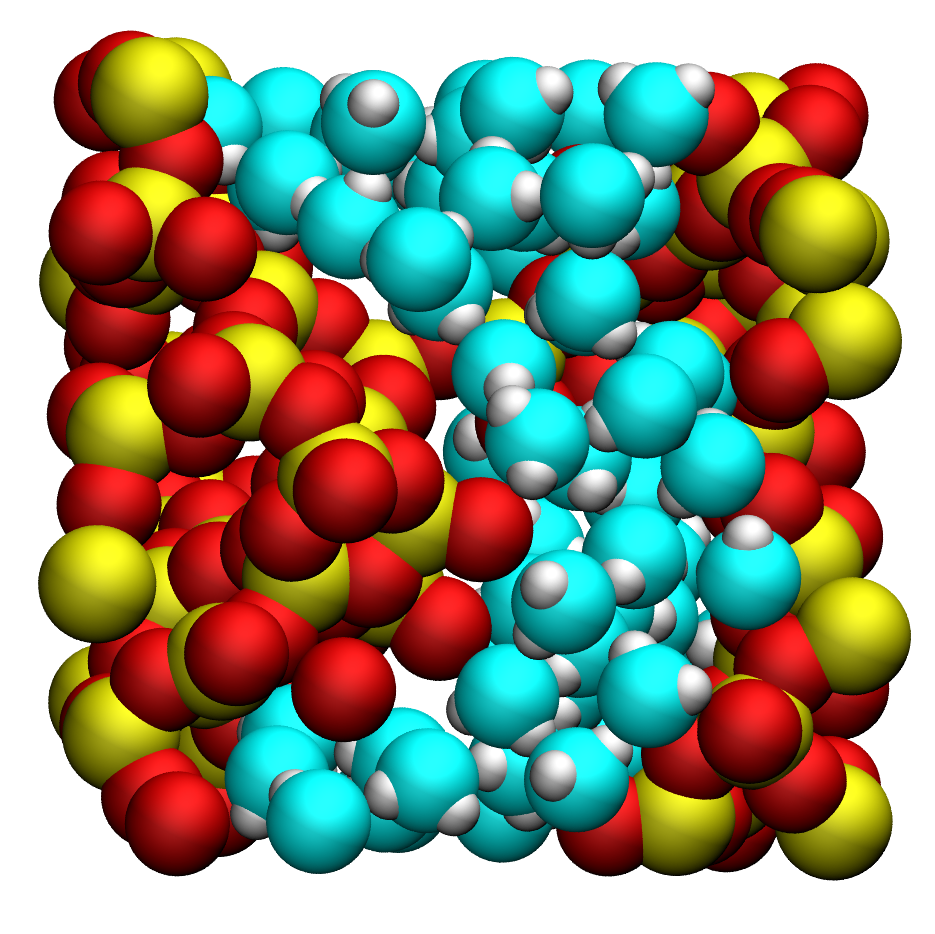}
\caption{Water molecules (H$_2$O) adsorbed in cracked silica (SiO$_2$) material as simulated
during \hyperref[gcmc-silica-label]{Tutorial 6}.  The oxygen atoms of the water
molecules are represented in cyan, the silicon atoms in yellow, and the oxygen
atoms of the solid in red.}
\label{fig:GCMC}
\end{figure}

\noindent The objective of this tutorial is to combine molecular dynamics and
grand canonical Monte Carlo simulations to compute the adsorption of water
molecules in cracked silica material (Fig.~\ref{fig:GCMC}).  This tutorial
illustrates the use of the grand canonical ensemble in molecular simulation, an
open ensemble where the number of atoms or molecules in the simulation box can vary.
By using this combination, we simulate water in a nanoporous
SiO$_2$ structure at a specified chemical potential.

\subsubsection{Generation of the silica block}

To begin this tutorial, select \guicmd{Start Tutorial 6} from the
\guicmd{Tutorials} menu of \lammpsgui{} and follow the instructions.
The editor should display the following content corresponding to \flecmd{generate.lmp}:
\begin{lstlisting}
units metal
boundary p p p
atom_style full
pair_style vashishta
neighbor 1.0 bin
neigh_modify delay 1
\end{lstlisting}
The main difference from some of the previous tutorials is the use of the \lmpcmd{Vashishta}
pair style.  The Vashishta potential implicitly models atomic bonds through
energy terms dependent on interatomic distances and angles~\cite{vashishta1990interaction}.

Let us create a box for two atom types, \lmpcmd{Si}
of mass 28.0855\,g/mol and \lmpcmd{O} of mass 15.9994\,g/mol.
Add the following lines to \flecmd{generate.lmp}:
\begin{lstlisting}
region box block -18.0 18.0 -9.0 9.0 -9.0 9.0
create_box 2 box
labelmap atom 1 Si 2 O
mass Si 28.0855
mass O 15.9994
create_atoms Si random 240 5802 box overlap 2.0 maxtry 500
create_atoms O random 480 1072 box overlap 2.0 maxtry 500
\end{lstlisting}
In line with what is done in previous tutorials, the
\lmpcmd{create\_atoms} commands are used to place
240 Si atoms and 480 O atoms, respectively, in the region previously defined.  This corresponds to
an initial density of approximately $2$\,g/cm$^3$, which is close
to the expected final density of amorphous silica at 300\,K.

Now, specify the potential parameters by indicating that the first atom type
is \lmpcmd{Si} and the second is \lmpcmd{O}:
\begin{lstlisting}
pair_coeff * * SiO.1990.vashishta Si O
\end{lstlisting}
Ensure that the \href{\filepath tutorial6/SiO.1990.vashishta}{\dwlcmd{SiO.1990.vashishta}}
file is located in the same directory as \flecmd{generate.lmp}.

\begin{figure}
\centering
\includegraphics[width=\linewidth]{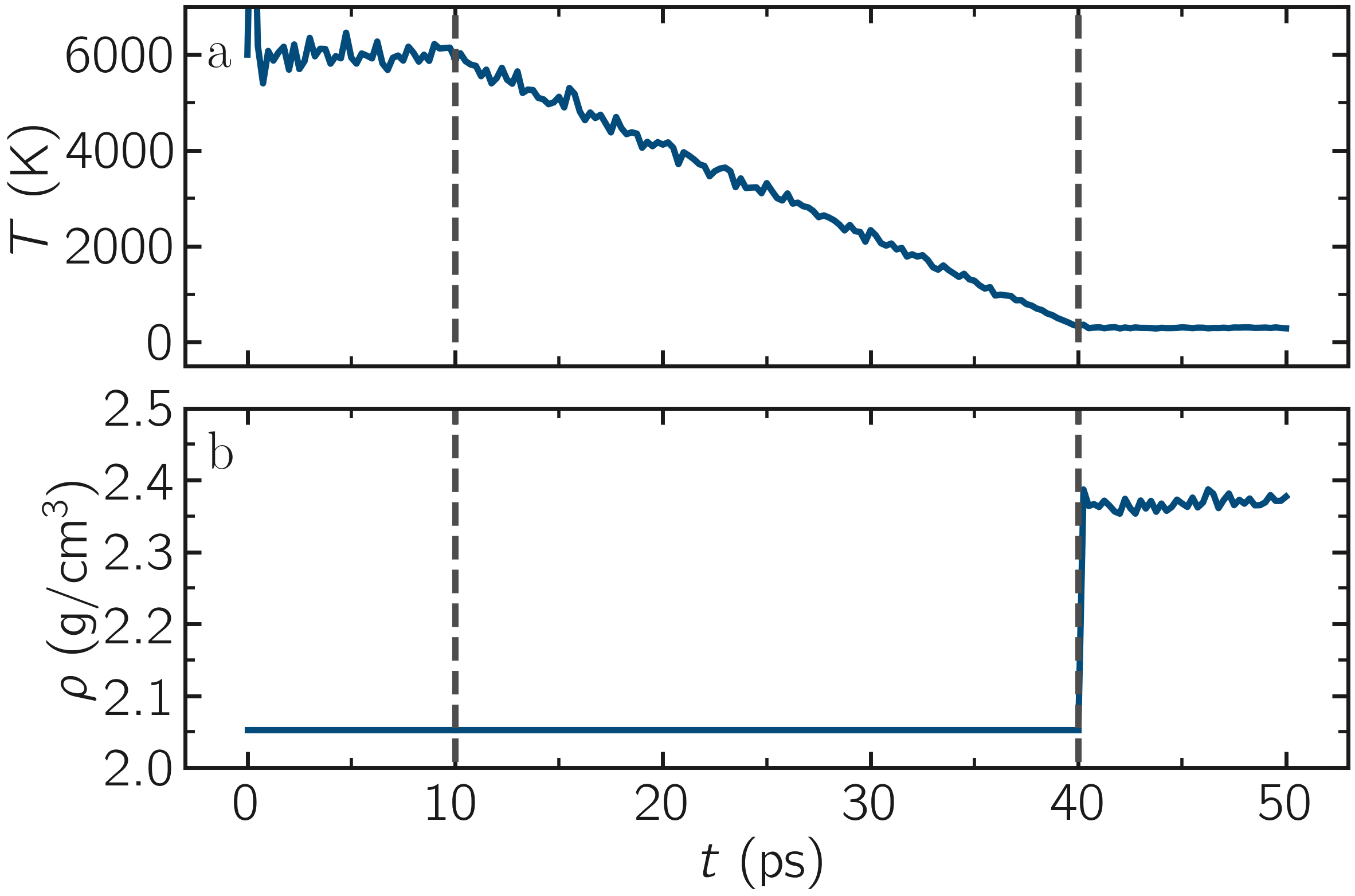}\\[-2ex]
\caption{a) Temperature, $T$, as a function of time, $t$, during the annealing
of the silica system from \hyperref[gcmc-silica-label]{Tutorial 6}.
b) System density, $\rho$, during the annealing process.  The vertical dashed lines
mark the transition between the different phases of the simulation.}
\label{fig:GCMC-dimension}
\end{figure}

Next, add a \lmpcmd{dump image} command to \flecmd{generate.lmp} to follow the
evolution of the system with time:
\begin{lstlisting}
dump viz all image 250 myimage-*.ppm type type &
  shiny 0.1 box no 0.01 view 180 90 zoom 3.4 size 1700 700
dump_modify viz backcolor white &
  acolor Si yellow adiam Si 2.5 &
  acolor O red adiam O 2
\end{lstlisting}
Let us also print the box volume and system density, alongside the
temperature and total energy:
\begin{lstlisting}
thermo 250
thermo_style custom step temp etotal vol density
\end{lstlisting}

Finally, let us implement the annealing procedure which
consists of three consecutive runs.  This procedure was inspired
by Ref.\,\cite{della1992molecular}.  First, to melt the system,
a $10\,\text{ps}$ run at $T = 6000\,\text{K}$ is performed:
\begin{lstlisting}
velocity all create 6000 8289 rot yes dist gaussian
fix mynvt all nvt temp 6000 6000 0.1
timestep 0.001
run 10000
\end{lstlisting}
Next, a second run, during which the system is cooled down from $T = 6000\,\text{K}$
to $T = 300\,\text{K}$, is implemented as follows:
\begin{lstlisting}
fix mynvt all nvt temp 6000 300 0.1
run 30000
\end{lstlisting}
In this case, the initial and final target temperatures
set for the Nos\'e-Hoover thermostat is different, causing it to evolve
linearly within the number of timesteps evoked in the \lmpcmd{run} command.
In the third run, the system is equilibrated at the final desired
conditions, $T = 300\,\text{K}$ and $p = 1\,\text{atm}$,
using an anisotropic pressure coupling:
\begin{lstlisting}
unfix mynvt

fix mynpt all npt temp 300 300 0.1 aniso 1 1 1
run 10000

write_data generate.data
\end{lstlisting}
Here, an anisotropic barostat is used.
As previously mentioned, anisotropic
barostats adjust the dimensions independently, which is
generally suitable for a solid phase.

\begin{figure}
\centering
\includegraphics[width=0.9\linewidth]{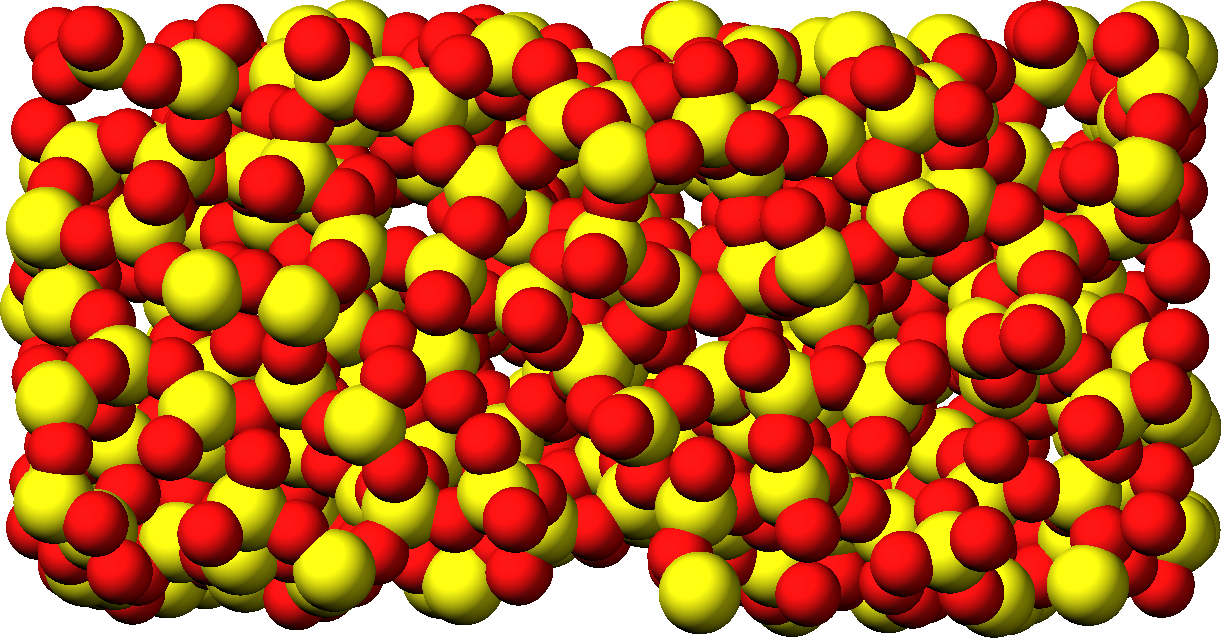}
\caption{Amorphous silica ($\text{SiO}_2$) simulated
during \hyperref[gcmc-silica-label]{Tutorial 6}.  Silicon atoms are
represented in yellow, and oxygen atoms in red.}
\label{fig:GCMC-snapshot}
\end{figure}

Run the simulation using LAMMPS.  From the \guicmd{Charts} window, the temperature
evolution can be observed, showing that it closely follows the desired annealing procedure (Fig.~\ref{fig:GCMC-dimension}\,a).
The evolution of the box dimensions over time confirms that the box
is deforming during the last stage of the simulation
(Fig.~\ref{fig:GCMC-dimension}\,b).  After the simulation completes, the final
microstate attained during the dynamics and the system topology
will be written to a LAMMPS data file called \flecmd{generate.data}
which will be located next to \flecmd{generate.lmp} (Fig.~\ref{fig:GCMC-snapshot}).

\subsubsection{Cracking the silica}

\begin{figure}
\centering
\includegraphics[width=\linewidth]{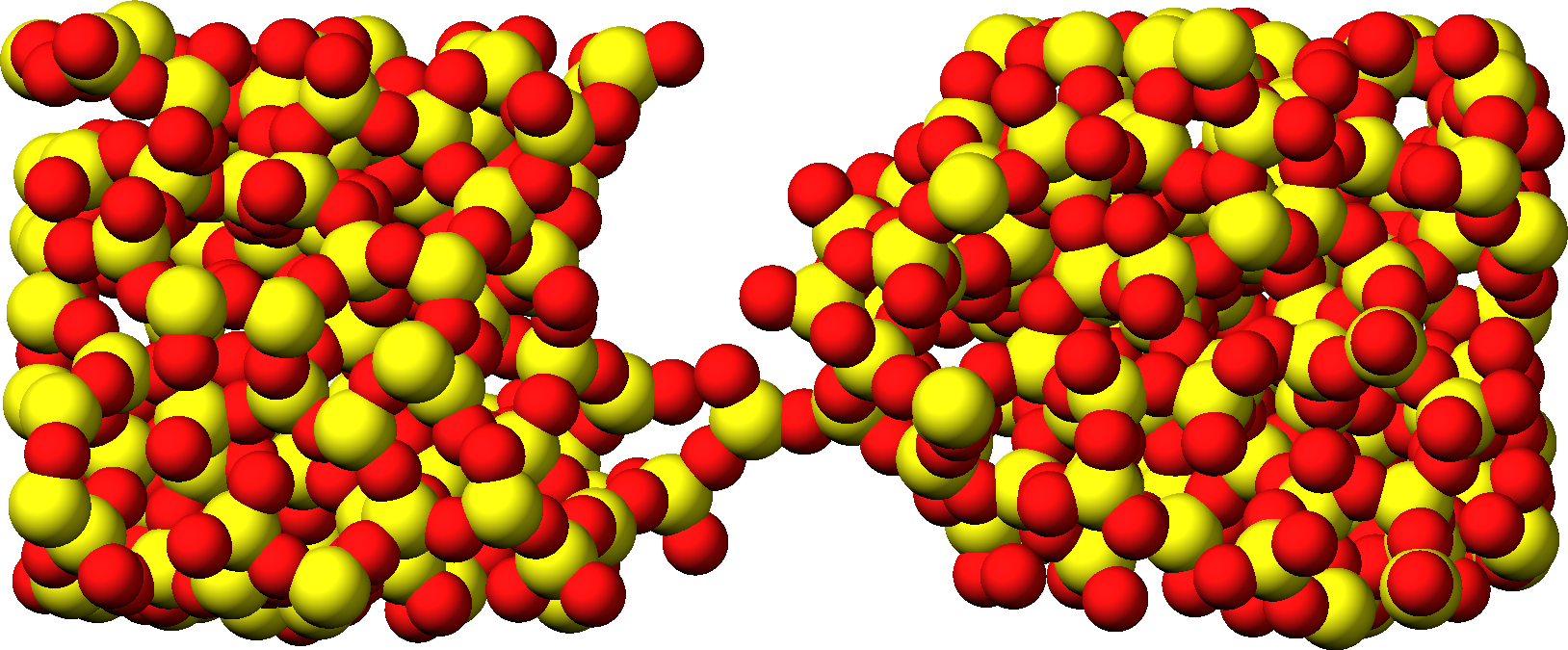}
\caption{Block of silica from \hyperref[gcmc-silica-label]{Tutorial 6}
after deformation.  Silicon atoms are represented in yellow,
and oxygen atoms in red.  The crack was induced by the
imposed deformation of the box along the $x$-axis (i.e.,~the horizontal axis).}
\label{fig:GCMC-cracked}
\end{figure}

Open the \flecmd{cracking.lmp} file, which must contain the following familiar lines:
\begin{lstlisting}
units metal
boundary p p p
atom_style full
pair_style vashishta
neighbor 1.0 bin
neigh_modify delay 1

read_data generate.data

pair_coeff * * SiO.1990.vashishta Si O

dump viz all image 250 myimage-*.ppm type type &
  shiny 0.1 box no 0.01 view 180 90 zoom 3.4 size 1700 700
dump_modify viz backcolor white &
  acolor Si yellow adiam Si 2.5 &
  acolor O red adiam O 2

thermo 250
thermo_style custom step temp etotal vol density
\end{lstlisting}
Let us progressively increase the size of the box in the $x$ direction,
forcing the silica to deform and eventually crack.  To achive this,
the \lmpcmd{fix deform} command is used, with a rate
of $0.005\,\text{ps}^{-1}$.  Add the following lines to
the \flecmd{cracking.lmp} file:
\begin{lstlisting}
timestep 0.001
fix nvt1 all nvt temp 300 300 0.1
fix mydef all deform 1 x erate 0.005
run 50000

write_data cracking.data
\end{lstlisting}
As discussed, the \lmpcmd{fix nvt} command integrates the Nosé-Hoover equations
of motion and is employed to control the temperature of the system.
As observed from the generated images, the atoms
progressively adjust to the changing box dimensions.  At some point,
bonds begin to break, leading to the appearance of
dislocations (Fig.~\ref{fig:GCMC-cracked}).

\begin{note}
  Although the Nosé-Hoover equations were originally formulated to sample the
  NVT ensemble, using the \lmpcmd{fix nvt} command does not guarantee that
  a simulation actually samples the NVT ensemble.
\end{note}

\subsubsection{Adding water}

To add the water molecules to the silica, we will employ the Monte Carlo
method in the grand canonical ensemble (GCMC).  In short, the system is
placed into contact with a virtual reservoir containing pure
water at a given thermodynamic state, and multiple attempts to insert
water molecules at random positions are made.  In the grand
canonical ensemble, each attempt is either accepted or rejected based on
internal energy and chemical potential, $\mu$
considerations.  For further details, please refer to
classical textbooks like Ref.~\citenum{frenkel2023understanding}.

\paragraph{Adapting the pair style}

For this next step, we need to specify the force field used
to model the interactions in the system. The TIP4P/2005 model is employed
for the water~\cite{abascal2005general}, while no interaction
within silica is defined, as it will be seen farther below.
This is because the atoms of the silica will
remain frozen during this part of the simulation.
Only the cross-interactions between water and silica need
to be defined. Open the \flecmd{gcmc.lmp} file, which should contain the following lines:
\begin{lstlisting}
units metal
boundary p p p
atom_style full
neighbor 1.0 bin
neigh_modify delay 1
pair_style lj/cut/tip4p/long OW HW OW-HW HW-OW-HW &
    0.1546 10
kspace_style pppm/tip4p 1.0e-5
bond_style harmonic
angle_style harmonic
\end{lstlisting}
The PPPM solver~\cite{luty1996calculating} is specified with the \lmpcmd{kspace}
command, and is used to compute the long-range Coulomb interactions associated
with \lmpcmd{tip4p/long}.  Finally, the form of the bond
and angle potentials of the water molecules are defined; however,
as previously discussed, these specifications are
not critical since TIP4P/2005 is a rigid water model.

\begin{note}
  In practice, it is possible to use both \lmpcmd{vashishta} and
  \lmpcmd{lj/cut/tip4p/long} pair styles at the same time by employing the
  \lmpcmd{pair\_style hybrid}
  command.  However, hybridizing force fields should be done with caution, as there
  is no guarantee that the resulting force field will produce meaningful results.
\end{note}

The water molecule template called \href{\filepath tutorial6/H2O.mol}{\dwlcmd{H2O.mol}}
must be downloaded and located next to \flecmd{gcmc.lmp}.

Before going further, we need to make a few changes to our data file.
Currently, the \flecmd{cracking.data} file includes only two atom types, but we require four.
Copy the previously generated \flecmd{cracking.data}, and name the duplicate \flecmd{cracking-mod.data}.
Make the following changes to the beginning of \flecmd{cracking-mod.data}
to ensure it matches the following format (with 4 atom types,
1 bond type, 1 angle type, the proper type labels, and four masses):
\begin{lstlisting}
720 atoms
4 atom types
1 bond types
1 angle types

2 extra bond per atom
1 extra angle per atom
2 extra special per atom

-22.470320800269317 22.470320800269317 xlo xhi
-8.579178758211475 8.579178758211475 ylo yhi
-8.491043517346204 8.491043517346204 zlo zhi

Atom Type Labels

1 Si
2 O
3 OW
4 HW

Bond Type Labels

1 OW-HW

Angle Type Labels

1 HW-OW-HW

Masses

1 28.0855
2 15.9994
3 15.9994
4 1.008

Atoms # full

(...)
\end{lstlisting}
Doing so, we anticipate that there will be 4 atom types in the simulations,
with the oxygens and hydrogens of $\text{H}_2\text{O}$ having
types \lmpcmd{OW} and \lmpcmd{HW}, respectively.  There
will also be 1 bond type (\lmpcmd{OW-HW}) and 1 angle type (\lmpcmd{OW-HW-HW}).
The \lmpcmd{extra bond}, \lmpcmd{extra angle}, and
\lmpcmd{extra special} lines are here for memory allocation.

We can now proceed to complete the \flecmd{gcmc.lmp} file by adding the system definition:
\begin{lstlisting}
read_data cracking-mod.data
molecule h2omol H2O.mol
create_atoms 0 random 3 3245 NULL mol h2omol 4585 &
  overlap 2.0 maxtry 50

group SiO type Si O
group H2O type OW HW
\end{lstlisting}
After reading the data file and defining the \lmpcmd{h2omol} molecule from the \flecmd{H2O.txt}
file, the \lmpcmd{create\_atoms} command is used to include three water molecules
in the system.  Then, add the following \lmpcmd{pair\_coeff} (and
\lmpcmd{bond\_coeff} and \lmpcmd{angle\_coeff}) commands
to \flecmd{gcmc.lmp} in order to set the potential parameters:
\begin{lstlisting}
pair_coeff * * 0 0
pair_coeff Si OW 0.0057 4.42
pair_coeff O OW 0.0043 3.12
pair_coeff OW OW 0.008 3.1589
pair_coeff HW HW 0.0 0.0
bond_coeff OW-HW 0 0.9572
angle_coeff HW-OW-HW 0 104.52
\end{lstlisting}
Pair coefficients for the \lmpcmd{lj/cut/tip4p/long} pair style are
defined between O($\text{H}_2\text{O}$) and between H($\text{H}_2\text{O}$)
atoms, as well as between O($\text{SiO}_2$)-O($\text{H}_2\text{O}$) and
Si($\text{SiO}_2$)-O($\text{H}_2\text{O}$).  Thus, the fluid-fluid and the
fluid-solid interactions will be adressed with by the \lmpcmd{lj/cut/tip4p/long} potential.
The \lmpcmd{bond\_coeff} and \lmpcmd{angle\_coeff} commands set the \lmpcmd{OW-HW}
bond length to 0.9572\,\AA, and the \lmpcmd{HW-OW-HW}
angle to $104.52^\circ$, respectively~\cite{abascal2005general}.

\begin{note}
  The pair coefficients for interactions between Si($\text{SiO}_2$)
  and O($\text{SiO}_2$) are set by the first command, \lmpcmd{pair\_coeff * * 0 0},
  which effectively means that they do not interact. This is acceptable here because
  the silica atoms remain frozen during this part of the tutorial.
\end{note}

Add the following lines to \flecmd{gcmc.lmp} as well:
\begin{lstlisting}
variable oxygen atom type==label2type(atom,OW)
group oxygen dynamic all var oxygen
variable nO equal count(oxygen)

fix shak H2O shake 1.0e-5 200 0 b OW-HW &
  a HW-OW-HW mol h2omol
\end{lstlisting}
The number of oxygen atoms from water molecules (i.e.~the number of molecules)
is calculated by the \lmpcmd{nO} variable.  As already discussed in other
tutorials, the SHAKE algorithm is used to
maintain the shape of the water molecules over time~\cite{ryckaert1977numerical, andersen1983rattle}.

\begin{note}
  Here, a variable of style `atom' is used.  Such variable
  defines a per-atom property, i.e., it evaluates the specified expression
  separately for each atom.  This is often used to select atoms based on
  their properties or types.
\end{note}

Finally, let us create images
of the system using \lmpcmd{dump image}:
\begin{lstlisting}
dump viz all image 250 myimage-*.ppm type type &
  shiny 0.1 box no 0.01 view 180 90 zoom 3.4 size 1700 700
dump_modify viz backcolor white &
  acolor Si yellow adiam Si 2.5 &
  acolor O red adiam O 2 &
  acolor OW cyan adiam OW 2 &
  acolor HW white adiam HW 1
\end{lstlisting}

\paragraph{GCMC simulation}

To prepare for the GCMC simulation, let us add the
following lines into \flecmd{gcmc.lmp}:
\begin{lstlisting}
compute ctH2O H2O temp
compute_modify thermo_temp dynamic/dof yes
compute_modify ctH2O dynamic/dof yes
fix mynvt H2O nvt temp 300 300 0.1
fix_modify mynvt temp ctH2O
timestep 0.001
\end{lstlisting}
Here, the \lmpcmd{fix nvt} applies only to the water molecules, so
the atoms in the silica remain fixed.  The \lmpcmd{compute\_modify} command with
the \lmpcmd{dynamic/dof yes} option is used for water to account for the fact
that the number of molecules is not constant.

\begin{figure}
\centering
\includegraphics[width=\linewidth]{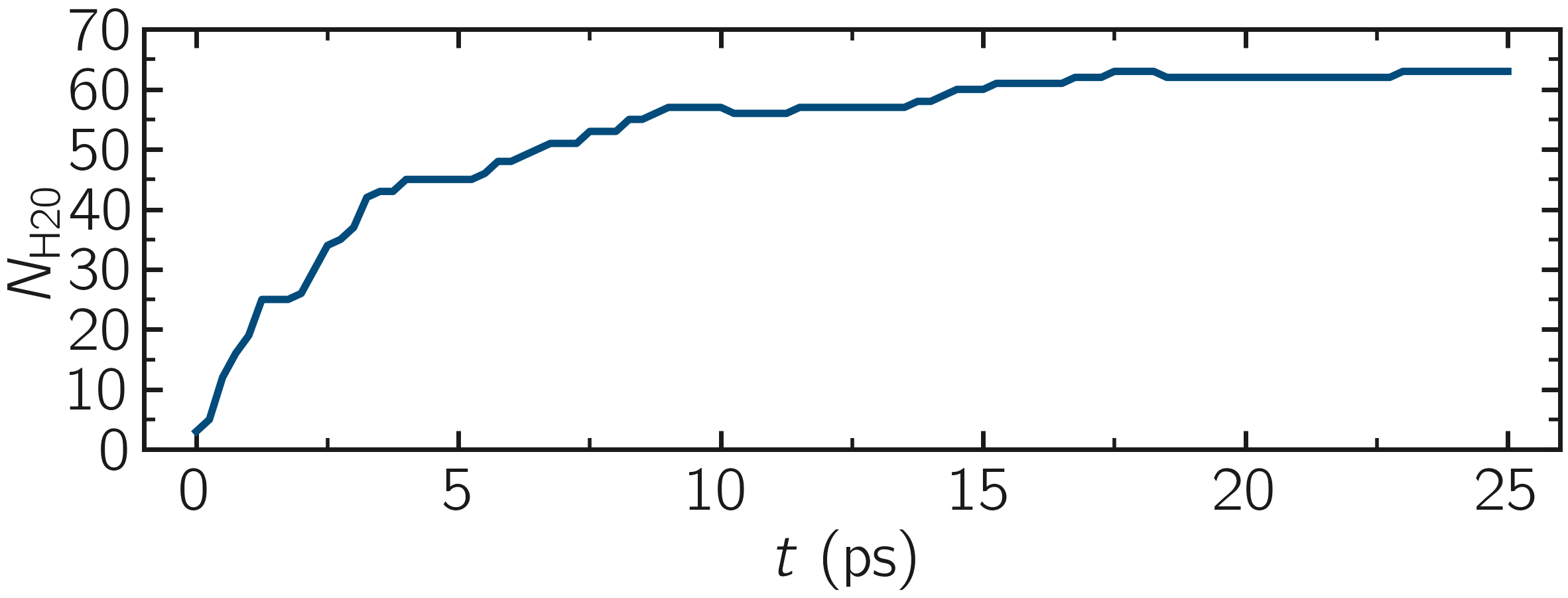}\\[-2ex]
\caption{Number of water molecules, $N_\text{H2O}$, as a function of time, $t$,
as extracted from \hyperref[gcmc-silica-label]{Tutorial 6}.}
\label{fig:GCMC-number}
\end{figure}

Finally, let us use the \lmpcmd{fix gcmc} and perform the grand canonical Monte
Carlo steps.  Add the following lines into \flecmd{gcmc.lmp}:
\begin{lstlisting}
variable tfac equal 5.0/3.0
fix fgcmc H2O gcmc 100 100 0 0 65899 300 -0.5 0.1 &
  mol h2omol tfac_insert ${tfac} shake shak &
  full_energy pressure 100
\end{lstlisting}
The \lmpcmd{fix gcmc} command performs grand canonical Monte Carlo
moves to insert, delete, or swap molecules. Here, 100 attempts are made every
100 steps.  The \lmpcmd{mol h2omol} keyword specifies the
molecule type being inserted/deleted, while \lmpcmd{shake shak} enforces rigid
molecular constraints during these moves. With the \lmpcmd{pressure 100} keyword,
a fictitious reservoir with a pressure of 100 atmospheres is used.
The \lmpcmd{tfac\_insert} option ensures the correct estimate for the temperature
of the inserted water molecules by taking into account the internal degrees of
freedom.

\begin{note}
  At a pressure of $p = 100\,\text{bar}$, the chemical potential of water vapor at $T = 300\,\text{K}$
  can be calculated using as $\mu = \mu_0 + RT \ln (\frac{p}{p_0}),$ where $\mu_0$ is the standard
  chemical potential (typically taken at a pressure $p_0 = 1 \, \text{bar}$), $R = 8.314\, \text{J/mol·K}$
  is the gas constant, $T = 300\,\text{K}$ is the temperature.
\end{note}

Finally, let us print some information and run for 25\,ps:
\begin{lstlisting}
thermo 250
thermo_style custom step temp etotal v_nO &
  f_fgcmc[3] f_fgcmc[4] f_fgcmc[5] f_fgcmc[6]

run 25000
\end{lstlisting}
The \lmpcmd{f\_} keywords extract the Monte Carlo move statistics
which is computed (and can be extracted) by the \lmpcmd{fix gcmc} command.

\begin{figure}
\centering
\includegraphics[width=\linewidth]{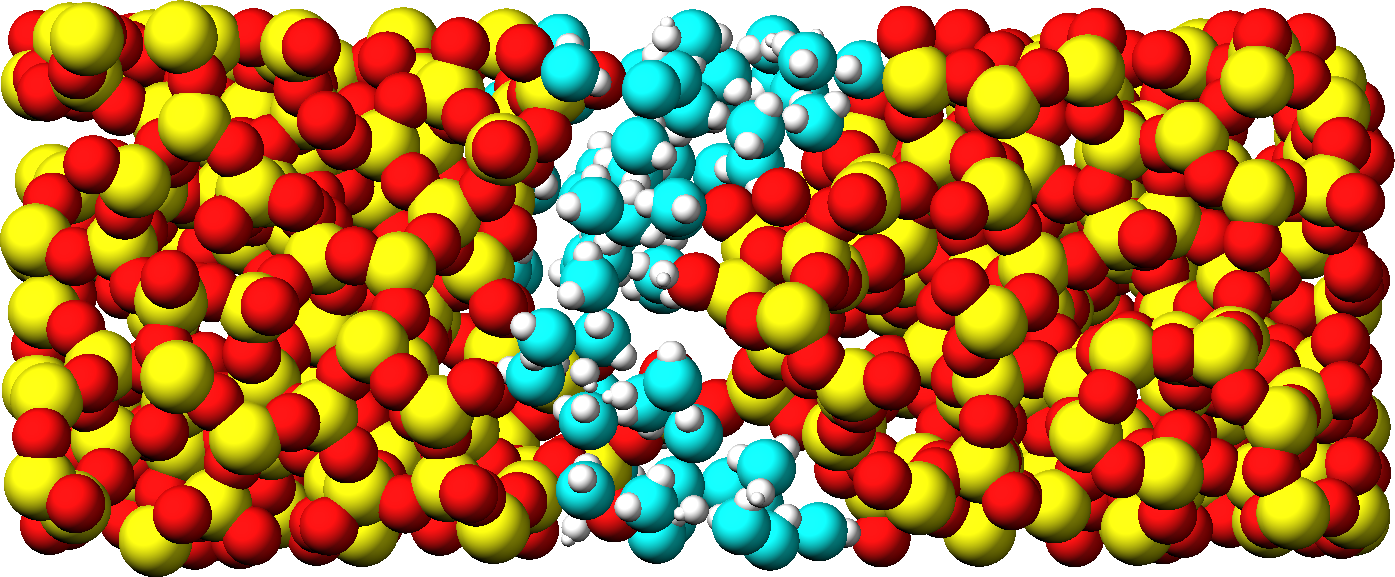}
\caption{Snapshot of the silica system after the adsorption of water molecules
during \hyperref[gcmc-silica-label]{Tutorial 6}.
The oxygen atoms of the water molecules are represented in cyan, the silicon
atoms in yellow, and the oxygen atoms of the solid in red.}
\label{fig:GCMC-solvated}
\end{figure}

\begin{note}
  When using the pressure argument, LAMMPS ignores the value of the
  chemical potential (here $\mu = -0.5\,\text{eV}$, which corresponds roughly to
  ambient conditions, i.e.~to a relative humidity $\text{RH} \approx 50\,\%$~\cite{gravelle2020multi}.)
  The large pressure value of 100\,bars was chosen to ensure that some successful
  insertions of molecules would occur during the short duration of this simulation.
\end{note}

Running this simulation using LAMMPS, one can see that
after a few GCMC steps, the number of molecules starts increasing.  Once the
crack is filled with water molecules, the total number of molecules reaches a plateau
(Figs.\,\ref{fig:GCMC-number}-\ref{fig:GCMC-solvated}).  The final number of
molecules depends on the imposed pressure, temperature, and the interaction
between water and silica (i.e.~its hydrophilicity).  Note that GCMC simulations
of such dense phases are usually slow to converge due to the very low probability
of successfully inserting a molecule.  Here, the short simulation duration was
made possible by the use of a high pressure.

\subsection{Tutorial 7: Free energy calculation}
\label{umbrella-sampling-label}

\begin{figure}
\centering
\includegraphics[width=0.7\linewidth]{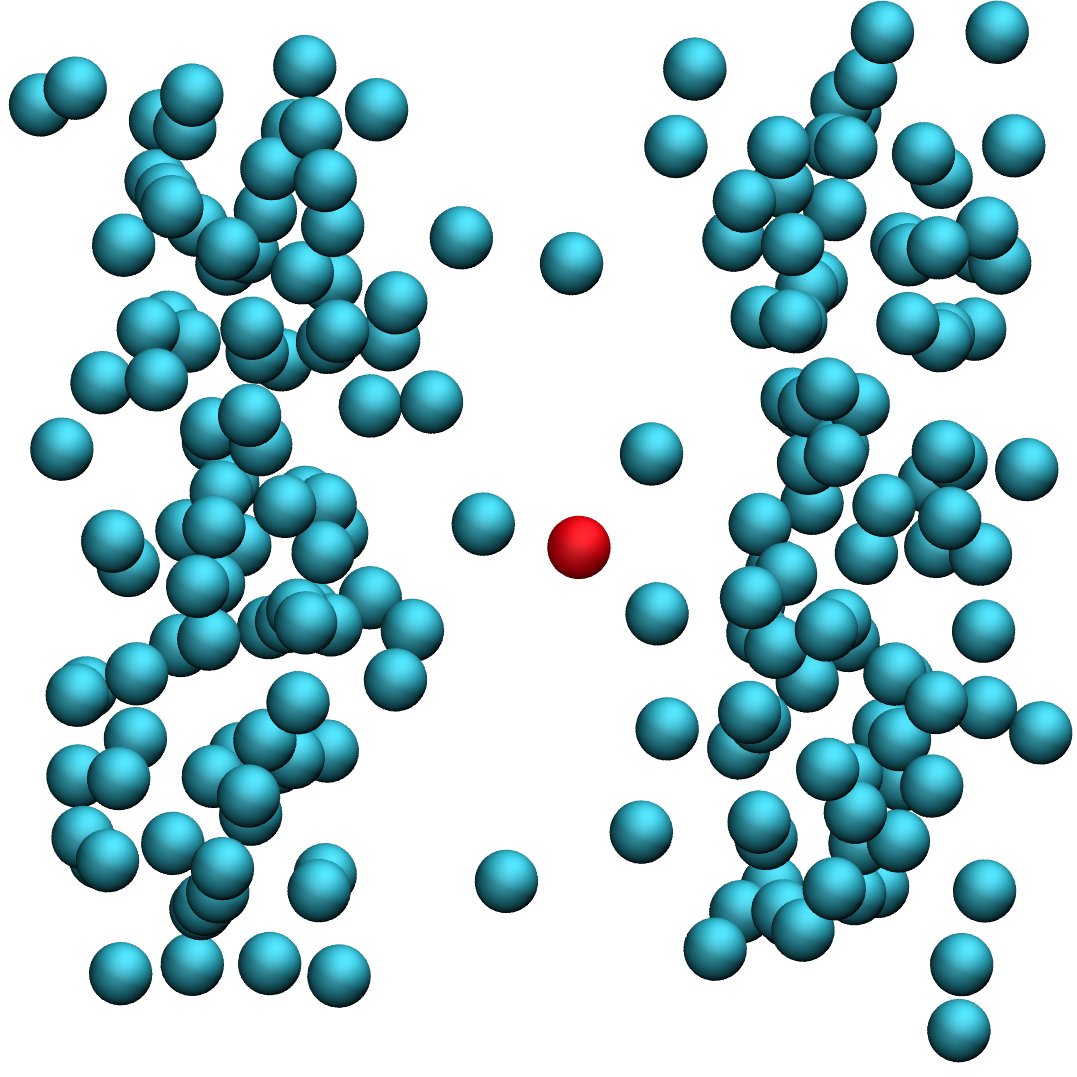}
\caption{System simulated during \hyperref[umbrella-sampling-label]{Tutorial 7}.
The pink atom explores the energetically unfavorable central area of the simulation
box thanks to the additional potential imposed during umbrella sampling.}
\label{fig:US}
\end{figure}

The objective of this tutorial is to measure the free energy profile of
particles through a barrier potential using two methods: free sampling
and umbrella sampling~\cite{kastner2011umbrella, allen2017computer,
  frenkel2023understanding} (Fig.~\ref{fig:US}).  To simplify the
process and minimize computation time, the barrier potential will be
imposed on the atoms using an additional force, mimicking the presence
of a repulsive area in the middle of the simulation box without needing
to simulate additional atoms.  The procedure is valid for more complex
systems and can be adapted to many other situations, such as measuring
adsorption barriers near an interface or calculating translocation
barriers through a membrane~\cite{wilson1997adsorption,
  makarov2009computer, gravelle2021adsorption, loche2022molecular,
  hayatifar2024probing}.

\subsubsection{Method 1: Free sampling}
The most direct way to estimate a free energy profile is to sample
the Boltzmann distribution using a classical (i.e.~unbiased) molecular dynamics
simulation, and compute relative Gibbs free energies from the relative probabilities
of states using
\begin{equation}
\Delta G = -RT \ln(\rho/\rho_0),
\label{eq:G}
\end{equation}
where $\Delta G$ is the free energy difference, $R$ is the gas constant, $T$
is the temperature, 
$\rho$ is the density, and $\rho_0$ is a reference density.
As an illustration, let us apply this method to a simple system
that consists of a particles in a box in the presence of a
position-dependent repulsive force that makes the center of the box a less
favorable area to explore.

\paragraph{Basic LAMMPS parameters}

To begin this tutorial, select \guicmd{Start Tutorial 7} from the
\guicmd{Tutorials} menu of \lammpsgui{} and follow the instructions.
The editor should display the following content corresponding to \flecmd{free-sampling.lmp}:
\begin{lstlisting}
variable sigma equal 3.405
variable epsilon equal 0.238
variable U0 equal 1.5*${epsilon}
variable dlt equal 1.0
variable x0 equal 10.0

units real
atom_style atomic
pair_style lj/cut $(2^(1/6)*v_sigma)
pair_modify shift yes
boundary p p p
\end{lstlisting}
Here, we begin by defining variables for the Lennard-Jones parameters
($\sigma$ and $\epsilon$) and for the repulsive potential parameters
$U$, which are $U_0$, $\delta$, and
$x_0$ [see Eqs.\,(\ref{eq:U}-\ref{eq:F}) below].  The cut-off value of
$ 2^{1/6} \sigma = 3.822$ was chosen to create a Weeks-Chandler-Andersen (WCA) potential,
which is a truncated and purely repulsive LJ potential~\cite{weeks1971role}.
The potential is also shifted to be equal to 0 at the cut-off
using the \lmpcmd{pair\_modify} command.  The unit system is
\lmpcmd{real}, in which energy is in kcal/mol, distance in Ångströms, or
time in femtosecond, has been chosen for practical reasons: the WHAM
algorithm used in the second part of the tutorial automatically assumes
the energy to be in kcal/mol.

\begin{note}
  The syntax \texttt{\$(...)}, where a dollar sign is followed by parentheses, allows
  you to evaluate a numeric formula immediately, without having to assign it
  to a named variable first.
\end{note}

\paragraph{System creation and settings}

Let us define the simulation box and randomly add atoms by addying the
following lines to \flecmd{free-sampling.lmp}:
\begin{lstlisting}
region myreg block -50 50 -15 15 -50 50
create_box 1 myreg
create_atoms 1 random 200 34134 myreg overlap 3 maxtry 50

mass * 39.95
pair_coeff * * ${epsilon} ${sigma}
\end{lstlisting}

\begin{note}
  In the \lmpcmd{pair\_coeff} command, the first two asterisks
  \lmpcmd{* *} indicate that the parameters apply to all atom types in the simulation.
\end{note}

\begin{figure}
\centering
\includegraphics[width=\linewidth]{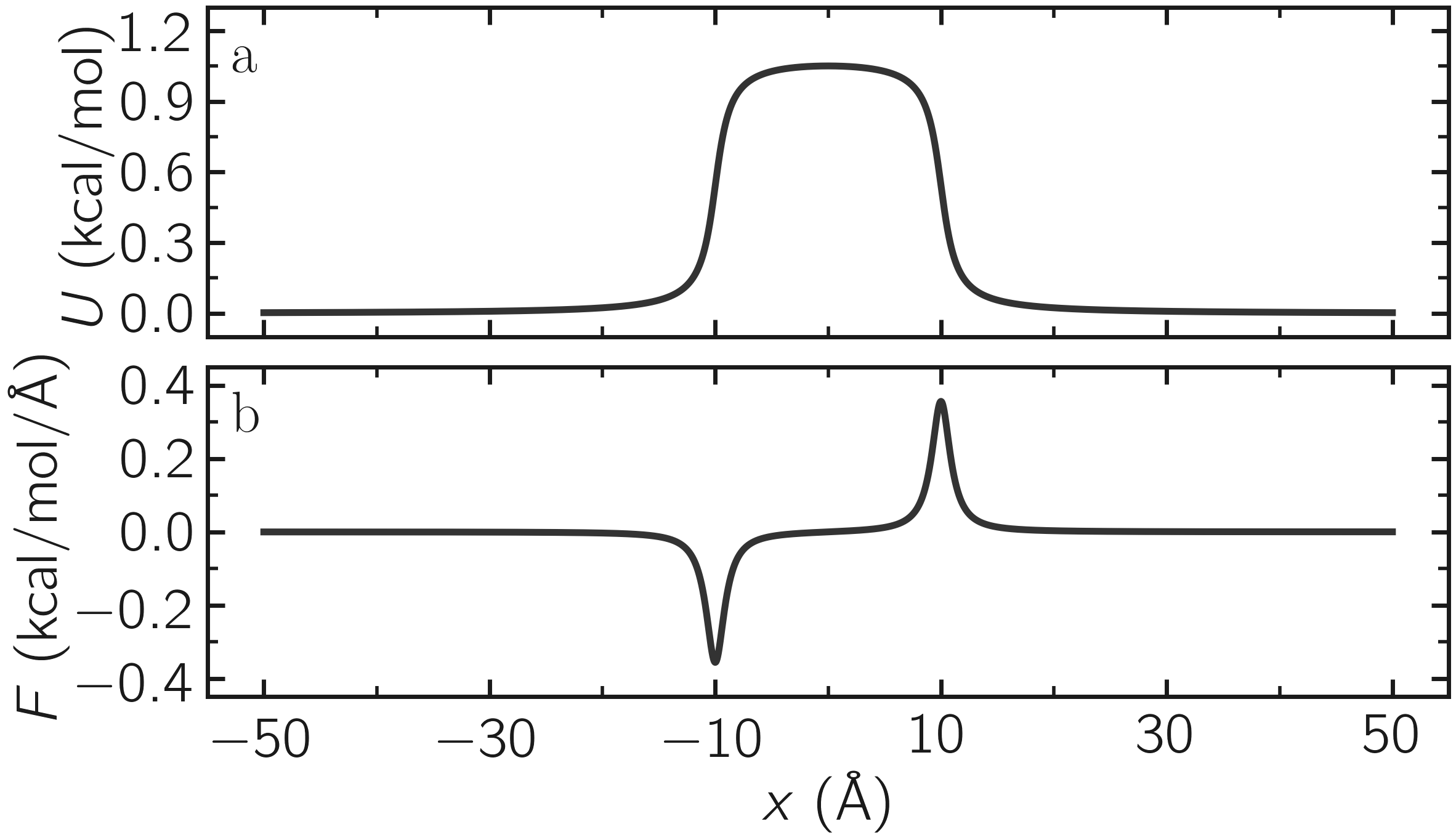}
\caption{Potential $U$ given in Eq.~\eqref{eq:U} (a) and force $F$ given in
Eq.~\eqref{eq:F} (b) as functions of the coordinate $x$. Here,
$U_0 = 0.36~\text{kcal/mol}$, $\delta = 1.0~\text{\AA{}}$, and $x_0 = 10~\text{\AA{}}$.}
\label{fig:potential}
\end{figure}

The variables $U_0$, $\delta$, and $x_0$, defined in the previous subsection, are
used here to create the repulsive potential, restricting the atoms from exploring
the center of the box:
\begin{equation}
U = U_0 \left[ \arctan \left( \dfrac{x+x_0}{\delta} \right)
- \arctan \left(\dfrac{x-x_0}{\delta} \right) \right].
\label{eq:U}
\end{equation}
Taking the derivative of the potential with respect to $x$, we obtain the expression
for the force that will be imposed on the atoms:
\begin{equation}
F = \dfrac{U_0}{\delta} \left[ \dfrac{1}{(x-x_0)^2/\delta^2+1}
- \dfrac{1}{(x+x_0)^2/\delta^2+1} \right].
\label{eq:F}
\end{equation}
Fig.~\ref{fig:potential} shows the potential $U$ and force $F$ along the $x$-axis.
With $U_0 = 1.5 \epsilon = 0.36\,\text{kcal/mol}$, $U_0$ is of the same order of magnitude as the
thermal energy $k_\text{B} T = 0.24\,\text{kcal/mol}$, where $k_\text{B} = 0.002\,\text{kcal/mol/K}$
is the Boltzmann constant and $T = 119.8\,\text{K}$ is the temperature
used in this simulation.  Under these conditions, particles are expected to
frequently overcome the energy barrier due to thermal agitation.

We impose the force $F(x)$ to the atoms in the simulation
using the \lmpcmd{fix addforce} command.  Add the following
lines to \flecmd{free-sampling.lmp}:
\begin{lstlisting}
variable U atom ${U0}*atan((x+${x0})/${dlt})&
    -${U0}*atan((x-${x0})/${dlt})
variable F atom ${U0}/((x-${x0})^2/${dlt}^2+1)/${dlt}&
    -${U0}/((x+${x0})^2/${dlt}^2+1)/${dlt}
fix myadf all addforce v_F 0.0 0.0 energy v_U
\end{lstlisting}
Next, we use the Newtonian equations of motion with a
Langevin thermostat by combining the \lmpcmd{fix nve} with a
\lmpcmd{fix langevin} command:
\begin{lstlisting}
fix mynve all nve
fix mylgv all langevin 119.8 119.8 500 30917
\end{lstlisting}
When combining these two commands, the MD simulation operates
in the NVT ensemble, maintaining a constant number of
atoms $N$, constant volume $V$, and a temperature $T$ that
fluctuates around a target value.

\begin{note}
  LAMMPS documentation suggests using damping constants for thermostats 
  that are approximately 100 times the timestep value.  In this case, a value of 500 
  is used, resulting in a relatively weak coupling to the thermostat.
\end{note}

\begin{figure}
\centering
\includegraphics[width=\linewidth]{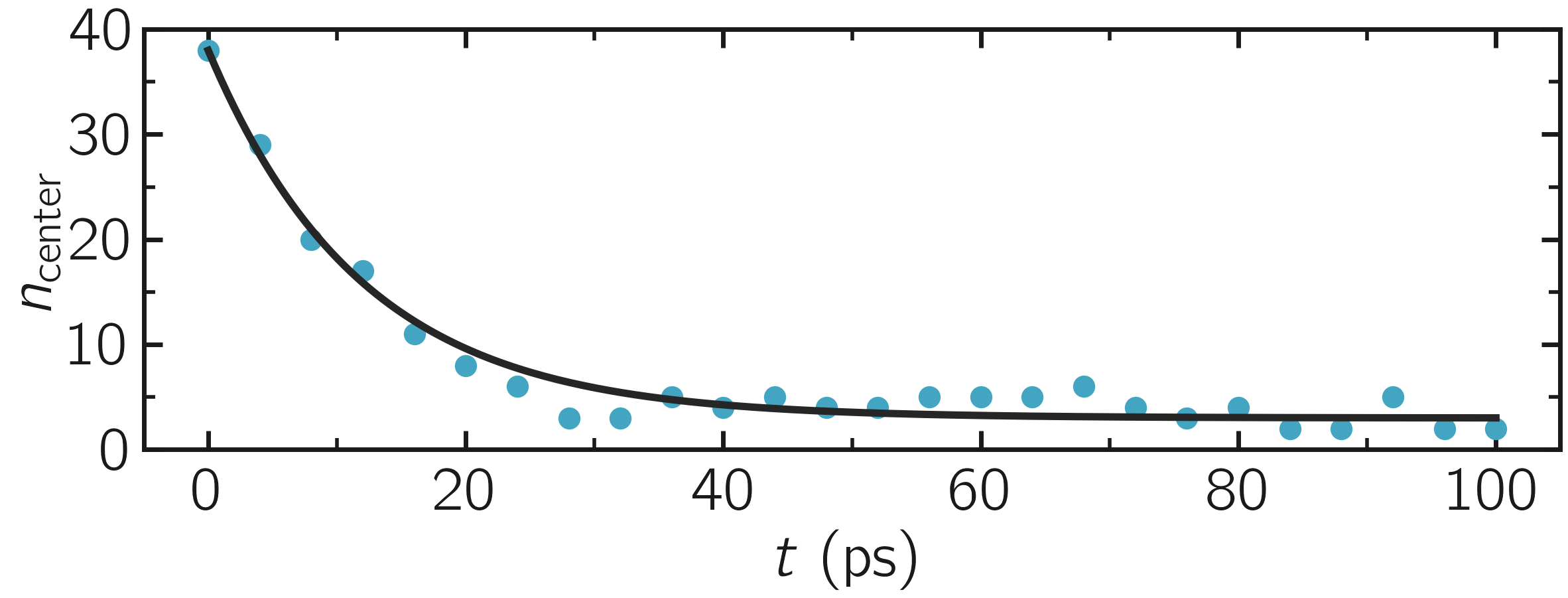}
\caption{Evolution of the number of atoms $n_\text{center}$ in the central
region \lmpcmd{mymes} as a function of time $t$ during equilibration.  The dark line
is $n_\text{center} = 22 \exp(-t/160)+5$ and serves as a guide for the eyes.
Here, $U_0 = 0.36~\text{kcal/mol}$, $\delta = 1.0~\text{\AA{}}$, and $x_0 = 10~\text{\AA{}}$.}
\label{fig:US-density-evolution}
\end{figure}

To ensure that the equilibration time is sufficient, we will track the evolution of
the number of atoms in the central - energetically unfavorable - region,
defined below under the name \lmpcmd{mymes}, using
the \lmpcmd{n\_center} variable:
\begin{lstlisting}
region mymes block -${x0} ${x0} INF INF INF INF
variable n_center equal count(all,mymes)
thermo_style custom step temp etotal v_n_center
thermo 10000

dump viz all image 5000 myimage-*.ppm type type &
    shiny 0.1 box yes 0.01 view 180 90 zoom 6 &
    size 1600 500 fsaa yes
dump_modify viz backcolor white acolor 1 cyan &
    adiam 1 3 boxcolor black
\end{lstlisting}
A \lmpcmd{dump image} command was also added for system visualization.
The other commands should also be familiar from previous tutorials.

\begin{figure}
\centering
\includegraphics[width=\linewidth]{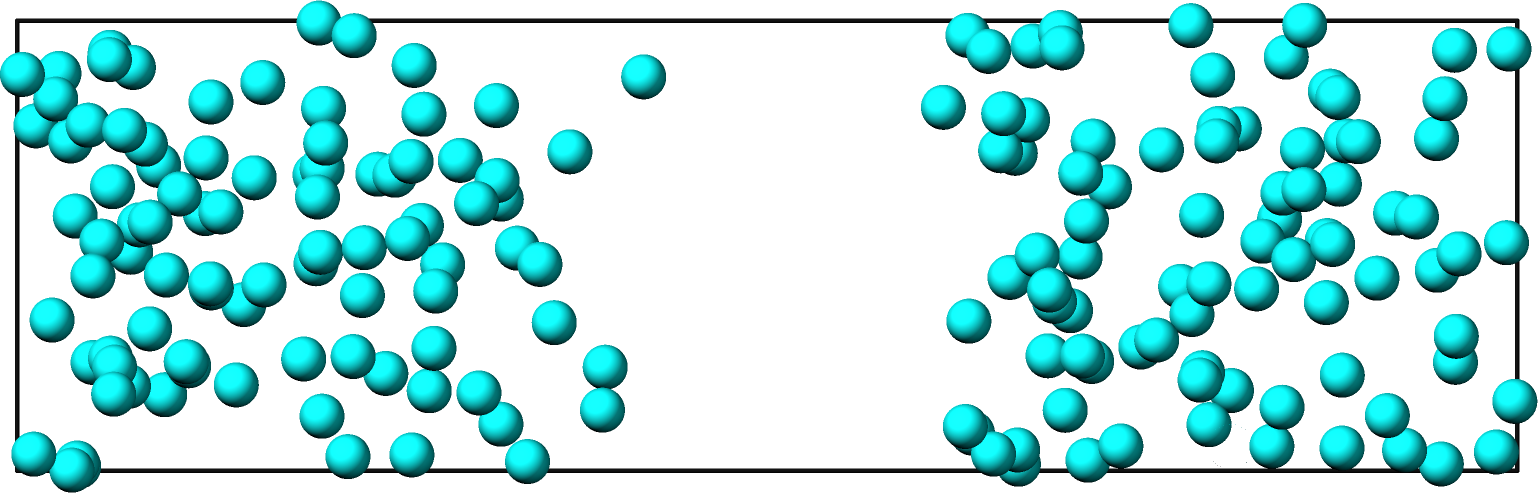}
\caption{Snapshot of the system simulated during the free sampling
step of \hyperref[umbrella-sampling-label]{Tutorial 7}.
The atoms density is the lowest in the central
part of the box, \lmpcmd{mymes}.  Here,
$U_0 = 0.36~\text{kcal/mol}$, $\delta = 1.0~\text{\AA{}}$, and $x_0 = 10~\text{\AA{}}$.}
\label{fig:US-system-unbiased}
\end{figure}

\noindent Finally, let us perform an equilibration of 50000 steps,
using a timestep of $2\,\text{fs}$, corresponding to a total duration of $100\,\text{ps}$:
\begin{lstlisting}
timestep 2.0
run 50000
\end{lstlisting}
Run the simulation with LAMMPS.  The number of atoms in the
central region, $n_\mathrm{center}$, reaches its equilibrium value after approximately $40\,\text{ps}$
(Fig.~\ref{fig:US-density-evolution}).  A snapshot of the equilibrated system is shown in Fig.~\ref{fig:US-system-unbiased}.

\paragraph{Run and data acquisition}

Once the system is equilibrated, we will record the density profile of
the atoms along the $x$-axis using the \lmpcmd{ave/chunk} command.
Add the following line to \flecmd{free-sampling.lmp}:
\begin{lstlisting}
reset_timestep 0

thermo 200000

compute cc1 all chunk/atom bin/1d x 0.0 2.0
fix myac all ave/chunk 100 20000 2000000 &
    cc1 density/number file free-sampling.dat

run 2000000
\end{lstlisting}
Here, the \lmpcmd{chunk/atom} command discretizes the simulation
domain into spatial bins of size 2~\AA{} along the $x$ direction,
and the \lmpcmd{fix ave/chunk} command outputs the number density of
atoms within each bin to the file \flecmd{free-sampling.dat}.
The step count is reset to 0 using \lmpcmd{reset\_timestep} to synchronize it
with the output times of \lmpcmd{fix ave/chunk}.  Run the simulation using
LAMMPS.

\paragraph{Data analysis}

\begin{figure}
\centering
\includegraphics[width=\linewidth]{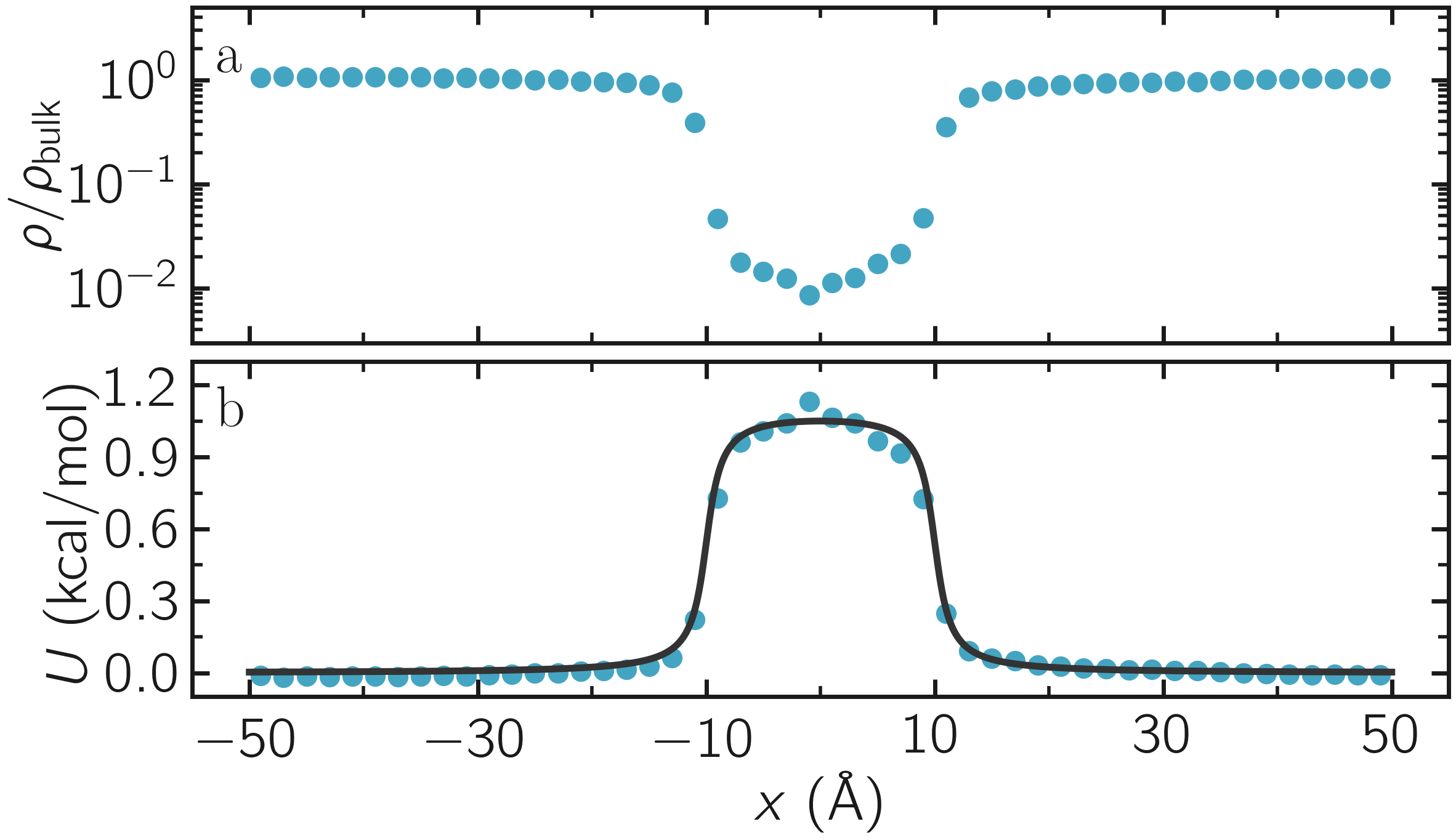}\\[-2ex]
\caption{a) Fluid density, $\rho$, along the $x$ direction.
b) Potential, $U$, as a function of $x$ measured using free sampling (blue disks)
compared to the imposed potential given in Eq.~\eqref{eq:U} (dark line).
Here, $U_0 = 0.36~\text{kcal/mol}$, $\delta = 1.0~\text{\AA{}}$, $x_0 = 10~\text{\AA{}}$,
and the measured reference density in the reservoir is $\rho_\text{bulk} = 0.0009~\text{\AA{}}^{-3}$.} 
\label{fig:US-density}
\end{figure}

Once the simulation is complete, the density profile from \flecmd{free-sampling.dat}
shows that the density in the center of the box is
about two orders of magnitude lower than inside the reservoir (Fig.~\ref{fig:US-density}\,a).
Next, we plot $-R T \ln(\rho/\rho_\mathrm{bulk})$, 
where $\rho/\rho_\mathrm{bulk}$ is the the density ratio, and compare it
with the imposed potential $U$ from Eq.~\eqref{eq:U} (Fig.~\ref{fig:US-density}\,b).
The reference density, $\rho_\text{bulk} = 0.0009~\text{\AA{}}^{-3}$,
was estimated by measuring the density of the reservoir from the density
profiles.  The agreement between the MD results and the imposed energy profile
is excellent, despite some noise in the central part, where fewer data points
are available due to the repulsive potential.

\paragraph{The limits of free sampling}

Increasing the value of $U_0$ reduces the average number of atoms in the central
region, making it difficult to achieve a high-resolution free energy profile
within reasonable simulation times.  For example, running the same
simulation with $U_0 = 10 \epsilon$, corresponding to $U_0 \approx 10 k_\text{B} T$, results in no atoms exploring
the central part of the simulation box during the simulation.
In such a case, employing an enhanced sampling method is recommended, as done in the next section.

\subsubsection{Method 2: Umbrella sampling}

Umbrella sampling is a biased molecular dynamics method in which
additional forces are added to a chosen atom to force it to explore the
more unfavorable areas of the system~\cite{kastner2011umbrella,
  allen2017computer, frenkel2023understanding}.  Here, to encourage one
of the atoms to explore the central region of the box, we apply a
potential $V$ and force it to move along the $x$-axis. The chosen path
is called the axis of reaction. Several simulations (called windows)
will be conducted with varying positions for the center of the applied
biasing. The results will be analyzed using the weighted histogram
analysis method (WHAM)~\cite{kumar1992weighted,kumar1995multidim}, which
allows for the removal of the biasing effect and ultimately deduces the
unbiased free energy profile.

\paragraph{LAMMPS input script}

Open the file named \flecmd{umbrella-sampling.lmp}, which should
contain the following lines:
\begin{lstlisting}
variable sigma equal 3.405
variable epsilon equal 0.238
variable U0 equal 10*${epsilon}
variable dlt equal 1.0
variable x0 equal 10
variable k equal 0.5

units real
atom_style atomic
pair_style lj/cut $(2^(1/6)*v_sigma)
pair_modify shift yes
boundary p p p
\end{lstlisting}
The first difference from the previous case is the larger value
for the repulsive potential parameter $U_0$, which makes the central area
of the system very unlikely to be visited by free particles.  The second
difference is the introduction of the variable $k$, which will be used for
the biasing potential.

\begin{figure}
\centering
\includegraphics[width=\linewidth]{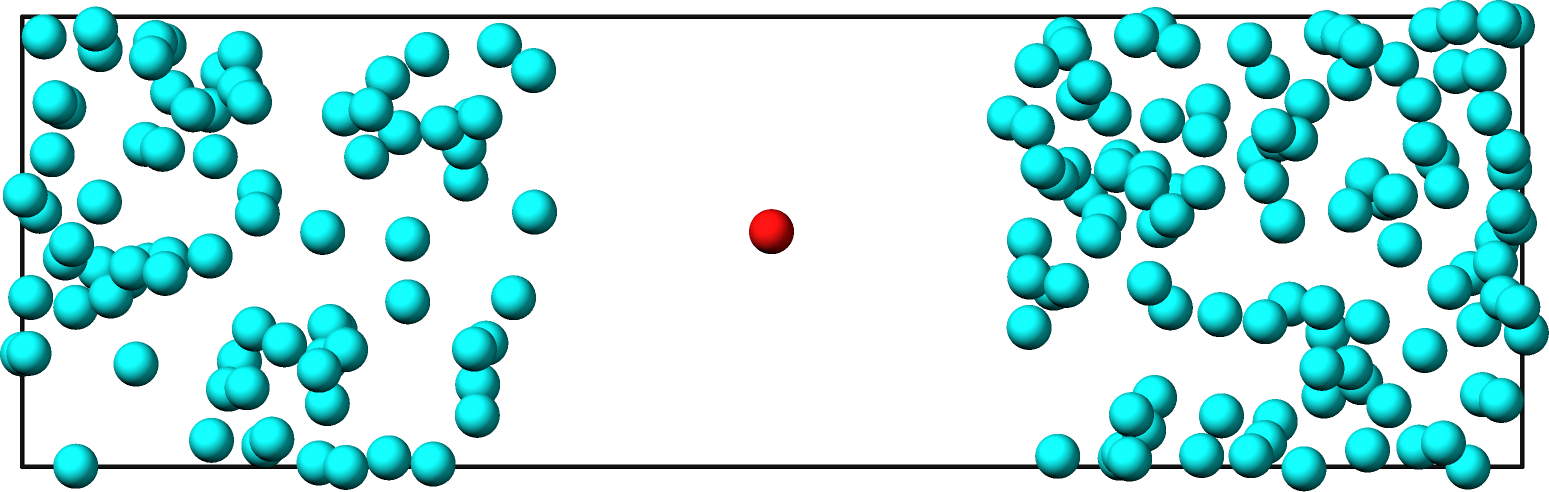}
\caption{Snapshot of the system simulated during the umbrella sampling
step of \hyperref[umbrella-sampling-label]{Tutorial 7}, showing type-1 atoms
in cyan and the type-2 atom in red.  Only the type-2 atom explores the central part of the box,
\lmpcmd{mymes}, due to the additional biasing potential $V$. Parameters are
$U_0 = 2.38~\text{kcal/mol}$, $\delta = 1.0~\text{\AA{}}$, and $x_0 = 10~\text{\AA{}}$.}
\label{fig:US-system-biased}
\end{figure}

Let us create a simulation box with two atom types, including a single particle of type 2,
by adding the following lines to \flecmd{umbrella-sampling.lmp}:
\begin{lstlisting}
region myreg block -50 50 -15 15 -50 50
create_box 2 myreg
create_atoms 2 single 0 0 0
create_atoms 1 random 199 34134 myreg overlap 3 maxtry 50
\end{lstlisting}
Next, we assign the same mass and LJ parameters to both atom types
1 and 2, and place the atoms of type 2 into a group named \lmpcmd{topull}:
\begin{lstlisting}
mass * 39.948
pair_coeff * * ${epsilon} ${sigma}
group topull type 2
\end{lstlisting}
Then, the same potential $U$ and force $F$ are applied to all the atoms,
together with the same \lmpcmd{fix nve} and \lmpcmd{fix langevin} commands:
\begin{lstlisting}
variable U atom ${U0}*atan((x+${x0})/${dlt})&
    -${U0}*atan((x-${x0})/${dlt})
variable F atom ${U0}/((x-${x0})^2/${dlt}^2+1)/${dlt}&
    -${U0}/((x+${x0})^2/${dlt}^2+1)/${dlt}
fix myadf all addforce v_F 0.0 0.0 energy v_U

fix mynve all nve
fix mylgv all langevin 119.8 119.8 500 30917
\end{lstlisting}
Next, we perform a brief equilibration to prepare for the
umbrella sampling run:
\begin{lstlisting}
thermo 5000

dump viz all image 5000 myimage-*.ppm type type &
    shiny 0.1 box yes 0.01 view 180 90 zoom 6 &
    size 1600 500 fsaa yes
dump_modify viz backcolor white acolor 1 cyan &
acolor 2 red adiam 1 3 adiam 2 3 boxcolor black

timestep 2.0
run 50000
\end{lstlisting}

So far, our code resembles that of Method 1, except for the additional particle
of type 2.  Particles of types 1 and 2 are identical.
However, the particle of type 2 will also
be exposed to the biasing potential $V$, which forces it to explore the
central part of the box (Fig.~\ref{fig:US-system-biased}), thus
justifying the definition of two atom types.

Now, we create a loop with 15 steps and progressively move the center of the
bias potential by increments of 0.4\,nm.  Add the following lines to \flecmd{umbrella-sampling.lmp}:
\begin{lstlisting}
variable a loop 15
label loop

variable xdes equal 4*${a}-32
variable xave equal xcm(topull,x)
fix mytth topull spring tether ${k} ${xdes} 0 0 0

run 20000

fix myat1 all ave/time 10 10 100 &
    v_xave v_xdes file umbrella-sampling.${a}.dat

run 200000
unfix myat1
next a
jump SELF loop
\end{lstlisting}
The definition of a variable of loop style serves the same purpose
as in (\hyperref[reactive-silicon-dioxide-label]{Tutorial 5}), and we highlight
here the particular utility of using its value
to distinguish the files written by the \lmpcmd{fix ave\_time} command for the
different bias potentials.
The \lmpcmd{spring} command imposes the additional harmonic potential $V$ with
the previously defined spring constant $k$ to the atoms in the group \lmpcmd{topull}.  The center of the harmonic
potential, $x_\text{des}$, successively takes values
from $-28\,\text{\AA{}}$ to $28\,\text{\AA{}}$.  For each value of $x_\text{des}$,
an equilibration step of 40\,ps is performed, followed by a step
of 400\,ps during which the position of the particle of
type 2 along the $x$-axis, $x_\text{ave}$, is saved in data files named \flecmd{umbrella-sampling.i.dat},
where $i$ ranges from 1 to 15.  Run the \flecmd{umbrella-sampling.lmp} file using LAMMPS.

\begin{note}
  The value of $k$ should be chosen with care:
  if $k$ is too small the particle won't follow the biasing potential,
  and if $k$ is too large there will be no overlapping between
  the different windows, leading to poor reconstruction of the free energy profile.
\end{note}

\paragraph{WHAM algorithm}

\begin{figure}
\centering
\includegraphics[width=\linewidth]{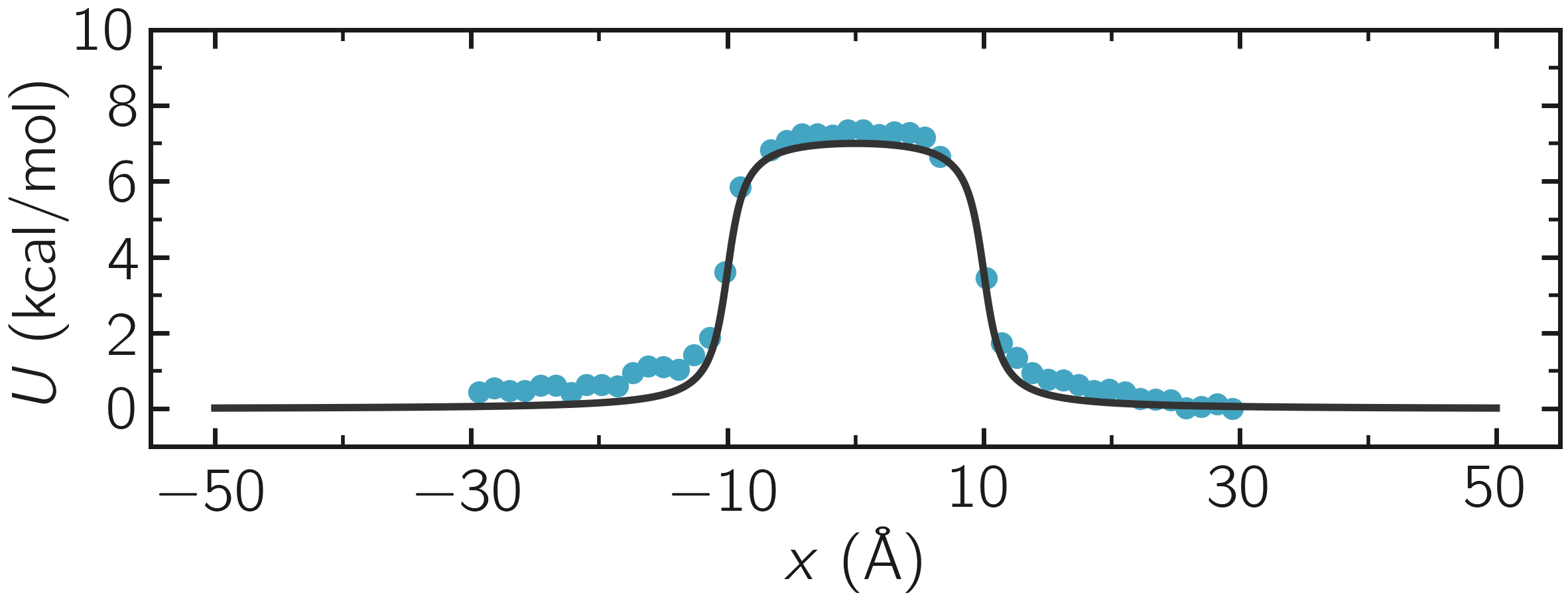}
\caption{The potential, $U$, as a function of $x$, measured using umbrella
sampling during \hyperref[umbrella-sampling-label]{Tutorial 7} (blue disks),
is compared to the imposed potential given in Eq.~\eqref{eq:U}
(dark line).  Parameters are $U_0 = 2.38~\text{kcal/mol}$, $\delta = 1.0~\text{\AA{}}$,
and $x_0 = 10~\text{\AA{}}$.}
\label{fig:US-freenergy}
\end{figure}

To generate the free energy profile from the particle positions saved in
the \flecmd{umbrella-sampling.i.dat} files, we use the
WHAM~\cite{kumar1992weighted,kumar1995multidim} algorithm as implemented
by Alan Grossfield~\cite{grossfieldimplementation}.  You can download it
from \href{http://membrane.urmc.rochester.edu/?page_id=126}{Alan
  Grossfield}'s website.  Make sure you download the WHAM code version
2.1.0 or later which introduces the \lmpcmd{units} command-line option
used below. The executable called \flecmd{wham} generated by following
the instructions from the website must be placed next to
\flecmd{umbrella-sampling.lmp}.  To apply the WHAM algorithm to our
simulation, we need a metadata file containing:
\begin{itemize}
\item the paths to all the data files,
\item the values of $x_\text{des}$,
\item the values of $k$.
\end{itemize}
Download the
\href{\filepath tutorial7/umbrella-sampling.meta}{\dwlcmd{umbrella-sampling.meta}}
file and save it next to \flecmd{umbrella-sampling.lmp}.  Then, run the
WHAM algorithm by typing the following command in the terminal:
\begin{lstlisting}
./wham units real -30 30 50 1e-8 119.8 0 \
    umbrella-sampling.meta umbrella-sampling.dat
\end{lstlisting}
where -30 and 30 are the boundaries, 50 is the number of bins, 1e-8 is the tolerance,
and 119.8 is the temperature in Kelvin.  A file called \flecmd{umbrella-sampling.dat} is created,
containing the free energy profile in kcal/mol.  The resulting PMF can be compared
with the imposed potential $U$, showing excellent agreement
(Fig.~\ref{fig:US-freenergy}).  Remarkably, this excellent agreement is achieved despite
the very short calculation time and the high value for the energy barrier.
Achieving similar results through free sampling would require performing extremely
long and computationally expensive simulations.

\subsection{Tutorial 8: Reactive Molecular Dynamics}
\label{bond-react-label}

\begin{figure}
\centering
\includegraphics[width=\linewidth]{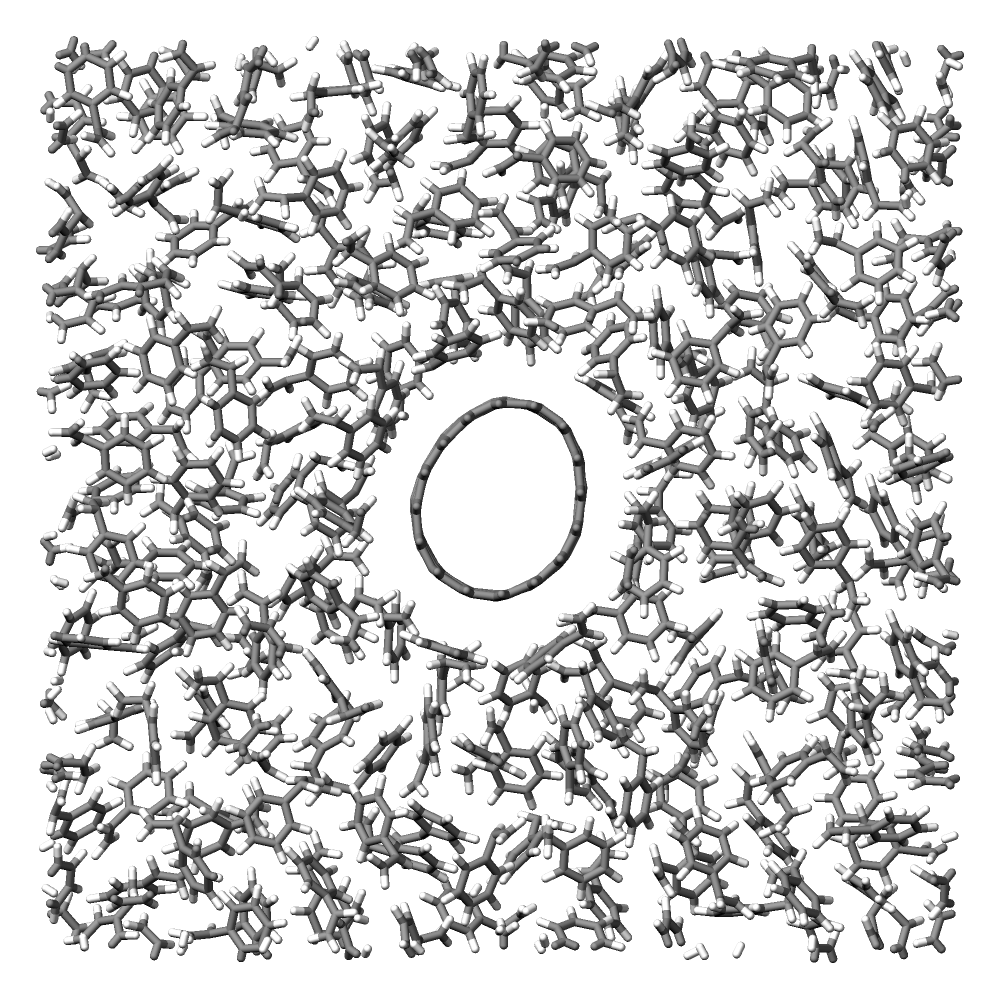}
\caption{Initial configuration for \hyperref[bond-react-label]{Tutorial 8}.
The system consists of 200 styrene molecules packed around a single-walled
CNT, with a mass density for the whole system of $0.9~\text{g/cm}^3$.}
\label{fig:REACT}
\end{figure}

The goal of this tutorial is to create a model of a carbon nanotube (CNT) embedded
in a polymer melt made of polystyrene (PS) (Fig.~\ref{fig:REACT}).  The
REACTER protocol is used to simulate the polymerization of styrene monomers, and the
polymerization reaction is followed in time~\cite{gissinger2017polymer, gissinger2020reacter, gissinger2024molecular}.
In contrast with AIREBO (\hyperref[carbon-nanotube-label]{Tutorial 2})
and ReaxFF (\hyperref[reactive-silicon-dioxide-label]{Tutorial 5}), the REACTER
protocol relies on the use of a \textit{classical} force field
that does not inherently model bond formation or breaking, but
instead couples with an external algorithm to simulate polymerization reactions.

\subsubsection{Creating the system}

To begin this tutorial, select \guicmd{Start Tutorial 8} from the
\guicmd{Tutorials} menu of \lammpsgui{} and follow the instructions.
The editor should display the following content corresponding to \flecmd{mixing.lmp}:
\begin{lstlisting}
units real
boundary p p p
atom_style full

kspace_style pppm 1.0e-5
pair_style lj/class2/coul/long 8.5
angle_style class2
bond_style class2
dihedral_style class2
improper_style class2

pair_modify tail yes mix sixthpower
special_bonds lj/coul 0 0 1
\end{lstlisting}
The \lmpcmd{class2} \lmpcmd{pair\_styles} compute a 6/9 Lennard-Jones potential~\cite{sun1998compass}.
The \textit{class2} bond, angle, dihedral, and improper styles are used as
well, see the documentation for a description of the respective potential
form they, each, prescribe.
The \lmpcmd{tail yes} option adds long-range van der Waals tail
corrections to the energy and pressure.
The \lmpcmd{mix sixthpower} imposes the following mixing rule for the calculation
of the cross coefficients:
\begin{eqnarray}
\nonumber
\sigma_{ij} & = & 2^{-1/6} (\sigma^6_i+\sigma_j^6)^{1/6}, ~ \text{and} \\
\nonumber
\epsilon_{ij} & = & \dfrac{2 \sqrt{\epsilon_i \epsilon_j} \sigma^3_i \sigma^3_j}{\sigma^6_i+\sigma_j^6}.
\end{eqnarray}

Let us read the \href{\filepath tutorial8/CNT.data}{\dwlcmd{CNT.data}} file, which
contains a periodic single-walled CNT.  Add the following line to \flecmd{mixing.lmp}:
\begin{lstlisting}
read_data CNT.data extra/special/per/atom 20
\end{lstlisting}
The CNT is approximately $1.1~\text{nm}$ in diameter and $1.6\,\text{nm}$ in length, oriented
along the $x$-axis. The simulation box is initially 12.0~nm in the two other dimensions before densification,
making it straightforward to fill the box with styrene.
To add 200 styrene molecules to the simulation box, we will use the
\href{\filepath tutorial8/styrene.mol}{\dwlcmd{styrene.mol}} molecule template file.
Include the following commands to \flecmd{mixing.lmp}:
\begin{lstlisting}
molecule styrene styrene.mol
create_atoms 0 random 200 8305 NULL overlap 2.75 &
    maxtry 500 mol styrene 7687
\end{lstlisting}
Finally, let us use the \lmpcmd{minimize} command to reduce the potential energy of the system:
\begin{lstlisting}
minimize 1.0e-4 1.0e-6 100 1000
reset_timestep 0
\end{lstlisting}
These commands were covered in earlier tutorials and should already be familiar.

Then, let us densify the system to a target value of $0.9~\text{g/cm}^3$
by imposing the shrinking of the simulation box at a constant rate.
The dimension parallel to the CNT axis is maintained fixed because the CNT is periodic in that direction.
Add the following commands to \flecmd{mixing.lmp}:
\begin{lstlisting}
velocity all create 530 9845 dist gaussian rot yes
fix mynvt all nvt temp 530 530 100

fix mydef all deform 1 y erate -0.0001 z erate -0.0001
variable rho equal density
fix myhal all halt 10 v_rho > 0.9 error continue

thermo 200
thermo_style custom step temp pe etotal press density

run 9000
\end{lstlisting}
The \lmpcmd{fix halt} command is used to stop the box shrinkage once the
target density is reached, and the other commands should be
familiar from previous tutorials.

\begin{figure}
\centering
\includegraphics[width=\linewidth]{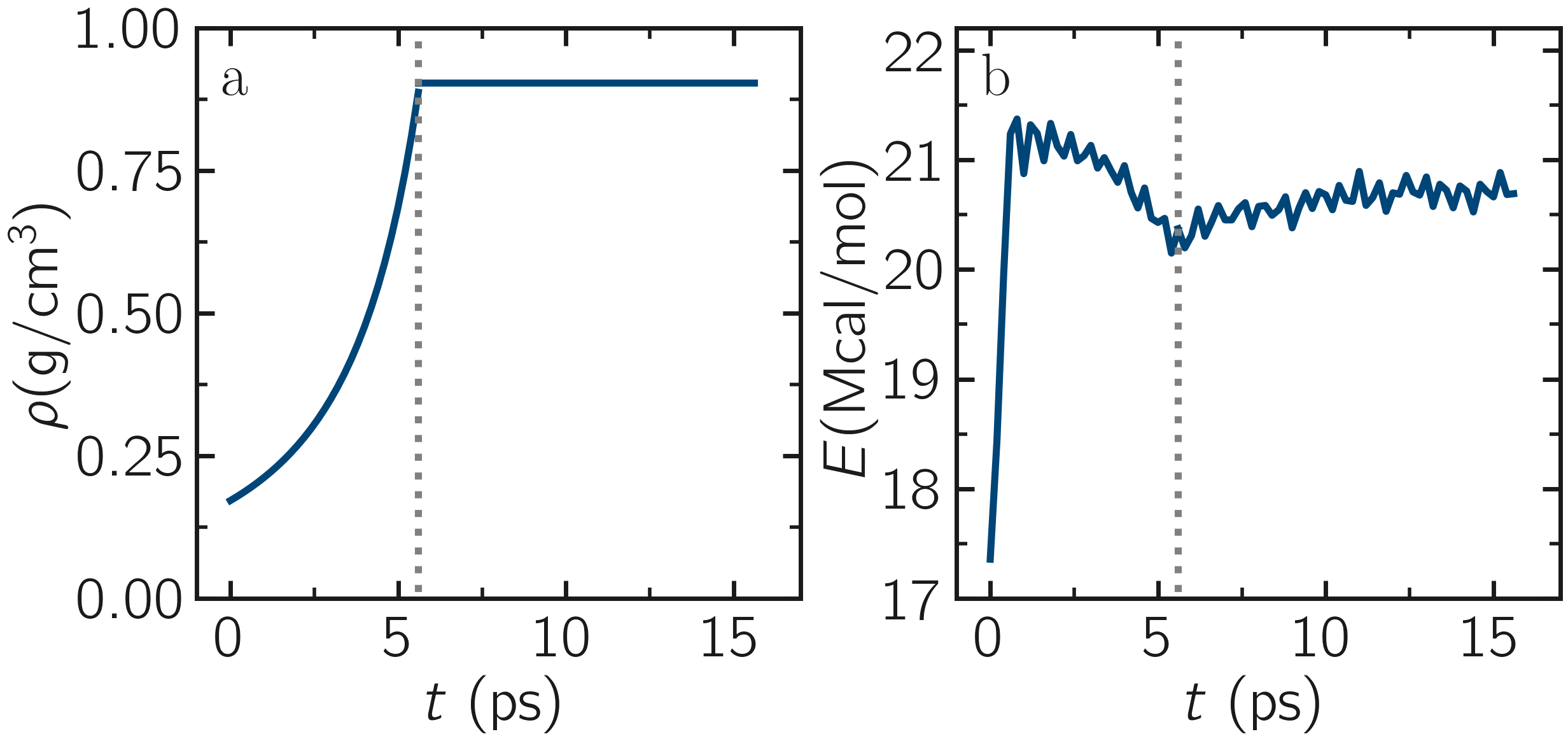}
\caption{a) Evolution of the density, $\rho$, as a function of the
time, $t$, during equilibration of the system from \hyperref[bond-react-label]{Tutorial 8}.
b) Evolution of the total energy, $E$, of the system.
The vertical dashed lines mark the transition between the different
phases of the simulation.}
\label{fig:evolution-density}
\end{figure}

For the next stage of the simulation, we will use \lmpcmd{dump image} to
output images every 200 steps:
\begin{lstlisting}
dump viz all image 200 myimage-*.ppm &
  type type shiny 0.1 box no 0.01 size 1000 1000 &
  view 90 0 zoom 1.8 fsaa yes bond atom 0.5
dump_modify viz backcolor white &
  acolor cp gray acolor c=1 gray &
  acolor c= gray acolor c1 deeppink &
  acolor c2 deeppink acolor c3 deeppink &
  adiam cp 0.3 adiam c=1 0.3 &
  adiam c= 0.3 adiam c1 0.3 &
  adiam c2 0.3 adiam c3 0.3 &
  acolor hc white adiam hc 0.15
\end{lstlisting}
For the following $10~\text{ps}$, let us equilibrate the densified system
in the constant-volume ensemble, and write the final state of the
system in a file named \flecmd{mixing.data}:
\begin{lstlisting}
unfix mydef
unfix myhal
reset_timestep 0

group CNT molecule 1
fix myrec CNT recenter NULL 0 0 units box shift all

run 10000

write_data mixing.data
\end{lstlisting}
For visualization purposes, the atoms of the CNT \lmpcmd{group} are moved
to the center of the box using \lmpcmd{fix recenter}. As the time progresses, the system density,
$\rho$, gradually converges toward the target value of $0.9$\,g/cm$^3$ (Fig.~\ref{fig:evolution-density}\,a).
Meanwhile, the total energy of the system initially evolves rapidly, reflecting the
densification process, and then eventually stabilizes (Fig.~\ref{fig:evolution-density}\,b).
The final state is shown in Fig.~\ref{fig:REACT}.

\subsubsection{Reaction templates}

The REACTER protocol enables the modeling of chemical reactions using
classical force fields.  The user must provide a molecule template for the reactants,
a molecule template for the products, and a \flecmd{reaction map} file that
provides an atom mapping between the two templates.  The reaction map file also includes
additional information, such as which atoms act as initiators for the reaction and which
serve as edge atoms to connect the rest of a long polymer chain in the simulation.

There are three reactions to define: (1) the polymerization of two styrene monomers,
(2) the addition of a styrene monomer to the end of a growing polymer chain, and (3) the
linking of two polymer chains.  Download the three files associated with each reaction.
The first reaction uses the prefix `M-M' for the pre-reaction template,
post-reaction template, and reaction map file:
\begin{itemize}
\item \href{\filepath tutorial8/M-M_pre.mol}{\dwlcmd{M-M$\_$pre.mol}},
\item \href{\filepath tutorial8/M-M_post.mol}{\dwlcmd{M-M$\_$post.mol}},
\item \href{\filepath tutorial8/M-M.rxnmap}{\dwlcmd{M-M.rxnmap}}.
\end{itemize}
The second reaction uses the prefix `M-P',
\begin{itemize}
\item \href{\filepath tutorial8/M-P_pre.mol}{\dwlcmd{M-P$\_$pre.mol}},
\item \href{\filepath tutorial8/M-P_post.mol}{\dwlcmd{M-P$\_$post.mol}},
\item \href{\filepath tutorial8/M-P.rxnmap}{\dwlcmd{M-P.rxnmap}}.
\end{itemize}
The third reaction uses the prefix `P-P',
\begin{itemize}
\item \href{\filepath tutorial8/P-P_pre.mol}{\dwlcmd{P-P$\_$pre.mol}},
\item \href{\filepath tutorial8/P-P_post.mol}{\dwlcmd{P-P$\_$post.mol}},
\item \href{\filepath tutorial8/P-P.rxnmap}{\dwlcmd{P-P.rxnmap}}.
\end{itemize}
Here, the file names for each reaction use the abbreviation `M' for monomer and `P'
for polymer.

\begin{note}
  The data stored in molecule templates include atom coordinates,
  partial charges, molecule IDs, atom types, and interaction types for bonds,
  angles, dihedrals and impropers.  The map files contain information about
  the reaction.  The first mandatory section of the map files begins with the
  keyword “InitiatorIDs” and lists the two atom IDs of the initiator atom pair
  in the pre-reacted molecule template.  The second mandatory section begins
  with the keyword “Equivalences” and lists a one-to-one correspondence between
  atom IDs of the pre- and post-reacted templates.  Some atoms in the pre-reacted
  template that are not reacting may have missing topology with respect to the
  simulation.  For example, the pre-reacted template may contain an atom that,
  in the simulation, is currently connected to the rest of a long polymer
  chain.  These are referred to as edge atoms, and are also specified in the
  map file in the “EdgeIDs” section.
\end{note}

\subsubsection{Simulating the reaction}

The first step, before simulating the reaction, is to import the previously
generated configuration.  Open the file named \flecmd{polymerize.lmp},
which should contain the following lines:
\begin{lstlisting}
units real
boundary p p p
atom_style full

kspace_style pppm 1.0e-5
pair_style lj/class2/coul/long 8.5
angle_style class2
bond_style class2
dihedral_style class2
improper_style class2

pair_modify tail yes mix sixthpower
special_bonds lj/coul 0 0 1

read_data mixing.data &
  extra/bond/per/atom 5  &
  extra/angle/per/atom 15 &
  extra/dihedral/per/atom 15 &
  extra/improper/per/atom 25 &
  extra/special/per/atom 25
\end{lstlisting}
Here, the \lmpcmd{read\_data} command is used to import the
previously generated \flecmd{mixing.data} file.  All other commands
have been introduced in earlier parts of the tutorial.

Then, let us import all six molecules templates using the \lmpcmd{molecule} command:
\begin{lstlisting}
molecule mol1 M-M_pre.mol
molecule mol2 M-M_post.mol
molecule mol3 M-P_pre.mol
molecule mol4 M-P_post.mol
molecule mol5 P-P_pre.mol
molecule mol6 P-P_post.mol
\end{lstlisting}
In order to follow the evolution of the reaction with time, let us generate images
of the system using \lmpcmd{dump image}:
\begin{lstlisting}
dump viz all image 200 myimage-*.ppm &
  type type shiny 0.1 box no 0.01 size 1000 1000 &
  view 90 0 zoom 1.8 fsaa yes bond atom 0.5
dump_modify viz backcolor white &
  acolor cp gray acolor c=1 gray &
  acolor c= gray acolor c1 deeppink &
  acolor c2 gray acolor c3 deeppink &
  adiam cp 0.3 adiam c=1 0.3 &
  adiam c= 0.3 adiam c1 0.3 &
  adiam c2 0.3 adiam c3 0.3 &
  acolor hc white adiam hc 0.15
\end{lstlisting}

Let us use \lmpcmd{fix bond/react} by adding the following
line to \flecmd{polymerize.lmp}:
\begin{lstlisting}
fix rxn all bond/react &
  stabilization yes statted_grp 0.03 &
  react R1 all 1 0 3.0 mol1 mol2 M-M.rxnmap &
  react R2 all 1 0 3.0 mol3 mol4 M-P.rxnmap &
  react R3 all 1 0 5.0 mol5 mol6 P-P.rxnmap
\end{lstlisting}
With the \lmpcmd{stabilization} keyword, the \lmpcmd{fix bond/react} command will
stabilize the atoms involved in the reaction using the \lmpcmd{fix nve/limit}
command with a maximum displacement of $0.03\,\mathrm{\AA{}}$.
The \lmpcmd{fix nve/limit} command functions similar to
\lmpcmd{fix nve}, but restricts how far atoms can move in a single time step, even with
very large forces.
By default, each reaction is stabilized for 60 time steps.  Each \lmpcmd{react} keyword
corresponds to a reaction, e.g.,~a transformation of \lmpcmd{mol1} into \lmpcmd{mol2}. 
Implementation details about each reaction, such as the reaction distance cutoffs
and the frequency with which to search for reaction sites, are also specified in this command.

\begin{figure}
\centering
\includegraphics[width=\linewidth]{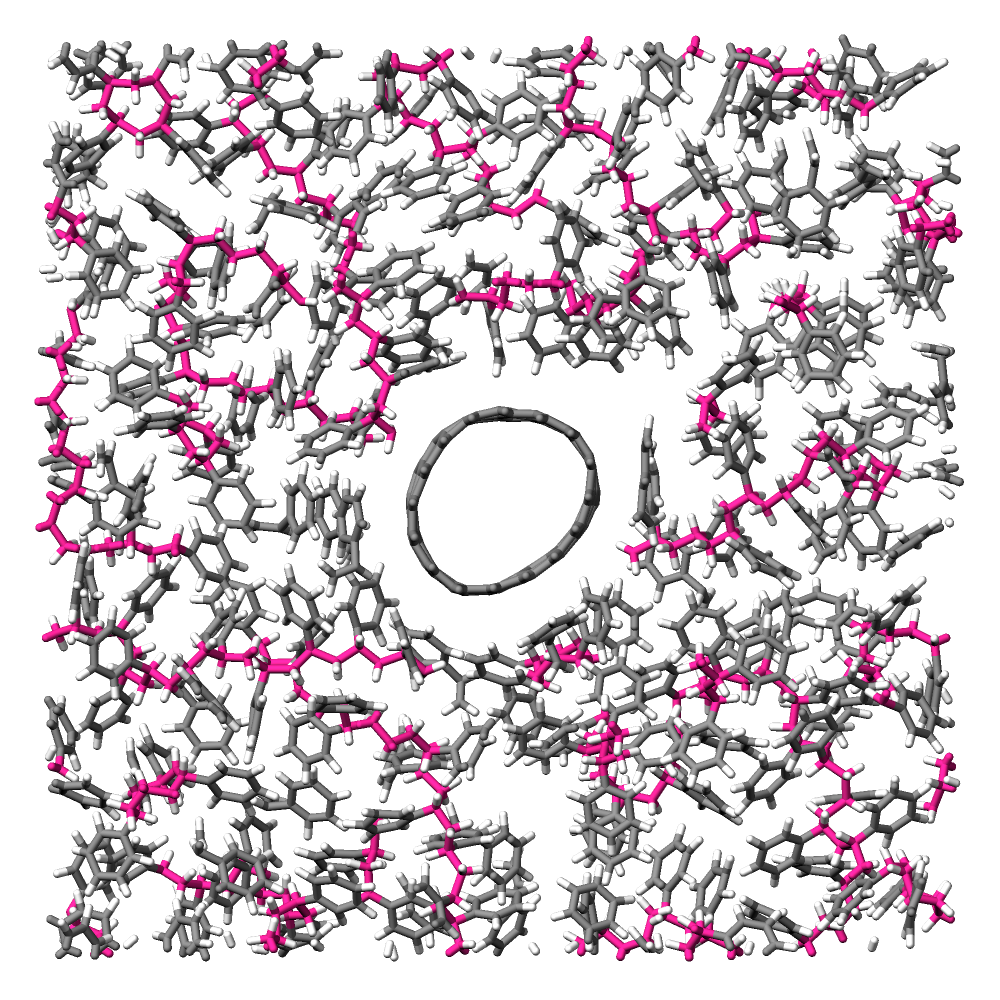}
\caption{Final configuration for \hyperref[bond-react-label]{Tutorial 8}.
The atoms from the formed polymer named \lmpcmd{c1}, \lmpcmd{c2}, and
\lmpcmd{c3} are colored in pink.}
\label{fig:REACT-final}
\end{figure}

\begin{note}
  The command \lmpcmd{fix bond/react} creates several groups of atoms that are dynamically updated
  to track which atoms are being stabilized and which atoms are undergoing
  dynamics with the system-wide time integrator (here, \lmpcmd{fix nvt}).
  When reaction stabilization is employed, there should not be a time integrator acting on
  the group \mbox{\lmpcmd{all}.}  Instead, the group of atoms not currently
  undergoing stabilization is named by appending `\_REACT' to the user-provided prefix.
\end{note}

\begin{figure}
\centering
\includegraphics[width=\linewidth]{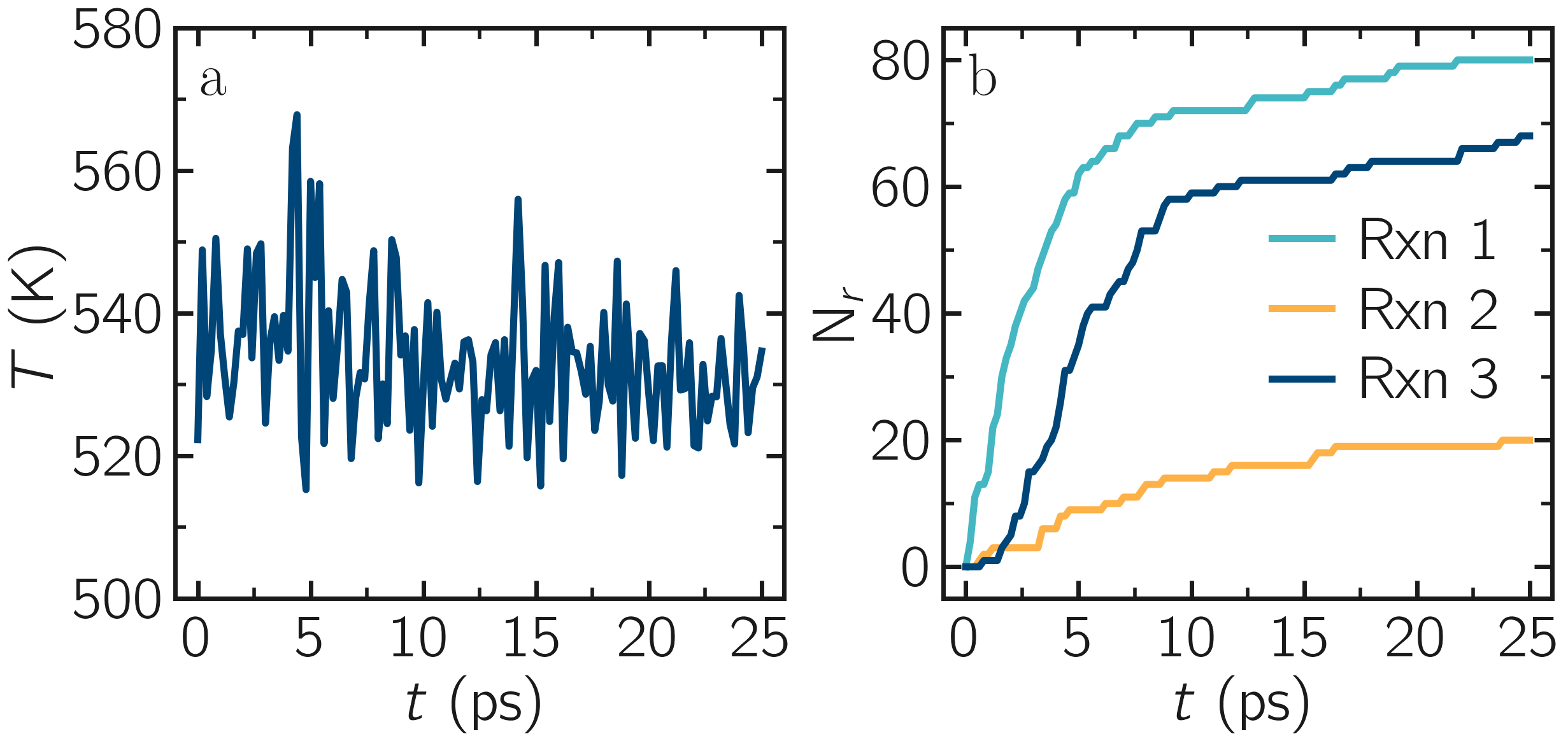}
\caption{a) Evolution of the system temperature, $T$,
as a function of the time, $t$, during the polymerization step of
\hyperref[bond-react-label]{Tutorial 8}.
b) Evolution of the three reaction counts, corresponding respectively to
the polymerization of two styrene monomers (Rxn~1), the  addition of a styrene
monomer to the end of a growing polymer chain (Rxn~2), and to the linking
of two polymer chains (Rxn~3).}
\label{fig:evolution-reacting}
\end{figure}

Add the following commands to \flecmd{polymerize.lmp} to
carry out the dynamics using a Nosé-Hoover thermostat
while ensuring that the CNT remains centered in the simulation box:
\begin{lstlisting}
fix mynvt statted_grp_REACT nvt temp 530 530 100
group CNT molecule 1 2 3
fix myrec CNT recenter NULL 0 0 shift all

thermo 1000
thermo_style custom step temp press density f_rxn[*]

run 25000
\end{lstlisting}
Here, the \lmpcmd{thermo\_style custom} command is used
to print the cumulative reaction counts which are calculated by \lmpcmd{fix rxn}
and thus can be extracted from it.
Run the simulation using LAMMPS.  As the simulation progresses, polymer chains are
observed forming (Fig.~\ref{fig:REACT-final}).  During this reaction process, the
temperature of the system remains well-controlled (Fig.~\ref{fig:evolution-reacting}\,a),
while the number of reactions, $N_r$, increases with time (Fig.~\ref{fig:evolution-reacting}\,b).

\section*{Author Contributions}

S.G. conceived and wrote the original online tutorials and underlying Sphinx documentation
for \href{https://lammpstutorials.github.io}{lammpstutorials.github.io}.  C.A. tested
the tutorials extensively, providing feedback that improved their clarity, accuracy, and usability.
J.G. is the principal author of \lmpcmd{fix bond/react} and \lmpcmd{type labels}
support in LAMMPS.  He revised the tutorials to incorporate type labels and wrote Tutorial 8.
A.K. developed the \lammpsgui{} software and assisted in revising the
tutorials for use with it.  All authors participated in the revision and finalization
of the manuscript.

\section*{Potentially Conflicting Interests}

There are no conflicting interests to declare.

\section*{Funding Information}

S.G. acknowledges funding from the European Union's Horizon 2020 research and
innovation programme under the Marie Skłodowska-Curie grant agreement $\text{N}^\circ\;101065060$.
A.K. acknowledges financial support by Sandia National Laboratories under
POs~2149742 and 2407526.  This work used Expanse at the San Diego Supercomputer
Center through allocation MAT240081 from the Advanced Cyberinfrastructure Coordination
Ecosystem: Services \& Support (ACCESS) program.

\section*{Author Information}
\makeorcid

\bibliography{journal-article}

\begin{thebibliography}{72}
\providecommand{\natexlab}[1]{#1}
\providecommand{\urlprefix}{}
\providecommand{\doiprefix}{https://doi.org/}

\bibitem[{Frenkel and Smit(2023)Frenkel, Daan and Smit,
  Berend}]{frenkel2023understanding}
\textbf{\color{LiveCoMSMediumGrey} Frenkel D}, Smit B.
\newblock Understanding Molecular Simulation: From Algorithms to Applications.
\newblock Elsevier; 2023.
\newblock \href{https://doi.org/10.1016/B978-0-12-267351-1.X5000-7}{\doiprefix
  \detokenize{10.1016/B978-0-12-267351-1.X5000-7}}.

\bibitem[{Allen and Tildesley(2017)Allen, Michael P and Tildesley, Dominic
  J}]{allen2017computer}
\textbf{\color{LiveCoMSMediumGrey} Allen MP}, Tildesley DJ.
\newblock Computer Simulation of Liquids.
\newblock Oxford university press; 2017.
\newblock \href{https://doi.org/10.1093/oso/9780198803195.001.0001}{\doiprefix
  \detokenize{10.1093/oso/9780198803195.001.0001}}.

\bibitem[{van Gunsteren et~al.(2008)van Gunsteren, Wilfred F and Dolenc,
  Jo{\v{z}}ica and Mark, Alan E}]{van2008molecular}
\textbf{\color{LiveCoMSMediumGrey} van Gunsteren WF}, Dolenc J, Mark AE.
\newblock Molecular Simulation as an Aid to Experimentalists.
\newblock Current Opinion in Structural Biology.  2008; 18(2):149--153.
\newblock \href{https://doi.org/10.1016/j.sbi.2007.12.007}{\doiprefix
  \detokenize{10.1016/j.sbi.2007.12.007}}.

\bibitem[{lam(????)}]{lammps_home}
LAMMPS Homepage;.
\newblock Accessed: 2024-07-15.
\newblock \url{https://www.lammps.org}.

\bibitem[{Thompson et~al.(2022)Thompson, Aidan P and Aktulga, H Metin and
  Berger, Richard and Bolintineanu, Dan S and Brown, W Michael and Crozier,
  Paul S and in't Veld, Pieter J and Kohlmeyer, Axel and Moore, Stan G and
  Nguyen, Trung Dac and others}]{thompson2022lammps}
\textbf{\color{LiveCoMSMediumGrey} Thompson AP}, Aktulga HM, Berger R,
  Bolintineanu DS, Brown WM, Crozier PS, in't Veld PJ, Kohlmeyer A, Moore SG,
  Nguyen TD, et~al.
\newblock {{LAMMPS}}-A Flexible Simulation Tool for Particle-Based Materials
  Modeling at the Atomic, Meso, and Continuum Scales.
\newblock Computer Physics Communications.  2022; 271:108171.
\newblock \href{https://doi.org/10.1016/j.cpc.2021.108171}{\doiprefix
  \detokenize{10.1016/j.cpc.2021.108171}}.

\bibitem[{lam(????)}]{lammps_docs}
LAMMPS Online Documentation for Latest Stable Version;.
\newblock Accessed: 2024-07-15.
\newblock \url{https://docs.lammps.org/stable}.

\bibitem[{Wong-Ekkabut and Karttunen(2016)Wong-Ekkabut, Jirasak and Karttunen,
  Mikko}]{wong2016good}
\textbf{\color{LiveCoMSMediumGrey} Wong-Ekkabut J}, Karttunen M.
\newblock The Good, the Bad and the User in Soft Matter Simulations.
\newblock Biochimica et Biophysica Acta (BBA)-Biomembranes.  2016;
  1858(10):2529--2538.
\newblock \href{https://doi.org/10.1016/j.bbamem.2016.02.004}{\doiprefix
  \detokenize{10.1016/j.bbamem.2016.02.004}}.

\bibitem[{van Gunsteren et~al.(2018)van Gunsteren, Wilfred F and Daura, Xavier
  and Hansen, Niels and Mark, Alan E and Oostenbrink, Chris and Riniker,
  Sereina and Smith, Lorna J}]{van2018validation}
\textbf{\color{LiveCoMSMediumGrey} van Gunsteren WF}, Daura X, Hansen N, Mark
  AE, Oostenbrink C, Riniker S, Smith LJ.
\newblock Validation of Molecular Simulation: An Overview of Issues.
\newblock Angewandte Chemie International Edition.  2018; 57(4):884--902.
\newblock \href{https://doi.org/10.1002/anie.201702945}{\doiprefix
  \detokenize{10.1002/anie.201702945}}.

\bibitem[{Prasad et~al.(2018)Prasad, S and Mobley, DL and Braun, E and Mayes,
  HB and Monroe, JI and Zuckerman, DM and others}]{prasad2018best}
\textbf{\color{LiveCoMSMediumGrey} Prasad S}, Mobley D, Braun E, Mayes H,
  Monroe J, Zuckerman D, et~al.
\newblock Best Practices for Foundations in Molecular Simulations [Article v1.
  0].
\newblock Living Journal of Computational Molecular Science.  2018; 1:1--28.
\newblock \href{https://doi.org/10.33011/livecoms.1.1.5957}{\doiprefix
  \detokenize{10.33011/livecoms.1.1.5957}}.

\bibitem[{Jorgensen et~al.(1996)Jorgensen, William L. and Maxwell, David S. and
  Tirado-Rives, Julian}]{jorgensenDevelopmentTestingOPLS1996}
\textbf{\color{LiveCoMSMediumGrey} Jorgensen WL}, Maxwell DS, Tirado-Rives J.
\newblock Development and Testing of the OPLS All-Atom Force Field on
  Conformational Energetics and Properties of Organic Liquids.
\newblock Journal of the American Chemical Society.  1996;
  118(45):11225--11236.
\newblock \href{https://doi.org/10.1021/ja9621760}{\doiprefix
  \detokenize{10.1021/ja9621760}}.

\bibitem[{Stuart et~al.(2000)Stuart, Steven J and Tutein, Alan B and Harrison,
  Judith A}]{stuart2000reactive}
\textbf{\color{LiveCoMSMediumGrey} Stuart SJ}, Tutein AB, Harrison JA.
\newblock A Reactive Potential for Hydrocarbons With Intermolecular
  Interactions.
\newblock The Journal of Chemical Physics.  2000; 112(14):6472--6486.
\newblock \href{https://doi.org/10.1063/1.481208}{\doiprefix
  \detokenize{10.1063/1.481208}}.

\bibitem[{Gissinger et~al.(2024)Gissinger, Jacob R and Nikiforov, Ilia and
  Afshar, Yaser and Waters, Brendon and Choi, Moon-ki and Karls, Daniel S and
  Stukowski, Alexander and Im, Wonpil and Heinz, Hendrik and Kohlmeyer, Axel
  and others}]{gissinger2024type}
\textbf{\color{LiveCoMSMediumGrey} Gissinger JR}, Nikiforov I, Afshar Y, Waters
  B, Choi Mk, Karls DS, Stukowski A, Im W, Heinz H, Kohlmeyer A, et~al.
\newblock Type Label Framework for Bonded Force Fields in LAMMPS.
\newblock The Journal of Physical Chemistry B.  2024; 128(13):3282--3297.
\newblock \href{https://doi.org/10.1021/acs.jpcb.3c08419}{\doiprefix
  \detokenize{10.1021/acs.jpcb.3c08419}}.

\bibitem[{Abascal and Vega(2005)Abascal, Jose LF and Vega,
  Carlos}]{abascal2005general}
\textbf{\color{LiveCoMSMediumGrey} Abascal JL}, Vega C.
\newblock A General Purpose Model for the Condensed Phases of Water:
  TIP4P/2005.
\newblock The Journal of Chemical Physics.  2005; 123(23).
\newblock \href{https://doi.org/10.1063/1.2121687}{\doiprefix
  \detokenize{10.1063/1.2121687}}.

\bibitem[{van Duin et~al.(2001)van Duin, Adri CT and Dasgupta, Siddharth and
  Lorant, Francois and Goddard, William A}]{van2001reaxff}
\textbf{\color{LiveCoMSMediumGrey} van Duin AC}, Dasgupta S, Lorant F, Goddard
  WA.
\newblock ReaxFF: A Reactive Force Field for Hydrocarbons.
\newblock The Journal of Physical Chemistry A.  2001; 105(41):9396--9409.
\newblock \href{https://doi.org/10.1021/jp004368u}{\doiprefix
  \detokenize{10.1021/jp004368u}}.

\bibitem[{Gissinger et~al.(2020)Gissinger, Jacob R and Jensen, Benjamin D and
  Wise, Kristopher E}]{gissinger2020reacter}
\textbf{\color{LiveCoMSMediumGrey} Gissinger JR}, Jensen BD, Wise KE.
\newblock REACTER: A Heuristic Method for Reactive Molecular Dynamics.
\newblock Macromolecules.  2020; 53(22):9953--9961.
\newblock \href{https://doi.org/10.1021/acs.macromol.0c02012}{\doiprefix
  \detokenize{10.1021/acs.macromol.0c02012}}.

\bibitem[{lam(????)}]{lammps_gui_docs}
LAMMPS-GUI Online Documentation for Latest Stable Version;.
\newblock Accessed: 2024-07-15.
\newblock \url{https://docs.lammps.org/stable/Howto_lammps_gui.html}.

\bibitem[{Barrat and Hansen(2003)Barrat, Jean-Louis and Hansen,
  Jean-Pierre}]{barrat2003basic}
\textbf{\color{LiveCoMSMediumGrey} Barrat JL}, Hansen JP.
\newblock Basic Concepts for Simple and Complex Liquids.
\newblock Cambridge University Press; 2003.
\newblock \href{https://doi.org/10.1017/CBO9780511606533}{\doiprefix
  \detokenize{10.1017/CBO9780511606533}}.

\bibitem[{Hansen and McDonald(2013)Hansen, Jean-Pierre and McDonald, Ian
  Ranald}]{hansen2013theory}
\textbf{\color{LiveCoMSMediumGrey} Hansen JP}, McDonald IR.
\newblock Theory of Simple Liquids: With Applications to Soft Matter.
\newblock Academic press; 2013.
\newblock \href{https://doi.org/10.1016/C2010-0-66723-X}{\doiprefix
  \detokenize{10.1016/C2010-0-66723-X}}.

\bibitem[{{SklogWiki contributors}(n.d.)}]{sklogwiki_main_page}
\textbf{\color{LiveCoMSMediumGrey} {SklogWiki contributors}}, Main Page; n.d.
\newblock Accessed: 2024-12-22.
\newblock \url{http://www.sklogwiki.org/SklogWiki/index.php/Main_Page}.

\bibitem[{Plimpton et~al.(2024)Plimpton, Steven J. and Kohlmeyer, Axel and
  Thompson, Aidan P. and Moore, Stan G. and Berger, Richard}]{lammps_code}
\textbf{\color{LiveCoMSMediumGrey} Plimpton SJ}, Kohlmeyer A, Thompson AP,
  Moore SG, Berger R, LAMMPS: Large-Scale Atomic/Molecular Massively Parallel
  Simulator.
\newblock Zenodo; 2024.
\newblock \href{https://doi.org/10.5281/zenodo.3726416}{\doiprefix
  \detokenize{10.5281/zenodo.3726416}}.

\bibitem[{lam(????)}]{lammps_github_release}
LAMMPS Releases Page on GitHub;.
\newblock Accessed: 2024-12-26.
\newblock \url{https://github.com/lammps/lammps/releases}.

\bibitem[{van Rossum and Drake~Jr(1995)van Rossum, Guido and Drake Jr, Fred
  L}]{van1995python}
\textbf{\color{LiveCoMSMediumGrey} van Rossum G}, Drake~Jr FL.
\newblock Python Reference Manual.
\newblock Centrum voor Wiskunde en Informatica Amsterdam; 1995.

\bibitem[{Hunter(2007)Hunter, J. D.}]{hunter2007Matplotlib}
\textbf{\color{LiveCoMSMediumGrey} Hunter JD}.
\newblock Matplotlib: A 2D Graphics Environment.
\newblock Computing in Science \& Engineering.  2007; 9(3):90--95.
\newblock \href{https://doi.org/10.1109/MCSE.2007.55}{\doiprefix
  \detokenize{10.1109/MCSE.2007.55}}.

\bibitem[{vmd(????)}]{vmd_home}
VMD Homepage;.
\newblock Accessed: 2024-07-15.
\newblock \url{https://www.ks.uiuc.edu/Research/vmd}.

\bibitem[{Humphrey et~al.(1996)Humphrey, William and Dalke, Andrew and
  Schulten, Klaus}]{humphrey1996vmd}
\textbf{\color{LiveCoMSMediumGrey} Humphrey W}, Dalke A, Schulten K.
\newblock {{VMD}}: Visual Molecular Dynamics.
\newblock Journal of Molecular Graphics.  1996; 14(1):33--38.
\newblock \href{https://doi.org/10.1016/0263-7855(96)00018-5}{\doiprefix
  \detokenize{10.1016/0263-7855(96)00018-5}}.

\bibitem[{ovi(????)}]{ovito_home}
OVITO Homepage;.
\newblock Accessed: 2024-07-15.
\newblock \url{https://ovito.org}.

\bibitem[{Stukowski(2009)Stukowski, Alexander}]{stukowski2009visualization}
\textbf{\color{LiveCoMSMediumGrey} Stukowski A}.
\newblock Visualization and Analysis of Atomistic Simulation Data With
  OVITO--The Open Visualization Tool.
\newblock Modelling and Simulation in Materials Science and Engineering.  2009;
  18(1):015012.
\newblock \href{https://doi.org/10.1088/0965-0393/18/1/015012}{\doiprefix
  \detokenize{10.1088/0965-0393/18/1/015012}}.

\bibitem[{Wang et~al.(2020)Wang, Xipeng and Ram{\'\i}rez-Hinestrosa, Sim{\'o}n
  and Dobnikar, Jure and Frenkel, Daan}]{wang2020lennard}
\textbf{\color{LiveCoMSMediumGrey} Wang X}, Ram{\'\i}rez-Hinestrosa S, Dobnikar
  J, Frenkel D.
\newblock The Lennard-Jones Potential: When (Not) to Use It.
\newblock Physical Chemistry Chemical Physics.  2020; 22(19):10624--10633.
\newblock \href{https://doi.org/10.1039/C9CP05445F}{\doiprefix
  \detokenize{10.1039/C9CP05445F}}.

\bibitem[{Fischer and Wendland(2023)Fischer, Johann and Wendland,
  Martin}]{fischer2023history}
\textbf{\color{LiveCoMSMediumGrey} Fischer J}, Wendland M.
\newblock On the History of Key Empirical Intermolecular Potentials.
\newblock Fluid Phase Equilibria.  2023; 573:113876.
\newblock \href{https://doi.org/10.1016/j.fluid.2023.113876}{\doiprefix
  \detokenize{10.1016/j.fluid.2023.113876}}.

\bibitem[{Hestenes et~al.(1952)Hestenes, Magnus Rudolph and Stiefel, Eduard and
  others}]{hestenes1952methods}
\textbf{\color{LiveCoMSMediumGrey} Hestenes MR}, Stiefel E, et~al.
\newblock Methods of Conjugate Gradients for Solving Linear Systems, vol.~49.
\newblock NBS Washington, DC; 1952.
\newblock \href{https://doi.org/10.6028/jres.049.044}{\doiprefix
  \detokenize{10.6028/jres.049.044}}.

\bibitem[{Schneider and Stoll(1978)Schneider, T and Stoll,
  E}]{schneider1978molecular}
\textbf{\color{LiveCoMSMediumGrey} Schneider T}, Stoll E.
\newblock Molecular-Dynamics Study of a Three-Dimensional One-Component Model
  for Distortive Phase Transitions.
\newblock Physical Review B.  1978; 17(3):1302.
\newblock \href{https://doi.org/10.1103/PhysRevB.17.1302}{\doiprefix
  \detokenize{10.1103/PhysRevB.17.1302}}.

\bibitem[{Grossfield and Zuckerman(2009)Grossfield, Alan and Zuckerman, Daniel
  M}]{grossfield2009quantifying}
\textbf{\color{LiveCoMSMediumGrey} Grossfield A}, Zuckerman DM.
\newblock Quantifying uncertainty and sampling quality in biomolecular
  simulations.
\newblock Annual Reports in Computational Chemistry.  2009; 5:23--48.

\bibitem[{Kohlmeyer and Vermaas(2021)Kohlmeyer, Axel and Vermaas,
  Josh}]{kohlmeyer2017topotools}
\textbf{\color{LiveCoMSMediumGrey} Kohlmeyer A}, Vermaas J, {{TopoTools}}:
  {{Release}} 1.9.
\newblock Zenodo; 2021.
\newblock \href{https://doi.org/10.5281/zenodo.598373}{\doiprefix
  \detokenize{10.5281/zenodo.598373}}.

\bibitem[{Nos{\'e}(1984)Nos{\'e}, Shuichi}]{nose1984unified}
\textbf{\color{LiveCoMSMediumGrey} Nos{\'e} S}.
\newblock A Unified Formulation of the Constant Temperature Molecular Dynamics
  Methods.
\newblock The Journal of Chemical Physics.  1984; 81(1):511--519.
\newblock \href{https://doi.org/10.1063/1.447334}{\doiprefix
  \detokenize{10.1063/1.447334}}.

\bibitem[{Hoover(1985)Hoover, William G}]{hoover1985canonical}
\textbf{\color{LiveCoMSMediumGrey} Hoover WG}.
\newblock Canonical Dynamics: Equilibrium Phase-Space Distributions.
\newblock Physical Review A.  1985; 31(3):1695.
\newblock \href{https://doi.org/10.1103/PhysRevA.31.1695}{\doiprefix
  \detokenize{10.1103/PhysRevA.31.1695}}.

\bibitem[{Schmid et~al.(2011)Schmid, Nathan and Eichenberger, Andreas P and
  Choutko, Alexandra and Riniker, Sereina and Winger, Moritz and Mark, Alan E
  and van Gunsteren, Wilfred F}]{schmid2011definition}
\textbf{\color{LiveCoMSMediumGrey} Schmid N}, Eichenberger AP, Choutko A,
  Riniker S, Winger M, Mark AE, van Gunsteren WF.
\newblock Definition and Testing of the GROMOS Force-Field Versions 54A7 and
  54B7.
\newblock European Biophysics Journal.  2011; 40:843--856.
\newblock \href{https://doi.org/10.1007/s00249-011-0700-9}{\doiprefix
  \detokenize{10.1007/s00249-011-0700-9}}.

\bibitem[{Wu et~al.(2006)Wu, Yujie and Tepper, Harald L and Voth, Gregory
  A}]{wu2006flexible}
\textbf{\color{LiveCoMSMediumGrey} Wu Y}, Tepper HL, Voth GA.
\newblock Flexible Simple Point-Charge Water Model With Improved Liquid-State
  Properties.
\newblock The Journal of Chemical Physics.  2006; 124(2).
\newblock \href{https://doi.org/10.1063/1.2136877}{\doiprefix
  \detokenize{10.1063/1.2136877}}.

\bibitem[{Luty and van Gunsteren(1996)Luty, Brock A and van Gunsteren, Wilfred
  F}]{luty1996calculating}
\textbf{\color{LiveCoMSMediumGrey} Luty BA}, van Gunsteren WF.
\newblock Calculating Electrostatic Interactions Using the Particle-Particle
  Particle-Mesh Method With Nonperiodic Long-Range Interactions.
\newblock The Journal of Physical Chemistry.  1996; 100(7):2581--2587.
\newblock \href{https://doi.org/10.1021/jp9518623}{\doiprefix
  \detokenize{10.1021/jp9518623}}.

\bibitem[{Liese et~al.(2017)Liese, Susanne and Gensler, Manuel and Krysiak,
  Stefanie and Schwarzl, Richard and Achazi, Andreas and Paulus, Beate and
  Hugel, Thorsten and Rabe, Jürgen P and Netz, Roland R}]{liese2017hydration}
\textbf{\color{LiveCoMSMediumGrey} Liese S}, Gensler M, Krysiak S, Schwarzl R,
  Achazi A, Paulus B, Hugel T, Rabe JP, Netz RR.
\newblock Hydration Effects Turn a Highly Stretched Polymer From an Entropic
  Into an Energetic Spring.
\newblock ACS Nano.  2017; 11(1):702--712.
\newblock \href{https://doi.org/10.1021/acsnano.6b07071}{\doiprefix
  \detokenize{10.1021/acsnano.6b07071}}.

\bibitem[{Oostenbrink et~al.(2004)Oostenbrink, Chris and Villa, Alessandra and
  Mark, Alan E and van Gunsteren, Wilfred F}]{oostenbrink2004biomolecular}
\textbf{\color{LiveCoMSMediumGrey} Oostenbrink C}, Villa A, Mark AE, van
  Gunsteren WF.
\newblock A Biomolecular Force Field Based on the Free Enthalpy of Hydration
  and Solvation: The GROMOS Force-Field Parameter Sets 53A5 and 53A6.
\newblock Journal of Computational Chemistry.  2004; 25(13):1656--1676.
\newblock \href{https://doi.org/10.1002/jcc.20090}{\doiprefix
  \detokenize{10.1002/jcc.20090}}.

\bibitem[{Berendsen et~al.(1981)Berendsen, H. J. C. and Postma, J. P. M. and
  van Gunsteren, W. F. and Hermans, J.}]{berendsen1981interaction}
\textbf{\color{LiveCoMSMediumGrey} Berendsen HJC}, Postma JPM, van Gunsteren
  WF, Hermans J.
\newblock In: Pullman B, editor. Interaction Models for Water in Relation to
  Protein Hydration Dordrecht: Springer Netherlands; 1981. p. 331--342.
\newblock \href{https://doi.org/10.1007/978-94-015-7658-1_21}{\doiprefix
  \detokenize{10.1007/978-94-015-7658-1_21}}.

\bibitem[{Ewald(1921)Ewald, Paul P}]{ewald1921berechnung}
\textbf{\color{LiveCoMSMediumGrey} Ewald PP}.
\newblock Die Berechnung Optischer und Elektrostatischer Gitterpotentiale.
\newblock Annalen der Physik.  1921; 369(3):253--287.
\newblock \href{https://doi.org/10.1002/andp.19213690304}{\doiprefix
  \detokenize{10.1002/andp.19213690304}}.

\bibitem[{Martyna et~al.(1994)Martyna, Glenn J and Tobias, Douglas J and Klein,
  Michael L}]{martyna1994constant}
\textbf{\color{LiveCoMSMediumGrey} Martyna GJ}, Tobias DJ, Klein ML.
\newblock Constant Pressure Molecular Dynamics Algorithms.
\newblock The Journal of Chemical Physics.  1994; 101(5):4177--4189.
\newblock \href{https://doi.org/10.1063/1.467468}{\doiprefix
  \detokenize{10.1063/1.467468}}.

\bibitem[{Malde et~al.(2011)Malde, Alpeshkumar K and Zuo, Le and Breeze,
  Matthew and Stroet, Martin and Poger, David and Nair, Pramod C and
  Oostenbrink, Chris and Mark, Alan E}]{malde2011automated}
\textbf{\color{LiveCoMSMediumGrey} Malde AK}, Zuo L, Breeze M, Stroet M, Poger
  D, Nair PC, Oostenbrink C, Mark AE.
\newblock An Automated Force Field Topology Builder (ATB) and Repository:
  Version 1.0.
\newblock Journal of Chemical Theory and Computation.  2011; 7(12):4026--4037.
\newblock \href{https://doi.org/10.1021/ct200196m}{\doiprefix
  \detokenize{10.1021/ct200196m}}.

\bibitem[{Fixman(1962)Fixman, Marshall}]{fixmanRadiusGyrationPolymer1962a}
\textbf{\color{LiveCoMSMediumGrey} Fixman M}.
\newblock Radius of {{Gyration}} of {{Polymer Chains}}.
\newblock The Journal of Chemical Physics.  1962; 36(2):306--310.
\newblock \href{https://doi.org/10.1063/1.1732501}{\doiprefix
  \detokenize{10.1063/1.1732501}}.

\bibitem[{Kadaoluwa~Pathirannahalage et~al.(2021)Kadaoluwa Pathirannahalage,
  Sachini P and Meftahi, Nastaran and Elbourne, Aaron and Weiss, Alessia CG and
  McConville, Chris F and Padua, Agilio and Winkler, David A and Costa Gomes,
  Margarida and Greaves, Tamar L and Le, Tu C and
  others}]{kadaoluwa2021systematic}
\textbf{\color{LiveCoMSMediumGrey} Kadaoluwa~Pathirannahalage SP}, Meftahi N,
  Elbourne A, Weiss AC, McConville CF, Padua A, Winkler DA, Costa~Gomes M,
  Greaves TL, Le TC, et~al.
\newblock Systematic Comparison of the Structural and Dynamic Properties of
  Commonly Used Water Models for Molecular Dynamics Simulations.
\newblock Journal of Chemical Information and Modeling.  2021;
  61(9):4521--4536.
\newblock \href{https://doi.org/10.1021/acs.jcim.1c00794}{\doiprefix
  \detokenize{10.1021/acs.jcim.1c00794}}.

\bibitem[{Ryckaert et~al.(1977)Ryckaert, Jean-Paul and Ciccotti, Giovanni and
  Berendsen, Herman JC}]{ryckaert1977numerical}
\textbf{\color{LiveCoMSMediumGrey} Ryckaert JP}, Ciccotti G, Berendsen HJ.
\newblock Numerical Integration of the Cartesian Equations of Motion of a
  System With Constraints: Molecular Dynamics of n-Alkanes.
\newblock Journal of Computational Physics.  1977; 23(3):327--341.
\newblock \href{https://doi.org/10.1016/0021-9991(77)90098-5}{\doiprefix
  \detokenize{10.1016/0021-9991(77)90098-5}}.

\bibitem[{Andersen(1983)Andersen, Hans C}]{andersen1983rattle}
\textbf{\color{LiveCoMSMediumGrey} Andersen HC}.
\newblock Rattle: A “Velocity” Version of the SHAKE Algorithm for Molecular
  Dynamics Calculations.
\newblock Journal of Computational Physics.  1983; 52(1):24--34.
\newblock \href{https://doi.org/10.1016/0021-9991(83)90014-1}{\doiprefix
  \detokenize{10.1016/0021-9991(83)90014-1}}.

\bibitem[{Mills(1955)Mills, Reginald}]{mills1955remeasurement}
\textbf{\color{LiveCoMSMediumGrey} Mills R}.
\newblock A Remeasurement of the Self-Diffusion Coefficients of Sodium Ion in
  Aqueous Sodium Chloride Solutions.
\newblock Journal of the American Chemical Society.  1955; 77(23):6116--6119.
\newblock \href{https://doi.org/10.1021/ja01628a008}{\doiprefix
  \detokenize{10.1021/ja01628a008}}.

\bibitem[{Gravelle et~al.(2021)Gravelle, Simon and Kamal, Catherine and Botto,
  Lorenzo}]{gravelle2021violations}
\textbf{\color{LiveCoMSMediumGrey} Gravelle S}, Kamal C, Botto L.
\newblock Violations of Jeffery's Theory in the Dynamics of Nanographene in
  Shear Flow.
\newblock Physical Review Fluids.  2021; 6(3):034303.
\newblock \href{https://doi.org/10.1103/PhysRevFluids.6.034303}{\doiprefix
  \detokenize{10.1103/PhysRevFluids.6.034303}}.

\bibitem[{{Wolde-Kidan} and Netz(2021){Wolde-Kidan}, Amanuel and Netz, Roland
  R.}]{wolde-kidanInterplayInterfacialViscosity2021}
\textbf{\color{LiveCoMSMediumGrey} {Wolde-Kidan} A}, Netz RR.
\newblock Interplay of {{Interfacial Viscosity}}, {{Specific-Ion}}, and
  {{Impurity Adsorption Determines Zeta Potentials}} of {{Phospholipid
  Membranes}}.
\newblock Langmuir.  2021; 37(28):8463--8473.
\newblock \href{https://doi.org/10.1021/acs.langmuir.1c00868}{\doiprefix
  \detokenize{10.1021/acs.langmuir.1c00868}}.

\bibitem[{Zou and van Duin(2012)Zou, Chenyu and van Duin,
  Adri}]{zou2012investigation}
\textbf{\color{LiveCoMSMediumGrey} Zou C}, van Duin A.
\newblock Investigation of Complex Iron Surface Catalytic Chemistry Using the
  ReaxFF Reactive Force Field Method.
\newblock Jom.  2012; 64:1426--1437.
\newblock \href{https://doi.org/10.1007/s11837-012-0463-5}{\doiprefix
  \detokenize{10.1007/s11837-012-0463-5}}.

\bibitem[{Vashishta et~al.(1990)Vashishta, P and Kalia, Rajiv K and Rino,
  Jos{\'e} P and Ebbsj{\"o}, Ingvar}]{vashishta1990interaction}
\textbf{\color{LiveCoMSMediumGrey} Vashishta P}, Kalia RK, Rino JP, Ebbsj{\"o}
  I.
\newblock Interaction Potential for SiO2: A Molecular-Dynamics Study of
  Structural Correlations.
\newblock Physical Review B.  1990; 41(17):12197.
\newblock \href{https://doi.org/10.1103/PhysRevB.41.12197}{\doiprefix
  \detokenize{10.1103/PhysRevB.41.12197}}.

\bibitem[{Rappe and Goddard~III(1991)Rappe, Anthony K and Goddard III, William
  A}]{rappe1991charge}
\textbf{\color{LiveCoMSMediumGrey} Rappe AK}, Goddard~III WA.
\newblock Charge Equilibration for Molecular Dynamics Simulations.
\newblock The Journal of Physical Chemistry.  1991; 95(8):3358--3363.
\newblock \href{https://doi.org/10.1021/j100161a070}{\doiprefix
  \detokenize{10.1021/j100161a070}}.

\bibitem[{Sulpizi et~al.(2012)Sulpizi, Marialore and Gaigeot, Marie-Pierre and
  Sprik, Michiel}]{sulpizi2012silica}
\textbf{\color{LiveCoMSMediumGrey} Sulpizi M}, Gaigeot MP, Sprik M.
\newblock The Silica-Water Interface: How the Silanols Determine the Surface
  Acidity and Modulate the Water Properties.
\newblock Journal of Chemical Theory and Computation.  2012; 8(3):1037--1047.
\newblock \href{https://doi.org/10.1021/ct2007154}{\doiprefix
  \detokenize{10.1021/ct2007154}}.

\bibitem[{Della~Valle and Andersen(1992)Della Valle, Raffaele Guido and
  Andersen, Hans C}]{della1992molecular}
\textbf{\color{LiveCoMSMediumGrey} Della~Valle RG}, Andersen HC.
\newblock Molecular Dynamics Simulation of Silica Liquid and Glass.
\newblock The Journal of Chemical Physics.  1992; 97(4):2682--2689.
\newblock \href{https://doi.org/10.1063/1.463056}{\doiprefix
  \detokenize{10.1063/1.463056}}.

\bibitem[{Gravelle and Dumais(2020)Gravelle, Simon and Dumais,
  Jacques}]{gravelle2020multi}
\textbf{\color{LiveCoMSMediumGrey} Gravelle S}, Dumais J.
\newblock A Multi-Scale Model for Fluid Transport Through a Bio-Inspired
  Passive Valve.
\newblock The Journal of Chemical Physics.  2020; 152(1).
\newblock \href{https://doi.org/10.1063/1.5126481}{\doiprefix
  \detokenize{10.1063/1.5126481}}.

\bibitem[{K{\"a}stner(2011)K{\"a}stner, Johannes}]{kastner2011umbrella}
\textbf{\color{LiveCoMSMediumGrey} K{\"a}stner J}.
\newblock Umbrella Sampling.
\newblock Wiley Interdisciplinary Reviews: Computational Molecular Science.
  2011; 1(6):932--942.
\newblock \href{https://doi.org/10.1002/wcms.66}{\doiprefix
  \detokenize{10.1002/wcms.66}}.

\bibitem[{Wilson and Pohorille(1997)Wilson, Michael A and Pohorille,
  Andrew}]{wilson1997adsorption}
\textbf{\color{LiveCoMSMediumGrey} Wilson MA}, Pohorille A.
\newblock Adsorption and Solvation of Ethanol at the Water Liquid--Vapor
  Interface: A Molecular Dynamics Study.
\newblock The Journal of Physical Chemistry B.  1997; 101(16):3130--3135.
\newblock \href{https://doi.org/10.1021/jp962629n}{\doiprefix
  \detokenize{10.1021/jp962629n}}.

\bibitem[{Makarov(2009)Makarov, Dmitrii E}]{makarov2009computer}
\textbf{\color{LiveCoMSMediumGrey} Makarov DE}.
\newblock Computer Simulations and Theory of Protein Translocation.
\newblock Accounts of Chemical Research.  2009; 42(2):281--289.
\newblock \href{https://doi.org/10.1021/ar800128x}{\doiprefix
  \detokenize{10.1021/ar800128x}}.

\bibitem[{Gravelle and Botto(2021)Gravelle, Simon and Botto,
  Lorenzo}]{gravelle2021adsorption}
\textbf{\color{LiveCoMSMediumGrey} Gravelle S}, Botto L.
\newblock Adsorption of Single and Multiple Graphene-Oxide Nanoparticles at a
  Water--Vapor Interface.
\newblock Langmuir.  2021; 37(45):13322--13330.
\newblock \href{https://doi.org/10.1021/acs.langmuir.1c01902}{\doiprefix
  \detokenize{10.1021/acs.langmuir.1c01902}}.

\bibitem[{Loche et~al.(2022)Loche, Philip and Bonthuis, Douwe J and Netz,
  Roland R}]{loche2022molecular}
\textbf{\color{LiveCoMSMediumGrey} Loche P}, Bonthuis DJ, Netz RR.
\newblock Molecular Dynamics Simulations of the Evaporation of Hydrated Ions
  From Aqueous Solution.
\newblock Communications Chemistry.  2022; 5(1):55.
\newblock \href{https://doi.org/10.1038/s42004-022-00669-5}{\doiprefix
  \detokenize{10.1038/s42004-022-00669-5}}.

\bibitem[{Hayatifar et~al.(2024)Hayatifar, Ardalan and Gravelle, Simon and
  Moreno, Beatriz D and Schoepfer, Valerie A and Lindsay, Matthew
  BJ}]{hayatifar2024probing}
\textbf{\color{LiveCoMSMediumGrey} Hayatifar A}, Gravelle S, Moreno BD,
  Schoepfer VA, Lindsay MB.
\newblock Probing Atomic-Scale Processes at the Ferrihydrite-Water Interface
  With Reactive Molecular Dynamics.
\newblock Geochemical Transactions.  2024; 25(1):10.
\newblock \href{https://doi.org/10.1186/s12932-024-00094-8}{\doiprefix
  \detokenize{10.1186/s12932-024-00094-8}}.

\bibitem[{Weeks et~al.(1971)Weeks, John D and Chandler, David and Andersen,
  Hans C}]{weeks1971role}
\textbf{\color{LiveCoMSMediumGrey} Weeks JD}, Chandler D, Andersen HC.
\newblock Role of Repulsive Forces in Determining the Equilibrium Structure of
  Simple Liquids.
\newblock The Journal of Chemical Physics.  1971; 54(12):5237--5247.
\newblock \href{https://doi.org/10.1063/1.1674820}{\doiprefix
  \detokenize{10.1063/1.1674820}}.

\bibitem[{Kumar et~al.(1992)Kumar, Shankar and Rosenberg, John M and Bouzida,
  Djamal and Swendsen, Robert H and Kollman, Peter A}]{kumar1992weighted}
\textbf{\color{LiveCoMSMediumGrey} Kumar S}, Rosenberg JM, Bouzida D, Swendsen
  RH, Kollman PA.
\newblock The Weighted Histogram Analysis Method for Free-Energy Calculations
  on Biomolecules. I. The Method.
\newblock Journal of Computational Chemistry.  1992; 13(8):1011--1021.
\newblock \href{https://doi.org/10.1002/jcc.540130812}{\doiprefix
  \detokenize{10.1002/jcc.540130812}}.

\bibitem[{Kumar et~al.(1995)Kumar, Shankar and Rosenberg, John M and Bouzida,
  Djamal and Swendsen, Robert H and Kollman, Peter A}]{kumar1995multidim}
\textbf{\color{LiveCoMSMediumGrey} Kumar S}, Rosenberg JM, Bouzida D, Swendsen
  RH, Kollman PA.
\newblock Multidimensional Free-Energy Calculations Using the Weighted
  Histogram Analysis Method.
\newblock Journal of Computational Chemistry.  1995; 16(11):1339--1350.
\newblock \href{https://doi.org/10.1002/jcc.540161104}{\doiprefix
  \detokenize{10.1002/jcc.540161104}}.

\bibitem[{Grossfield(2025)Grossfield, Alan}]{grossfieldimplementation}
\textbf{\color{LiveCoMSMediumGrey} Grossfield A}, An Implementation of WHAM:
  The Weighted Histogram Analysis Method Version 2.1.0; 2025.
\newblock \urlprefix\url{http://membrane.urmc.rochester.edu/content/wham/},
  accessed: 2025-03-16.

\bibitem[{Gissinger et~al.(2017)Jacob R. Gissinger and Benjamin D. Jensen and
  Kristopher E. Wise}]{gissinger2017polymer}
\textbf{\color{LiveCoMSMediumGrey} Gissinger JR}, Jensen BD, Wise KE.
\newblock Modeling Chemical Reactions in Classical Molecular Dynamics
  Simulations.
\newblock Polymer.  2017; 128:211--217.
\newblock \href{https://doi.org/10.1016/j.polymer.2017.09.038}{\doiprefix
  \detokenize{10.1016/j.polymer.2017.09.038}}.

\bibitem[{Gissinger et~al.(2024)Gissinger, Jacob R and Jensen, Benjamin D and
  Wise, Kristopher E}]{gissinger2024molecular}
\textbf{\color{LiveCoMSMediumGrey} Gissinger JR}, Jensen BD, Wise KE.
\newblock Molecular Modeling of Reactive Systems With REACTER.
\newblock Computer Physics Communications.  2024; 304:109287.
\newblock \href{https://doi.org/10.1016/j.cpc.2024.109287}{\doiprefix
  \detokenize{10.1016/j.cpc.2024.109287}}.

\bibitem[{Sun(1998)Sun, Huai}]{sun1998compass}
\textbf{\color{LiveCoMSMediumGrey} Sun H}.
\newblock COMPASS: An Ab Initio Force-Field Optimized for Condensed-Phase
  Applications Overview With Details on Alkane and Benzene Compounds.
\newblock The Journal of Physical Chemistry B.  1998; 102(38):7338--7364.
\newblock \href{https://doi.org/10.1021/jp980939v}{\doiprefix
  \detokenize{10.1021/jp980939v}}.

\bibitem[{fla(????)}]{flatpak_home}
Flatpak Homepage;.
\newblock Accessed: 2024-08-09.
\newblock \url{https://flatpak.org}.

\bibitem[{lam(????)}]{lammps_run_docs}
Run LAMMPS Online Documentation for Latest Stable Version;.
\newblock Accessed: 2024-12-14.
\newblock \url{https://docs.lammps.org/stable/Run_head.html}.

\end{thebibliography}


\begin{appendices}
\section{Using \lammpsgui{}}
\label{using-lammps-gui-label}

\begin{note}
  For simplicity, these tutorials reference keyboard shortcuts
  based on the assignments for Linux and Windows.  {macOS} users should
  use the ``Command'' key (\cmd) in place of the
  ``Ctrl'' key when using keyboard shortcuts.
\end{note}

\subsection{Installation}

Pre-compiled versions of \lammpsgui{} are available for Linux, {macOS},
and Windows on the LAMMPS GitHub Release
page~\cite{lammps_github_release}.  The Linux version is provided in two
formats: as compressed tar archive (.tar.gz) and as a Flatpak
bundle~\cite{flatpak_home}.  The {macOS} version is distributed as a
.dmg installer image, while the Windows version comes as an executable
installer package.

\subsubsection{Installing the Linux .tar.gz Package}

Download the archive (e.g.,~LAMMPS-Linux-x86\_64-GUI-22Jul2025.tar.gz)
and unpack it.  This will create a folder named \lammpsgui{} containing the
included commands, which can be launched directly using ``./lammps-gui'' or
``./lmp'', for example.  Adding this folder to the PATH environment
variable will make these commands accessible from everywhere, without the
need for the ``./'' prefix.

\subsubsection{Installing the Linux Flatpak Bundle}

You have to have Flatpak support installed on Linux machine to be able
to use the Flatpak bundle.  Download the bundle file
(e.g.,~LAMMPS-Linux-x86\_64-GUI-22Jul2025.flatpak) and then
install it using the following command:
\begin{lstlisting}[language=tcl]
flatpak install --user \
    LAMMPS-Linux-x86_64-GUI-22Jul2025.flatpak
\end{lstlisting}
This will integrate \lammpsgui{} into your desktop environment
(e.g.,~GNOME, KDE, XFCE) where it should appear in the ``Applications''
menu under ``Science''.  Additionally, the ``.lmp'' file extension will be
registered to launch \lammpsgui{} when opening a file with this
extension in the desktop's file manager.

You can also launch \lammpsgui{} from the command-line using the following command:
\begin{lstlisting}[language=tcl]
flatpak run org.lammps.lammps-gui
\end{lstlisting}
Similarly, for launching the LAMMPS command-line executable, use:
\begin{lstlisting}[language=tcl]
flatpak run --command=lmp org.lammps.lammps-gui -in in.lmp
\end{lstlisting}

\subsubsection{Installing the macOS Application Bundle}

After downloading the macOS app bundle image file
(e.g.,~LAMMPS-macOS-multiarch-GUI-22Jul2025.dmg), double-click
on it.  In the dialog that opens drag the \lammpsgui{} app bundle into
the Applications folder.  To enable command-line access, follow the
instructions in the README.txt file.  These macOS app-bundles contain
native executables for both, Intel and Apple CPUs.

After installation, you can launch \lammpsgui{} from the Applications
folder.  Additionally, you can drag an input file onto the app or open
files with the ``.lmp'' extension.  Note that the \lammpsgui{} app bundle is
currently not cryptographically signed, so macOS may initially prevent
it from launching.  If this happens, you need to adjust the settings in
the ``Security \& Privacy" system preferences dialog to allow access.

\subsubsection{Installing the Windows package}

Download the \lammpsgui{} installer for Windows
(e.g.,~LAMMPS-Win10-64bit-GUI-22Jul2025.exe).  Windows may warn
you that the file is from an unknown developer and was downloaded from
the internet.  This happens because neither the installer nor the
\lammpsgui{} application (or any other included applications) have been
cryptographically signed.  You will need to choose to keep the file, and
when launching the installer, confirm that you want to run it despite
the warning.

After installation, a new entry should appear in the Start menu.
Additionally, the ``.lmp'' file extension should be registered with
Windows File Explorer to open \lammpsgui{} when opening a file with the
``.lmp`` extension.  The ``lammps-gui'' and ``lmp'' commands should also
be available in the command-line.

\subsection{Opening, Editing, and Saving Files}

\lammpsgui{} can be launched from the command-line, as explained above, where you
can either launch it without arguments or provide one file name as an argument.  All
other arguments will be ignored.  For example:
\begin{lstlisting}[language=tcl]
lammps-gui input.lmp
\end{lstlisting}
Files can also be opened from the ``File'' menu.  You can select a
file through a dialog and then open it.  Additionally, a history of
the last five opened files is maintained, with entries to open them directly.
Finally, the \texttt{Ctrl-O} keyboard shortcut can also be used to open a file.
When integrated into a desktop environment, it is also possible to open
files with a ``.lmp'' extension or use drag-and-drop.

For the most part, the editor window behaves like other graphical
editors.  You can enter, delete, or copy and paste text.   When entering
text, a pop-up window will appear with possible completions after typing
the first two characters of the first word in a line.  You can
navigate the highlighted options using the up and down arrow keys, and select a
completion by pressing the Enter key or using the mouse.  You can also continue
typing, and the selection in the pop-up will be refined.  For some
commands, there will be completion pop-ups for their
keywords or when a filename is expected, in which case,
the pop-up will list files in the current folder.

As soon as \lammpsgui{} recognizes a command, it applies syntax
highlighting according to built-in categories.  This can help
detect typos, since those may cause \lammpsgui{} not to
recognize the syntax and thus not apply or partially apply
the syntax highlighting.  When you press the \texttt{Tab} key, the line will be
reformatted.  Consistent formatting can improve the readability of
input files, especially long and complex ones.

If the file in the editor has unsaved changes, the word
``*modified*'' will appear in the window title.  The current input
buffer can be saved by selecting ``Save'' or ``Save As...'' from the
``File'' menu.  You can also click the ``Save'' icon on the left side
of the status bar, or use the \texttt{Ctrl-S} keyboard shortcut.

\begin{note}
  When \lammpsgui{} opens a file, it will \emph{switch} the working directory
  to the folder that contains the input file.  The same happens when saving to
  a different folder than the current working directory.  The current working
  directory can be seen in the status bar at the bottom right.  This is important
  to note because LAMMPS input files often require additional files for reading and may
  write output files (such as images, trajectory dumps, or averaged data files),
  which are typically expected to be in the same folder as the input file.
\end{note}


\subsection{Running LAMMPS}
\label{running-lammps-label}


From within the \lammpsgui{} main window, LAMMPS can be started either from
the \guicmd{Run} menu by selecting the \guicmd{Run LAMMPS from Editor Buffer} entry,
using the keyboard shortcut Ctrl-Enter (Command-Enter on macOS), or by clicking the
green \guicmd{Run} button in the status bar.  While LAMMPS is running, a message on
the left side indicates that LAMMPS is active, along with the number of active threads.
On the right side, a progress bar is displayed, showing the estimated progress
of the current \lmpcmd{run} or \lmpcmd{minimize} command.

\subsection{Creating Snapshot Images}

Open the \guicmd{Image Viewer} using either the \guicmd{Create Image} option
from the \guicmd{Run} menu, the \guicmd{Ctrl-I} keyboard shortcut,
or click on the (right) palette button in the status bar.  The image
can be saved using the \guicmd{Save As...} option from the \guicmd{File} menu.

\subsection{The Output Window}

By default, when starting a run, the \guicmd{Output} window opens to display the screen
output of the running LAMMPS calculation.  The text in the Output window is
read-only and cannot be modified, but keyboard shortcuts for selecting and
copying all or part of the text can be used to transfer it to another program:
The keyboard shortcut \guicmd{Ctrl-S} (or \guicmd{Command-S} on {macOS}) can
be used to save the Output buffer to a file.  Additionally, the \guicmd{Select All}
and \guicmd{Copy} functions, along with a \guicmd{Save Log to File} option, are available
through the context menu, which can be accessed by right-clicking within the text area of the
\guicmd{Output} window.

\subsection{The Charts Window}

By default, when starting a run, a \guicmd{Charts} window opens to display
a plot of the thermodynamic output from the LAMMPS calculation.  From the \guicmd{File}
menu in the top-left corner, you can save an image of the
currently displayed plot or export the data in various formats:
plain text columns (for use with plotting tools like Gnuplot or XmGrace),
CSV data (suitable for processing in Microsoft Excel, LibreOffice Calc,
or Python with Pandas), or YAML (which can be imported into Python using PyYAML or Pandas).
You can use the mouse to zoom in on the graph by holding the left button and dragging
to select an area.  To zoom out, right-click anywhere on the graph.  You can reset the view
by clicking the \guicmd{lens} button located next to the data drop-down menu.

\subsection{Preferences}

The Preferences dialog allows customization of the behavior and appearance of
\lammpsgui{}.  Among other options:
\begin{itemize}
\item In the \guicmd{General Settings} tab, the \guicmd{Data update interval} setting
allows you to define the time interval, in milliseconds, between data updates during
a LAMMPS run.  By default, the data for the \guicmd{Charts} and \guicmd{Output}
windows is updated every 10 milliseconds.  Set this to 100 milliseconds or more
if \lammpsgui{} consumes too many resources during a run.  The \guicmd{Charts update interval}
controls the time interval between redrawing the plots in the \guicmd{Charts} window, in milliseconds.
\item The \guicmd{Accelerators} tab enables you to select an accelerator package
for LAMMPS to use.  Only settings supported by the LAMMPS library and local hardware
are available.  The \guicmd{Number of threads} field allows you to set the maximum
number of threads for accelerator packages that utilize threading.
\item The \guicmd{Editor Settings} tab allows you to adjust the settings of the editor
window.  Select the \guicmd{Auto-save on Run and Quit} option to automatically save changes
made to the \flecmd{.lmp} file upon closing \lammpsgui{}.
\end{itemize}
See Ref.\,\citenum{lammps_gui_docs} for a full list of options.

\section{Running LAMMPS on the Command-Line without the GUI}
\label{command-line-label}

LAMMPS can also be executed from the command-line on Linux, macOS, and
Windows without using the GUI.  This is the more common way to run LAMMPS.
Both, the \lammpsgui{} program and the LAMMPS command-line executable
utilize the same LAMMPS library and thus no changes to the input file are required.

First, open a terminal or command-line prompt window and navigate to the
directory containing the \flecmd{input.lmp} file.  Then execute:
\begin{lstlisting}[language=tcl]
lmp -in input.lmp
\end{lstlisting}
where \flecmd{lmp} is the command-line LAMMPS command.

For parallel execution with 4 processors (via OpenMP threads where supported
by the OPENMP package), use:
\begin{lstlisting}[language=tcl]
lmp -in input.lmp -pk omp 4 -sf omp
\end{lstlisting}

\begin{note}
  Running in parallel via MPI requires a specially compiled LAMMPS
  package and is not supported by the GUI.  On supercomputers or HPC
  clusters, pre-compiled LAMMPS executables are typically provided
  by the facility's user support team.  For more information, please
  refer to the facility's documentation or contact its user support staff.
\end{note}

See Ref.\,\citenum{lammps_run_docs} for a complete description on how to
run LAMMPS.

\end{appendices}

\end{document}